\pdfoutput=1

\documentclass[12pt]{article}
\usepackage{graphicx}
\usepackage{epsfig}
\usepackage{amsfonts}
\usepackage[lofdepth,lotdepth]{subfig}
\usepackage{url}
\usepackage{setspace}

\begin{document}

\newcommand{\brk}{\vspace*{0.18in}}

% No page number on the title page
\thispagestyle{empty}

% Center the whole title page
\begin{center}

\brk

% Large font and bold face for the headline. Try to keep it at one or
% two lines. Headlines over two lines will mess up the spacing, and you have to
% manually finetune it. Note that the line break in the SOURCE CODE
% does not affect the line breaking in the output file. If you want
% hardcoded line breaks, you have to mark them with a double backslash (\\)

   {\large
	\textbf{
	Nonclassical Structures within the $N$-qubit Pauli Group \footnote{This is the doctoral thesis of Mordecai Waegell, advised by P.K. Aravind}
	}
   }

%
%\brk
%by

\brk
% insert your name here.
Mordecai Waegell \footnote{caiw@wpi.edu}

\brk

{\it Worcester Polytechnic Institute}

{\it Worcester, MA, USA}

%
%% All this is constant:
%\brk\brk
%A Thesis
%
%\brk
%Submitted to the Faculty
%
%\brk
%of the
%
%\brk
%WORCESTER POLYTECHNIC INSTITUTE
%	
%\brk
%In partial fulfillment of the requirements for the
%
%\brk
%Degree of Doctor of Philosophy
%
%\brk
%in
%
%\brk
%Physics
%
%\brk
%by
%
%% This is how LaTeX draws lines :) It's where your signature goes.
%\brk\brk
%\rule{3in}{1.2pt}
%
%% Adjust this to your preferred month and year
%\brk
%May 2013
%
%
\end{center}
%	
%\vfill
%\vfill
%APPROVED:
%\vfill
%\vfill
%%
%%\vspace{0.5in}
%%\rule{3in}{0.8pt}
%%
%%Professor P.K. Aravind, Doctoral Thesis Advisor
%%
%%\vspace{0.5in}
%%\rule{3in}{0.8pt}
%%
%%Professor Germano S. Iannacchione, Head of Department	
%
%\begin{center}
%\noindent\begin{tabular}{ll}
%\makebox[2.5in]{\hrulefill} & \makebox[2.5in]{\hrulefill}\\
%Professor P.K. Aravind, & Professor David Cyganski,\\
%Doctoral Thesis Advisor & Thesis Committee Member \\[8ex]% adds space between the two sets of signatures
%\makebox[2.5in]{\hrulefill} & \makebox[2.5in]{\hrulefill}\\
%Professor Germano S. Iannacchione, & Professor L.R. Ram-Mohan\\
%Head of Department & Thesis Committee Member \\
%\end{tabular}
%\end{center}
%
%
%% end of titlepage
%\newpage
%

\begin{abstract}
Structures that demonstrate nonclassicality are of foundational interest in quantum mechanics, and can also be seen as resources for numerous applications in quantum information processing - particularly in the Hilbert space of $N$ qubits.  The theory of entanglement, quantum contextuality, and quantum nonlocality within the $N$-qubit Pauli group is further developed in this thesis.  The Strong Kochen-Specker theorem and the structures that prove it are introduced and explored in detail.  The pattern of connections between structures that show entanglement, contextuality, and nonlocality is explained.  Computational search algorithms and related tools were developed and used to perform complete searches for minimal nonclassical structures within the $N$-qubit Pauli group up to values of $N$ limited by our computational resources.  Our results are surveyed and prescriptions are given for using the elementary nonclassical structures we have found to construct more complex types of such structures.  Families of nonclassical structures are presented for all values of $N$, including the most compact family of projector-based parity proofs of the Kochen-Specker theorem yet discovered in all dimensions of the form $2^N$, where $N\geq2$.  The applications of our results and their connection with other work is also discussed.
\end{abstract}

\pagenumbering{roman}

\clearpage
\tableofcontents
\listoffigures
\listoftables
\clearpage

\pagenumbering{arabic}
\setcounter{page}{1}

%\documentclass[12pt]{article}
%\usepackage{graphicx}
%\usepackage{epsfig}
%\usepackage{amsfonts}
%\usepackage[lofdepth,lotdepth]{subfig}
%
%\begin{document}

\section{Introduction}

The purpose of this thesis is to present and classify structures that can be used, according to the laws of quantum physics, to demonstrate phenomena that are entirely nonclassical in nature.  What we mean by nonclassical is that these phenomena have the ability to violate some laws of classical physics, where classical physics refers generally to a collection of physical theories that make exact predictions, and which also tend to conform to the notion of reality as perceived through human experience.  One might say they are the `common sense' theories.  

Realism is perhaps the most fundamental of the assumptions in classical theories, and is the notion that for any object, all observable physical properties of that object have a definite value, and that a measurement on that object simply reveals some of those values.  As a corollary we also assume that the values revealed by measurement of one physical property should be independent of what other physical properties are being measured simultaneously - we refer to this assumption as noncontextuality, where a complete set of simultaneously observable properties is called a context.  Relativity is a less intuitive realist theory than classical or statistical mechanics, but it also makes exact predictions, and among these is that causality must be obeyed.

Quantum mechanics is our most precise physical theory and no flaws have yet been discovered that threaten its internal consistency or veracity.  Throughout this text, we will introduce many structures that demonstrate how quantum mechanics can violate the laws and assumptions of classical theories.  Only by demanding that our naive classical intuition about how the universe works must be satisfied do we arrive at interpretational conflicts.  If we simply accept what nature tells us, then what is called for is an entirely different way to see the universe.

In his seminal work of 1964-67 \cite{Bell1,Bell2}, John S. Bell sounded the death knell of the local hidden variable theories proposed much earlier by Einstein, Podolsky, and Rosen \cite{EPR}.  He showed that one could construct systems of measurements whose outcomes could not be satisfactorily explained by any theory that required local causality.  His original demonstration depended on an inequality distinguishing the probabilities of measurement outcomes as predicted by any local theory from the probabilities as predicted by quantum mechanics.  In 1967 Kochen and Specker proved one of Bell's results without using an inequality \cite{KS}.  This theorem, which has become known as the Bell-Kochen-Specker (BKS) theorem, shows that particular systems of measurements cannot be simultaneously assigned noncontextual hidden variables such that one and only one outcome is predicted for each measurement.  This theorem makes no direct claims about locality, but instead presents us with another type of nonclassical phenomenon, now referred to by most as simply quantum contextuality.  In some ways quantum contextuality is already familiar to those versed in quantum physics, in that it is well understood that the list of possible outcomes for a measurement will depend explicitly on the choice of context for that measurement.  Classically we would expect that for any context we choose, the measurement simply reveals the pre-existing value of the measured observable, but the BKS theorem shows us that pre-existing values of this sort are internally inconsistent, and simply cannot be fixed for all contexts at once.  Hence, not just the list of possible outcomes, but the specific outcome that obtains must depend in some way on the choice of context.

The BKS theorem can be proved by finding a particular geometry of measurement contexts to which noncontexual hidden variables cannot be assigned.  The proof originally given by Kochen and Specker involved 117 projectors in a Hilbert space of dimension 3.  Since then, many other proofs have been given in various dimensions, and using far fewer projectors, and indeed a sort of competition developed to find the most compact system of projectors and contexts that prove the BKS theorem.  The current title-holder, obtained by Cabello and others \cite{Cabello18_9}, is a set of 18 projectors which form 9 orthogonal bases, or measurement contexts, in a Hilbert space of 4 dimensions.  In a Hilbert space of 3 dimensions, the smallest proof, given by Kochen and Conway, requires just 31 projectors \cite{peres1993quantum}.  It is likely that the 18 projectors of Cabello constitute the smallest proof of the BKS theorem possible in any dimension.  Any of these particular geometries can also be extended into a proof of Bell's nonlocality theorem, by constructing an experimental setup where parts of an entangled state are shared by two spacelike separated observers \cite{cabello2001bell, aravind2002bell}.  This method also has the virtue of proving Bell's original theorem without the use of inequalities, and proofs of this type have also come to be called proofs of the Greenberger, Horne, and Zeilinger (GHZ) theorem \cite{GHZ}.  It is worth noting that not all proofs of Bell's theorem without inequalities follow from BKS proofs, and that more compact structures like the ones first given by GHZ and Shimony in 1990 \cite{GHSZ} can suffice.

These minimal structures highlight what we consider to be the central issue of this thesis: proofs of the BKS and GHZ theorems depend entirely on geometry in Hilbert spaces, and constitute a family of nonclassical geometric objects.  Since any of these geometric objects might find a use in quantum information processing, it is natural to ask how many distinct nonclassical structures there are in Hilbert space.  Do the stringent geometric requirements only allow a small class of these structures, or do they exist in many and varied forms?  As we have shown in previous publications \cite{WA_60Rays,WA_600cell_2,MP_600cell,600cellWebsite}, the latter answer is the correct one.  Even if we restrict ourselves to critical structures (by critical we mean that the proof would fail if any context were removed from the set), we find that the number and variety of nonclassical geometries is staggering.  Just within the family of 60 projectors and 105 contexts that are generated from the 2-qubit Pauli group, we estimate that there could be as many as $10^{12}$ distinct critical proofs of the BKS theorem.  So we can say with confidence, that rather than being rare freaks of geometry, Hilbert space abounds with these nonclassical structures.  Unfortunately, their sheer plentitude also foils any hope we might have of classifying and counting all of them, even within this finite system of 2-qubits.

There is, however, an alternate type of structure that can be used to prove the BKS theorem without making any explicit use of projectors.  In these proofs, we only make use of the observables of the $N$-qubit Pauli group, in which each set of mutually commuting observables is a context.  The two well-known cases \cite{Mermin_SquareStar} are the Mermin Square (related to the 24 projectors of Peres \cite{Peres24}), and the Mermin Star (related to GHZ's proof of Bell's theorem without inequalities).  As we will examine in detail throughout this text, these structures can be decomposed into different types of constituent pieces, the most important being sets of mutually commuting observables whose overall product is $\pm I$ in the space of all $N$ qubits.  We call these sets {\it Identity Products} (IDs), and we have developed the general theory of how all BKS proofs of this alternate type can be built from sets of IDs in the space of $N$ qubits.

Furthermore, we will show that the number of distinct IDs that we need to consider is much more manageable, at least for relatively small numbers of qubits, and so in these cases we are able to count and classify all of the IDs of the $N$-qubit Pauli group.  Using these IDs in conjunction with several procedures we have developed, we obtain many distinct BKS proofs of this alternate projector-free form.  This is the main result of our research: there exists a fairly small family of minimal nonclassical structures from which a vast class of BKS proofs can be assembled according to fairly simple prescriptions.  This text is designed to introduce the readers to the relevant details of these structures and prescriptions, such that they will be enabled to build their own proofs of this type, as needed for any given application in quantum information processing.

As we will see, the structure of an ID is related to the type of entanglement in the joint eigenstates of all of the observables in the ID.  IDs are then the elementary nonclassical structures of the $N$-qubit Pauli group, in the sense that entanglement is a nonclassical phenomenon.  Furthermore, entangled IDs are the necessary ingredient for all Observable-based BKS proofs, and so they are also the elementary nonclassical structures used to rule out local and/or noncontextual hidden variables theories.  We also see that entanglement itself is the necessary ingredient for all such proofs with $N\geq 2$ qubits.

We focus on the nonclassical nature of these structures because this is precisely the criterion by which we qualify a quantum phenomenon as being a useful resource for quantum information processing.  After reviewing the elementary structures that were introduced by other researchers, we will present an enormous variety of genuinely new structures that can be used to prove the BKS and GHZ theorems, some of which we have discussed in previous publications \cite{WA_24Rays, WA_3qubits, WA_4qubits, WA_Nqubits}. and we will show how different types nonclassical structures are related to one another.  We provide here the theoretical tools to construct an enormous variety of these resources for computations involving $N$ qubits, using only the observables of the $N$-qubit Pauli group.

The remainder of this text is organized as follows.  In Chapter \ref{sec:q2}, we examine structures for 2 qubits, beginning with the Peres-Mermin set.  We then move on to examine the full 2-qubit Pauli group, as well as the 600-cell, a regular polytope from $\mathfrak{R}^4$ which can be projected into the Hilbert space of 2 qubits.  In Chapter \ref{sec:q3}, we examine a number of minimal structures within the 3-qubit Pauli group, and in Chapter \ref{sec:q4}, structures within the 4-qubit Pauli group.  Throughout the preceding Chapters, we will use a number of examples to gradually introduce the important features of the structures we are developing.  In Chapter \ref{sec:Structures}, we will review and formalize the definitions and details of these structures, which include IDs, Kernels, Composite Kernel Structures, and the Observable-based KS proofs discussed above.  In Chapter \ref{sec:Families}, we will introduce several important families of Observable-based KS proofs for all numbers of qubits $N\geq 2$.  Within one of these families we also introduce what we believe to be the most compact projector-based proofs of the BKS theorem in the Hilbert space of $N$-qubits, beginning with the 18 rays of Cabello for 2 qubits.  At the end of this Chapter, we will also discuss a number of other particularly
interesting structures.  In Chapter \ref{sec:Algorithms} we will discuss the computational algorithms we used to obtain and verify many of our results.  Finally, in Chapter \ref{sec:Conclusions} we will review our results and discuss some of the interesting questions for which we do not yet have answers.  We will also discuss how the huge class of nonclassical resources we have developed here may be useful in many quantum information processing applications.

As a final note, we should point out that while this text contains what we believe is a fairly comprehensive set of examples, the full results of our computational searches are simply too large to be included here.  We have made this additional data freely available to any interested parties on a website \cite{MainWebsite} hosted by WPI.  We also leave some details not central to the topics at hand to papers we have already published.

%
%
%\end{document}

%\documentclass[12pt]{article}
%\usepackage{graphicx}
%\usepackage{epsfig}
%\usepackage{amsfonts}
%\usepackage[lofdepth,lotdepth]{subfig}
%
%\begin{document}

\section{Two qubits}\label{sec:q2}

In this chapter, we discuss several interesting structures in the $d=4$ Hilbert space of two qubits.  We will begin by discussing the 24 rays of Peres and the Mermin Square, and how they are related to various proofs of the Bell-Kochen-Specker (BKS) Theorem.  We use this case as a guide to introduce various new concepts that will be discussed in other chapters.  Next we discuss a proof we call the 2-qubit Whorl, and the associated set of 40 rays.  Subsequently we will discuss the 60 ray system that arises from the complete 2-qubit Pauli Group, which contains the 24 rays of Peres as a subset, as well as the 40 rays of the 2-qubit Whorl.  We will then discuss the 60 real rays that arise from the 600-cell, a well known regular polytope in $\mathfrak{R}^4$, and show how it gives BKS proofs that are quite different from those yielded by the 60 complex rays of the 2-qubit Pauli group.  Finally, we will discuss the Pentagon Inequalities that can be obtained using the various ray-sets of this chapter.

\subsection{The 24 Rays of Peres and the Mermin Square; Observable-based KS proofs and Kernels}\label{sec:MerminSquare}
In 1991 \cite{Peres24} Asher Peres gave a set of 24 real rays in Hilbert space of $d=4$ dimensions (the space of two qubits) that prove the KS theorem \cite{KS}.  These rays are actually generated by a proof of the KS theorem that uses only the observables of the 2-qubit Pauli Group - the familiar Mermin Square \cite{Mermin_SquareStar} shown in Table \ref{MerminSquare}.  Each row and column of the Mermin Square is a set of 3 mutually commuting observables from the 2-qubit Pauli group ($\mathfrak{G}^2$) whose joint eigenbasis consists of 4 rays.  All 6 eigenbases taken together give all 24 rays, and this is the sense in which these rays are generated by the Mermin Square.  Here and throughout this text we will omit the implicit tensor product symbols for $N$ qubit observables of the Pauli group (for example $ZZX \equiv Z\otimes Z\otimes X$).
\begin{table}[h]
\begin{center} {
\begin{tabular}{|c|c|c|}
\hline
 $ZI$ & $IX$ & $ZX$ \\
 \hline
 $IZ$ & $XI$ & $XZ$ \\
 \hline
 $ZZ$ & $XX$ & $YY$ \\
\hline
\end{tabular} }
\end{center}
\caption[2-qubit Mermin Square]{The Mermin Square.  Each row and column is a set of 3 mutually commuting 2-qubit Pauli observables, where the $Z,X,Y,$ and $I$ denote the single-qubit Pauli matrices and identity.  The product of the 3 observables in any row or column is +I (the 2-qubit identity), except for the last row, whose product is -I.}
\label{MerminSquare}
\end{table}

The Mermin square rules out Noncontextual Hidden Variables Theories (NCHVTs) in the following way:  Each of the observables in the Mermin Square has eigenvalues $\pm1$.  We define the quantity $A$ to be the product of all 3 rows and all 3 columns of the Mermin Square.  In an NCHVT theory, we must assign one of these eigenvalues as a preexisting truth value to each of the 9 observables, which gives us a noncontextual value for $A$,
\begin{equation}
A_{NC} \equiv \lambda_1^2 \lambda_2^2 \lambda_3^2 \lambda_4^2 \lambda_5^2 \lambda_6^2 \lambda_7^2 \lambda_8^2 \lambda_9^2 = 1,
\end{equation}
where the subscripts indicate the 9 observables, and each eigenvalue is squared because each observable appears in one row and one column.  However, the quantum prediction is $A_Q = -1$, since there are 5 mutually commuting sets with product $+I$ and 1 with product $-I$.  The quantum prediction has been borne out by many experiments \cite{kirchmair2009state,bartosik2009experimental,amselem2009state,moussa2010testing}, which seems to convincingly rule out NCHVTs, proving the KS theorem.

This geometric structure is of pivotal importance because it is the smallest and simplest in a family of structures that prove the KS theorem for $N$ qubits.  This family contains a large but finite number of geometric structures which can be used to prove the KS and GHZ theorems for $N$ qubits, and which may have many other interesting applications in quantum information processing.  Just as there are a finite number of regular polyhedra in $\mathfrak{R}^3$, there are a finite number of these KS structures in $\mathfrak{G}^N$.  It is one of the main purposes of this text to characterize these structures, which we will return to in later chapters.\newline

We will take a detour here to deconstruct the Mermin Square into its fundamental components and to introduce several new structures that we will be using throughout this text.  In this case the components are the 3 rows and 3 columns, or 6 sets of 3 mutually commuting observables, each with overall product $\pm I$.  We call each of these sets an {\it Identity Product} (or ID for short), and we say that the Mermin Square is composed of 6 IDs.

Let us consider what properties must be satisfied by this set of IDs in order to guarantee that it proves the KS theorem.  First, the IDs themselves are essential components for the success of the KS proof because the product of the observed eigenvalues in an ID is state-independent.  Furthermore, an odd number of the IDs in the proof need to have product $-I$ in order to guarantee $A_Q = -1$.  Finally, each of the 2-qubit observables must appear in an even number of the IDs in order to guarantee $A_{NC} = 1$.  Any set of IDs with these properties gives a proof of the KS theorem, and as we will see the Mermin Square is just one of many such sets.  We will refer to sets of this type as {\it Observable-based KS proofs.}  We introduce the compact symbol $O-I$ to describe sets containing $O$ Observables and $I$ IDs, and the expanded symbol $O_x-I_y$ to indicate that each observable appears in $x$ IDs, while each ID contains $y$ observables.  The Mermin Square then has compact symbol $9-6$ and expanded symbol $9_2 - 6_3$.  We will say that an Observable-based KS proof is critical if no subset of qubits and/or IDs can be deleted such that the remaining set is still an Observable-based KS proof.

We have discovered a method for generating Observable-based KS proofs from $\mathfrak{G}^N$ from more compact sets of IDs that we call {\it Kernels}.  A given Kernel can be used to generate many distinct Observable-based KS proofs.  We further conjecture that all possible Observable-based KS proofs in $\mathfrak{G}^N$ are generated by at least one Kernel, since we have never found a counterexample.

A Kernel is any set of IDs that satisfies two conditions:  First, an odd number of the IDs must be negative (product $-I$).  Second, for each individual qubit, all single-qubit Pauli observables must appear an even number of times throughout the entire Kernel.  The Kernel that generates the Mermin square is given in Table \ref{q2kernel}.
\begin{table}[h]
\begin{center} {
\begin{tabular}{|c|}
\hline
$ZZ$\\
$XX$\\
$YY$\\
\hline
$ZX$\\
$XZ$\\
$YY$\\
\hline
\end{tabular} }
\end{center}
\caption[2-qubit Kernel]{The 2-qubit Kernel.  The top block is a negative ID, while the bottom block is a positive ID.  Each column represents a distinct qubit, and for each qubit the single-qubit Pauli observables each appear twice.}
\label{q2kernel}
\end{table}

These conditions guarantee that all Kernels can be used to generate a complete Observable-based KS proof, but the Kernel alone also has physical significance.

The usual formulation of the KS theorem, and of $N$-qubit Observable-based KS proofs, requires that noncontextual truth-values $\pm1$ be assigned to every $N$-qubit Observable, as in the case of the Mermin Square.  If we choose to demand that things be even more `classical' we can instead require that noncontextual truth-values $\pm1$ be assigned to every single-qubit observable ($Z_i$, $X_i$, $Y_i$ - $i$ denotes the qubit index) in a collection of $N$ qubits.  Any Kernel gives a proof that quantum mechanics rules out the existence of these truth-values, which can be seen from the following contradiction: on the one hand, because every single-qubit observable occurs an even number of times, in any noncontextual theory the overall product of the IDs in the Kernel must be +1 , while on the other hand, because an odd number of the IDs are negative, the quantum prediction (and experimental result) will always be -1.

We will refer to this as the Strong KS theorem.  The simplest proof of this theorem is given in Table \ref{q2kernel} and requires an experimental setup with only 2 qubits and 2 measurement settings.  The Strong KS theorem essentially proves that the $N$ qubits are entangled, and rules out NCHVTS {\it without} entanglement.  To see the distinction here, note that the same experiment fails to prove the usual KS theorem (in which only 2-qubit observables are assigned truth-values) - the Kernel alone is not an Observable-based KS proof, but it is an Observable-based Strong KS proof.

In the special case of single-ID Kernels (which exist only for $N\geq3$ qubits), an experiment can be set up that uses locality to require noncontextual truth-values be assigned to the single-qubit observables \cite{GHZ, GHSZ} in order to prove Bell's theorem without inequalities (also called the GHZ theorem).  The simplest example of such an ID is shown in Table \ref{GHZKernel}.

Now we return to the assertion that each Kernel can generate Observable-based KS proofs.  In general a given Kernel can be used in many different ways to build such proofs by adding more IDs to the set, but there is one trivial procedure by which any Kernel can be extended into an Observables-based KS proof.  To do it we take each observable in the Kernel that appears an odd number of times and supplement it with its own single-qubit decomposition to form a new positive ID.  For example, we take $ZZ$, and supplement it with $ZI$ and $IZ$ to form a new ID.  It should be easy to see that applying this procedure to the Kernel of Table \ref{q2kernel} leads to the Mermin Square of Table \ref{MerminSquare}.  Because of how Kernels are defined, this procedure always results in a set of IDs that satisfy the requirements to be an Observable-based KS proof.  For the Kernel in Table \ref{q2kernel} there is actually only one other type of 2-qubit Observable-based KS proof that can be constructed, as shown in Figure \ref{q2Whorl}.\newline

\begin{figure}
  \centering
  \includegraphics[width=5in]{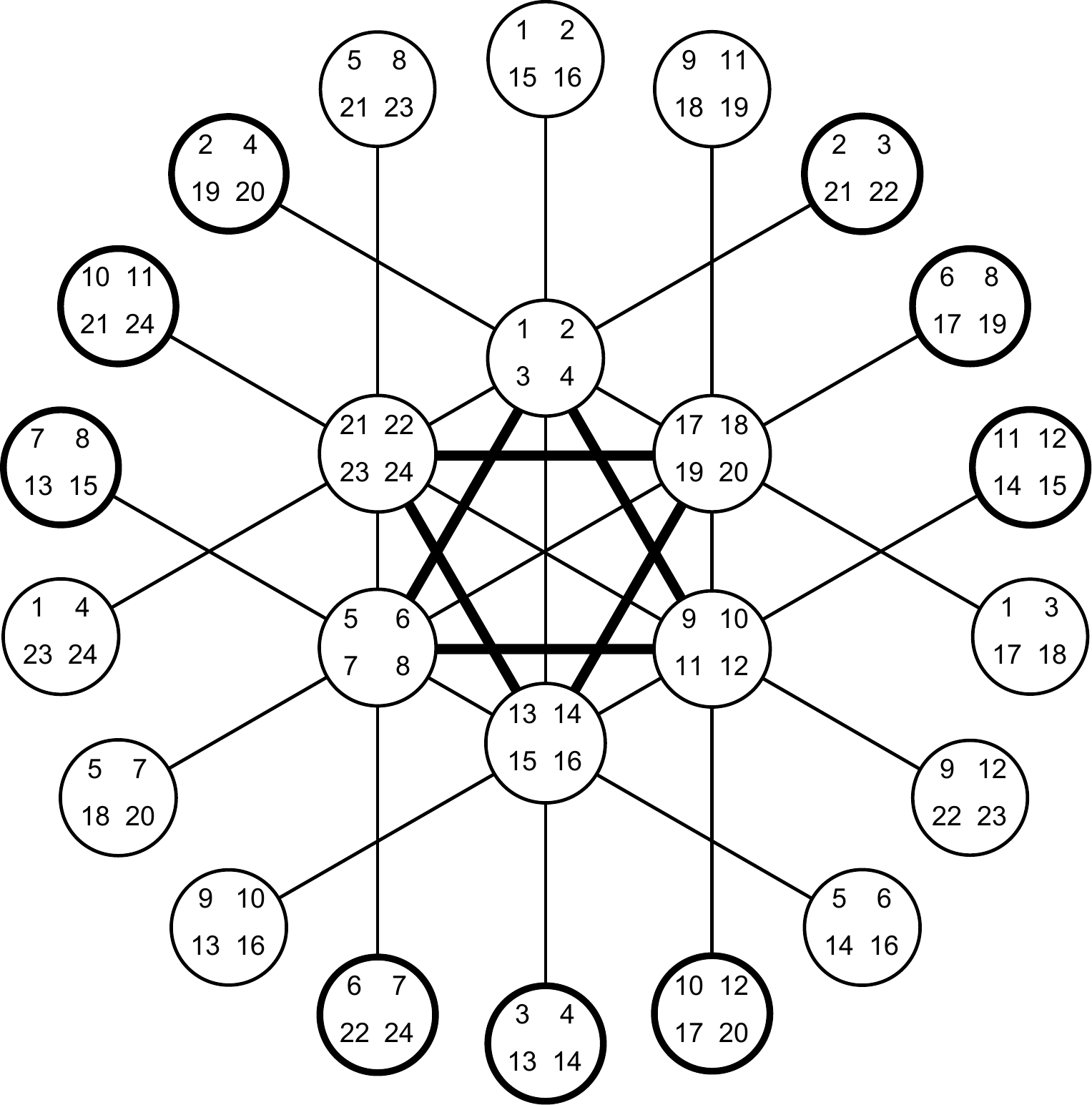}\\
  \caption[Basis Diagram of Peres $24-24$ set]{This figure shows the 24 bases of the Peres set, with the rays indexed 1 through 24.  The 6 inner bases are the eigenbases of the Mermin Square, and the triples of these bases joined by the thick triangles are mutually unbiased.  The 18 outer bases are hybrid bases which appear in 9 complementary pairs.  Each hybrid basis is formed of two rays from each of two different non-unbiased eigenbases, while the complementary basis is composed of the other two rays from each eigenbasis, as shown by the thin lines.  The 9 bases shown in thick circles are one example of an $18-9$ parity proof, while the 15 bases shown in thin circles are the complementary $24-15$ parity proof.}\label{Peres24_24}
\end{figure}

That concludes the detour.  We now transition from discussion of the Mermin Square to discussion of the 24 rays of Peres that are generated by it.  The 24 rays are the eigenbases of the 6 IDs of the Mermin square, but orthogonality between some of these rays also causes them to form 18 additional bases, which we call hybrid bases.  Each pair of IDs in the Mermin Square that share an observable in common generate one complementary pair of hybrid bases in addition to their pair of eigenbases.  One of the hybrid bases will have 2 rays from one eigenbases and 2 from other eigenbases.  The complementary hybrid basis has the other 2 rays from each eigenbasis.  This can be seen in Figure \ref{Peres24_24}, where the 6 inner bases are the eigenbases of the Mermin Square, and the 18 outer bases are the hybrid bases they form, with complementary pairs joined to their generating eigenbases by lines.

So altogether, the 24 rays of Peres form 24 bases.  We now introduce a system of symbols that we will use to describe sets of rays and bases of this type.  The most compact symbol we will use is simply $R-B$, where R and B are the number of rays and bases respectively.  Thus, the Peres set has the compact symbol $24-24$.  The simplest expanded symbol is of the form $R_m-B_d$ where the {\it multiplicity} $m$ is the number of bases each ray occurs in, and $d$ is the number of rays in each basis.  The expanded symbol for this set is $24_4-24_4$.  If a set has rays of different multiplicities then they will be separated in the number before the dash.  For example, the set $18_2 2_4 -11_4$ has 11 bases and 20 rays, with 18 rays appearing in 2 of the bases, and 2 rays appearing in 4 of the bases.

Now, this $24-24$ set is a proof of KS theorem because the 24 individual projectors cannot be assigned simultaneous (noncontextual) truth-values 0 and 1 such that each basis has exactly one ray valued 1 and all other rays valued 0.  The direct proof of this is slightly cumbersome, though it does follow directly from the Mermin Square.  This set is not critical, however we can obtain critical KS sets by discarding some of the bases in this set.  By critical here we mean that if any complete basis were to be discarded from the set, it would no longer prove the KS theorem.  Our exhaustive computer searches have found that there are in fact $2^9 = 512$ critical KS sets within the $24-24$ \cite{WA_24Rays, pavivcic2010new}.  The most compact proof, first discovered by Cabello et al, has the symbol $18_2-9_4$, and is almost certainly the most compact KS proof possible in any dimension.  Table \ref{Peres24_24_KS Sets} shows the symbols for the different types of KS sets among these 512.  Each of these sets has an odd number of bases, but all rays have even multiplicity.  We call KS sets with this property {\it parity proofs}, because they make the proof of the KS theorem transparently obvious.  Because the rays all have even multiplicities, a noncontextual truth-value assignment must have an even number of 1s.  However according to QM each of the (odd) number of bases will have exactly one ray with truth value 1, so the truth-value assignment would require an odd number of 1s.  The incompatibility of these two conditions shows that such an assignment is impossible and proves the KS theorem.

\begin{table}[h]
\begin{center} {
\begin{tabular}{|ccc|}
\hline
Compact Symbol & Expanded Symbol & \# of Proofs \\
\hline
$18-9$ & $18_2-9_4$ & 16 \\
$20-11$ & $18_2 2_4 -11_4$ & 240 \\
$22-13$ & $18_2 4_4 -13_4$ & 240 \\
$24-15$ & $18_2 6_4 -15_4$ & 16 \\
\hline
\end{tabular} }
\end{center}
\caption[Critical Parity Proofs of the $24-24$ set of Peres]{The 512 Critical Parity Proofs of the KS Theorem within the $24-24$ set of Peres.  These occur in 256 complementary pairs, in the sense that if the bases of one proof are removed from the full set of 24 bases, then the remaining bases form the complementary proof.  Each $18-9$ proof is complementary to a $24-15$ proof, and each $20-11$ proof to a $22-13$ proof.}
\label{Peres24_24_KS Sets}
\end{table}

We have further discovered a particularly simple way to construct all $2^9$ proofs in this set, which also in some sense explains why there are this number of KS sets.  Recall that the 18 hybrid bases can be divided into 9 complementary pairs.  There are exactly $2^9$ ways to choose one member of each pair, and it turns out that each of these choices leads to one of the KS sets.  Once these 9 hybrid bases are chosen, some of the eigenbases may need to be added to complete the set.  Eigenbases are added for all rays that have odd multiplicity among the 9 hybrid bases, until all rays have even multiplicity.  The result of this process gives all 512 critical parity proofs, and also makes the complementary pairs of proofs quite obvious.  It also explains why all 512 proofs contain exactly 9 hybrid bases.

To see how a proof is built explicitly, consider Figure \ref{Peres24_24}.  For each of the 9 thin lines, choose one of the two hybrid bases at the ends of the line - there are $2^9$ ways to make this choice.  Now, among the 9 bases you have chosen, find all rays that appear once or thrice, and add all eigenbases containing those rays to the set.  This process always results in a complete parity proof.

Finally let us turn to a discussion of the geometry of the Peres set.  Each of the 24 rays is a real-valued vector in a projective Hilbert space of $d=4$.  As such, each ray corresponds to a pair of antiparallel vectors in $\mathfrak{R}^4$, and so the full set contains 48 real vectors.  We can divide the rays into two groups of 12, taking each group to be the set of rays belonging to each of the triple of mutually unbiased bases in Figure \ref{Peres24_24}.  Each of these sets of 12 rays corresponds to 24 vectors in $\mathfrak{R}^4$, and each of these sets of vectors corresponds to a well known regular polytope called the 24-cell.  The 24-cell is a self-dual polytope (The dual of a 4-polytope is obtained by placing a vertex at the center of each cell), meaning that the dual polytope of a 24-cell is another 24-cell.  The two groups of 24 vectors are exactly such a dual-pair of polytopes.  It is worth noting that each 24-cell is a vertex-transitive figure (a property of all regular polytopes), as is the full $24-24$ set, which implies in some sense that this is a maximally symmetric set.  The symmetry group of the 24-24 has 1152 elements, and many of the proofs in this set are simply replicas of a few distinct types under the symmetry of the set.

The 12 rays of each 24-cell can also be associated with the well known Reyes Configuration, by defining {\it lines} as sets of 3 linearly dependent rays.  For detailed discussion of how the Reyes Configuration is related to the geometry of the KS parity proofs found within this $24-24$ set, refer to the paper \cite{WA_24Rays}.

\pagebreak

\subsection{The 40 Rays of the 2-qubit Whorl}\label{sec:q2Whorl}
\begin{figure}[h]
  \centering
  \includegraphics[width=4.5in]{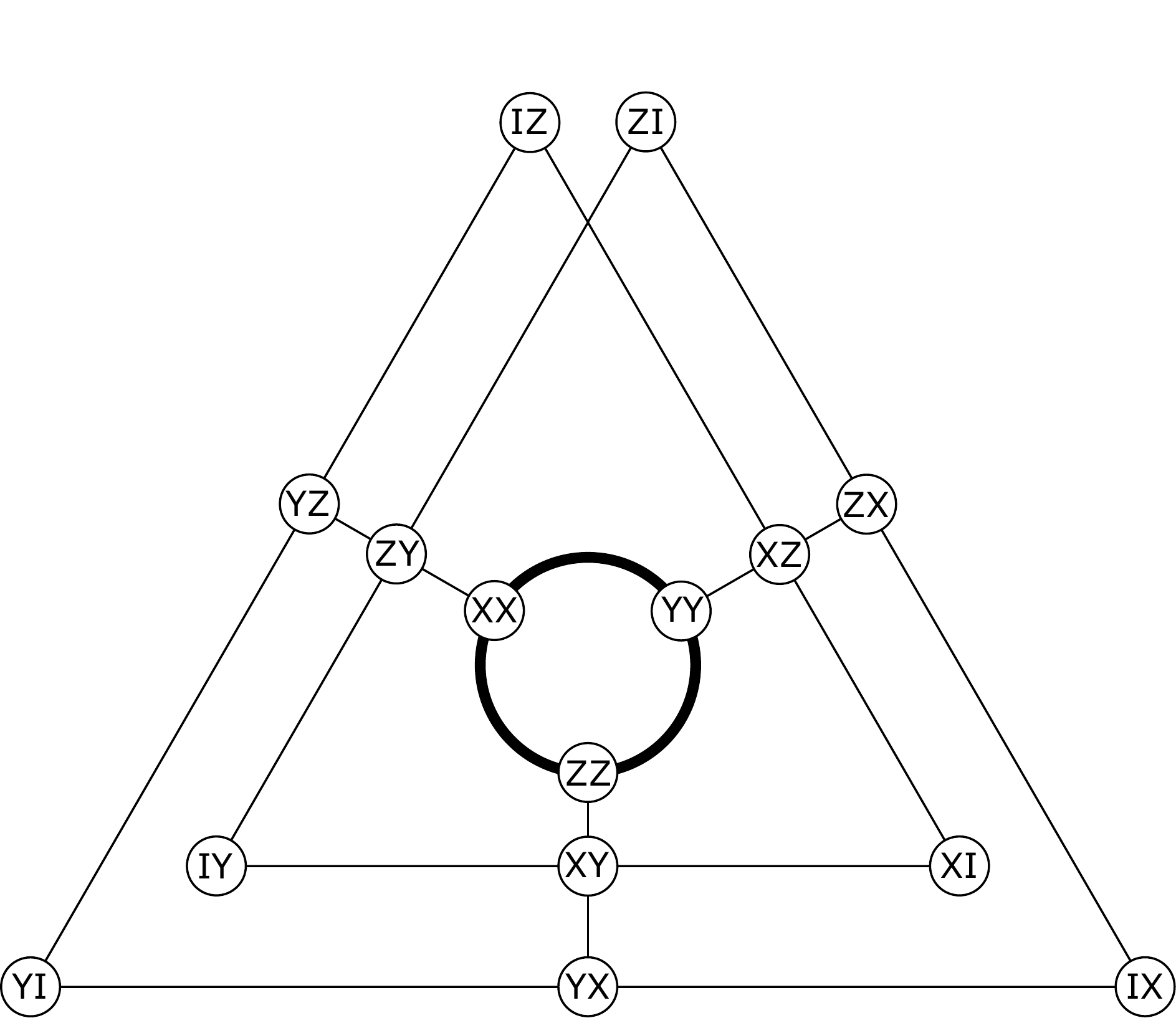}\\
  \caption[2-qubit Whorl]{The 2-qubit Whorl is a $15_2 - 10_3$ Observable-based KS proof.  In this diagram, the straight lines are positive IDs and the circle is a negative ID.  If $A$ is defined as the product of all IDs in the set, then QM predicts $A_Q = -1$ because only one ID is negative, while an NCHVT predicts $A_{NC} = 1$ because each observable appears in two IDs, which proves the KS theorem.}\label{q2Whorl}
\end{figure}
\begin{figure}[ht]
  \centering
  \includegraphics[width=3.5in]{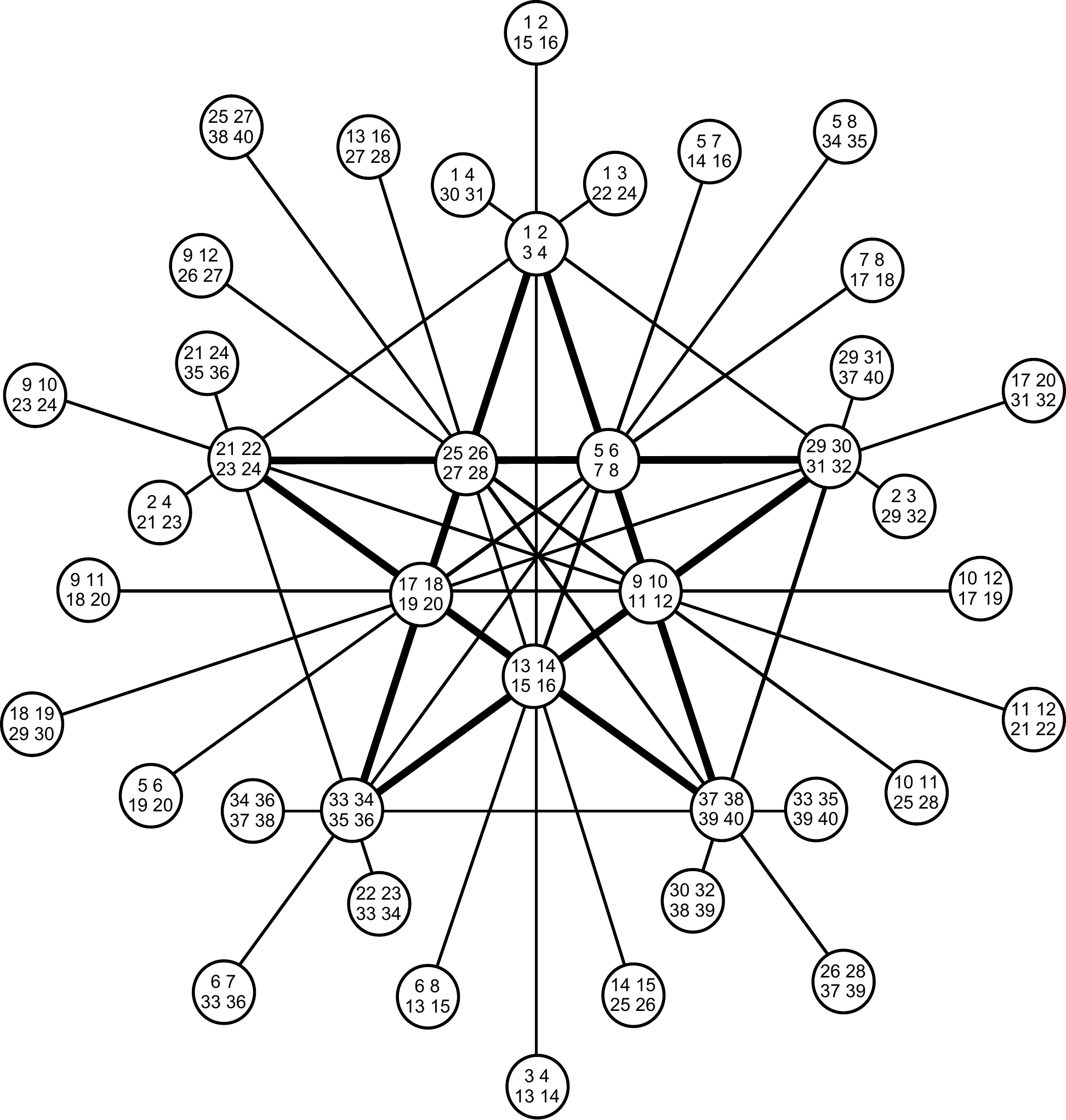}\\
  \caption[Basis Diagram for the 2-qubit Whorl]{This figure shows the 40 bases of the $40-40$ set, with the rays indexed 1 through 40.  The 10 inner bases are the eigenbases of the 2-qubit Whorl, and the 4 bases joined by each thick line are mutually unbiased.  The 30 outer bases are hybrid bases which appear in 15 complementary pairs.  Each hybrid basis is formed of two rays from each of two different non-unbiased eigenbases, while the complementary basis is composed of the other two rays from each eigenbasis, as shown by the thin lines.}\label{BasisDiagram_40_40}
\end{figure}

Next we will discuss the only other type of Observable-based KS proof that exists within the 2-qubit Pauli Group.  This set contains all 15 observables of the 2-qubit Pauli Group, and 10 IDs, and has symbol $15_2 - 10_3$.  We call this set the {\it 2-qubit Whorl} because of the structure of its diagram, as seen in Figure \ref{q2Whorl}.  It is worth noting that both the Mermin square and 2-qubit Whorl can be generated from the same Kernel (Table \ref{q2kernel}).  As we will show, the Whorl is essentially a less compact sibling of the Mermin square, with a striking number of identical features.

We will take a moment here to generally introduce the diagrams that we will use throughout this text to depict Observable-based KS proofs.  The rules for the diagrams are simple, and guarantee that the observables they contain prove the KS theorem:  Each ID is represented by a line, arc, or circle that connects all of the observables in the ID.  Thin lines denote a positive ID, while thick lines denote a negative ID.  An odd number of lines must be thick, and each observable must appear on an even number of lines.  The diagram of Figure \ref{q2Whorl} is the first of many diagrams in this thesis that conforms to these rules and so provides a proof of the KS theorem.\newline

Next we consider the set of 40 rays that are generated by the 2-qubit Whorl \cite{WA_60Rays}.  To begin we will examine all of the hybrid bases that are formed by this set of 10 eigenbases.  For each pair of IDs in the set that share an observable in common, we will obtain a single complementary pair of hybrid bases, just as before.  Each of the 15 observables appears in just one such pair of IDs, and so there are 15 complementary pairs of hybrid bases.  So this set of 40 rays form 40 bases, with symbol $40_4 - 40_4$, as shown in Figure \ref{BasisDiagram_40_40}.  Again, direct proof that this is a KS set is somewhat cumbersome, though it does follow from the 2-qubit Whorl.  Again, the set is not critical, but we can obtain critical KS sets by discarding some of the bases from the set.  An exhaustive computer search shows there are exactly $2^{15} = 32,768$ KS sets within the $40_4-40_4$, and that all of them are parity proofs.  Furthermore, these proofs break down into $2^{14}$ complementary pairs, in the same sense as the proofs of the $24-24$ set.  Table \ref{Whorl40_40_KS_Sets} shows the symbols of the proofs and their complements.
\begin{table}[h]
\begin{center} {
\begin{tabular}{|c|c|c|}
\hline
Parity Proof & Complementary Proof & \# of Proofs \\
\hline
$30-15$ $(30_2-15_4)$ & $40-25$ $(30_2 10_4 -25_4)$ & 64 \\
$32-17$ $(30_2 2_4-17_4)$ & $38-23$ $(30_2 8_4-23_4)$ & 2,880 \\
$34-19$ $(30_2 4_4 -19_4)$ & $36-21$ $(30_2 6_4-21_4)$ & 13,440 \\
\hline
\end{tabular} }
\end{center}
\caption[Critical Parity Proofs of the 2-qubit Whorl]{The 32,768 Critical Parity Proofs of the KS Theorem within the $40-40$ set.  Each set of proofs is shown alongside the set of complementary proofs, with the expanded symbols in parentheses.}
\label{Whorl40_40_KS_Sets}
\end{table}

Again, all $2^{15}$ parity proofs can be constructed by the same simple method we used for the $2^9$ proofs of the $24-24$ set:  There are $2^{15}$ ways to choose one member of each of the 15 complementary pairs of hybrid bases in this set, and each of these choices leads to a parity proof.  Once these 15 hybrid bases are chosen, some of the eigenbases may need to be added to complete the set.  Eigenbases are added for all rays that have odd multiplicity among the 15 hybrid bases, until all rays have even multiplicity.  The result of this process gives all 32,768 critical parity proofs, and again makes the complementary pairs of proofs quite obvious.  It also explains why all 32,768 proofs have exactly 15 hybrid bases.

Finally, the $40-40$ set has a symmetry group with 1,920 elements, and many of the proofs are simply replicas of a relatively small number of types, under the symmetry of the set.

\pagebreak

\subsection{The 60 Rays of the 2-qubit Pauli Group}\label{sec:q2Pauli}

The complete 2-qubit Pauli Group contains 15 observables, which form 15 mutually commuting triplets - all of which are IDs.  In the previous sections, we discussed the Mermin square, an Observable-based KS proof composed of 6 of these 15 IDs, as well as the 2-qubit Whorl, which is composed of 10 of the 15 IDs.  In this section we will discuss the complete $15_3 - 15_3$ set, and the 60 rays that it generates.  To begin, we will examine all of the hybrid bases that are formed by this set of 15 eigenbases.  Again each pair of IDs in the set that share an observable in common give us a complementary pair of hybrid bases.  Each observable appears in 3 of the 15 IDs, and so belongs to 3 such pairs of IDs, giving us a total of 45 complementary pairs of hybrid bases.  So the set of 60 rays form a total of 105 bases.  This set has expanded symbol $60_7 - 105_4$, and is clearly a (noncritical) KS set because it contains the $18-9$ proof as a subset.\newline

The $60-105$ set is also vertex-transitive, and has a symmetry group of 11,520 elements.  It is worth noting that this set contains 10 symmetric replicas of the $24-24$ set as subsets, as well as 6 symmetric replicas of the $40-40$.  Likewise, there are 10 distinct ways to build a symmetric version of the Mermin Square using the 2-qubit Pauli Operators, and 6 distinct ways to build the 2-qubit Whorl.  Refer to our paper \cite{WA_60Rays} for a more detailed account of other geometric structures and related subsets within the $60-105$ set.\newline

We conducted a large, but nowhere near exhaustive, computer search of parity proofs within this set, and we found critical parity proofs ranging from the $18-9$ all the way up to $60-41$.  An alternate approach was used to find some additional KS sets that are not parity proofs, but none with less than 9 bases or more than 41 bases.  We enumerated roughly $1.5 \times 10^8$ distinct parity proofs in our search, and estimate that there are at least $10^9$ in total.  Files containing this data and additional details about this set can be obtained from this website \cite{MainWebsite}.

\pagebreak

\subsection{The 600-cell}\label{sec:600cell}

The 600-cell is a well known regular polytope in $\mathfrak{R}^4$.  It is composed of 120 vertices which come in 60 antipodal pairs.  Taking the set of vectors that point from the center of the 600-cell to its vertices, we obtain 60 pairs of antiparallel vectors, which correspond to 60 real rays in a projective Hilbert space of $d=4$.  This set is vertex transitive, and has a symmetry group with 7200 elements.  The 60 rays and their components are given in Table \ref{RayTable60_75}.

\begin{table}[h]
\begin{center} {
\begin{tabular}{|c c c c|} % centered columns (4 columns)
\hline % inserts single horizontal line
1 = $2 0 0 0 $ & 2 = $0 2 0 0 $ & 3 = $0 0 2 0 $ & 4 = $0 0 0 2 $ \\
5 = $1 1 1 1 $ & 6 = $1 1 \overline{1} \overline{1} $ & 7 = $1 \overline{1} 1 \overline{1} $ & 8 = $1 \overline{1} \overline{1} 1 $ \\
9 = $1 \overline{1} \overline{1} \overline{1} $ & 10 = $1 \overline{1} 1 1 $ & 11 = $1 1 \overline{1} 1 $ & 12 = $1 1 1 \overline{1} $ \\
\hline
13 = $\kappa 0 \overline{\tau} \overline{1} $ & 14 = $0 \kappa 1 \overline{\tau} $ & 15 = $\tau \overline{1} \kappa 0 $ & 16 = $1 \tau 0 \kappa $ \\
17 = $\tau \kappa 0 \overline{1} $ & 18 = $1 0 \kappa \tau $ & 19 = $\kappa \overline{\tau} \overline{1} 0 $ & 20 = $0 1 \overline{\tau} \kappa $ \\
21 = $1 \kappa \tau 0 $ & 22 = $\tau 0 \overline{1} \kappa $ & 23 = $0 \tau \overline{\kappa} \overline{1} $ & 24 = $\kappa \overline{1} 0 \overline{\tau} $ \\
\hline
25 = $\tau 0 1 \kappa $ & 26 = $0 \tau \overline{\kappa} 1 $ & 27 = $1 \overline{\kappa} \overline{\tau} 0 $ & 28 = $\kappa 1 0 \overline{\tau} $ \\
29 = $0 \kappa 1 \tau $ & 30 = $\tau 1 \overline{\kappa} 0 $ & 31 = $\kappa 0 \tau \overline{1} $ & 32 = $1 \overline{\tau} 0 \kappa $ \\
33 = $\tau \overline{\kappa} 0 \overline{1} $ & 34 = $0 1 \overline{\tau} \overline{\kappa} $ & 35 = $1 0 \overline{\kappa} \tau $ & 36 = $\kappa \tau 1 0 $ \\
\hline
37 = $\tau 0 \overline{1} \overline{\kappa} $ & 38 = $0 \tau \kappa \overline{1} $ & 39 = $1 \overline{\kappa} \tau 0 $ & 40 = $\kappa 1 0 \tau $ \\
41 = $\tau 1 \kappa 0 $ & 42 = $0 \kappa \overline{1} \overline{\tau} $ & 43 = $1 \overline{\tau} 0 \overline{\kappa} $ & 44 = $\kappa 0 \overline{\tau} 1 $ \\
45 = $0 1 \tau \kappa $ & 46 = $\tau \overline{\kappa} 0 1 $ & 47 = $\kappa \tau \overline{1} 0 $ & 48 = $1 0 \kappa \overline{\tau} $ \\
\hline
49 = $\kappa 0 \tau 1 $ & 50 = $0 \kappa \overline{1} \tau $ & 51 = $\tau \overline{1} \overline{\kappa} 0 $ & 52 = $1 \tau 0 \overline{\kappa} $ \\
53 = $1 0 \overline{\kappa} \overline{\tau} $ & 54 = $\tau \kappa 0 1 $ & 55 = $0 1 \tau \overline{\kappa} $ & 56 = $\kappa \overline{\tau} 1 0 $ \\
57 = $\tau 0 1 \overline{\kappa} $ & 58 = $1 \kappa \overline{\tau} 0 $ & 59 = $\kappa \overline{1} 0 \tau $ & 60 = $0 \tau \kappa 1 $ \\
\hline %inserts single line
\end{tabular} }
\end{center}
\caption[Rays of the 600-cell]{The components of 60 Rays of the 600-cell.  $\tau = (1+\sqrt{5})/2$ is the Golden Ratio, and $\kappa$ its inverse.  Commas are omitted between components, and the overbar on a number denotes its negative.}\label{RayTable60_75} % is used to refer this table in the text
\end{table}

The 600-cell is not directly related to the 2-qubit Pauli Group, but its 60 rays still form 75 orthogonal bases in total, and the set has symbol $60_5 - 75_4$.  This set is also geometrically related to the $24-24$ set.  Recall that those 24 rays can be divided into two groups of 12 that each correspond to a 24-cell, another regular 4-polytope.  In a similar fashion, the 60 rays can be divided into 25 groups of 12 that each correspond to a 24-cell, but no 2 of these 25 are a dual-pair.  Instead, it is possible to form 10 different tilings of the 60 rays, each composed of 5 mutually disjoint 24-cells.  Table \ref{BasisTable60_75} shows the 75 basis broken down into 25 blocks of 3, with each block corresponding to a 24-cell.  The rows and columns of blocks show the 10 tilings.

\begin{table}[h]
\begin{center} {
\begin{tabular}{|c c c c | c c c c | c c c c | c c c c | c c c c| c c c c|} % centered columns (4 columns)
\hline % inserts single horizontal line

 1 & 2 & 3 & 4 & 31 & 42 & 51 & 16 & 22 & 60 & 39 & 28 & 57 & 23 & 27 & 40 & 44 & 29 & 15 & 52 \\
 5 & 6 & 7 & 8 & 38 & 24 & 58 & 25 & 18 & 47 & 33 & 55 & 36 & 53 & 20 & 46 & 59 & 26 & 37 & 21 \\
 9 & 10 & 11 & 12 & 56 & 45 & 17 & 35 & 13 & 32 & 50 & 41 & 43 & 49 & 30 & 14 & 34 & 19 & 48 & 54 \\
 \hline
 13 & 14 & 15 & 16 & 43 & 54 & 3 & 28 & 34 & 12 & 51 & 40 & 9 & 35 & 39 & 52 & 56 & 41 & 27 & 4 \\
 17 & 18 & 19 & 20 & 50 & 36 & 10 & 37 & 30 & 59 & 45 & 7 & 48 & 5 & 32 & 58 & 11 & 38 & 49 & 33 \\
 21 & 22 & 23 & 24 & 8 & 57 & 29 & 47 & 25 & 44 & 2 & 53 & 55 & 1 & 42 & 26 & 46 & 31 & 60 & 6 \\
\hline
 25 & 26 & 27 & 28 & 55 & 6 & 15 & 40 & 46 & 24 & 3 & 52 & 21 & 47 & 51 & 4 & 8 & 53 & 39 & 16 \\
 29 & 30 & 31 & 32 & 2 & 48 & 22 & 49 & 42 & 11 & 57 & 19 & 60 & 17 & 44 & 10 & 23 & 50 & 1 & 45 \\
 33 & 34 & 35 & 36 & 20 & 9 & 41 & 59 & 37 & 56 & 14 & 5 & 7 & 13 & 54 & 38 & 58 & 43 & 12 & 18 \\
\hline
 37 & 38 & 39 & 40 & 7 & 18 & 27 & 52 & 58 & 36 & 15 & 4 & 33 & 59 & 3 & 16 & 20 & 5 & 51 & 28 \\
 41 & 42 & 43 & 44 & 14 & 60 & 34 & 1 & 54 & 23 & 9 & 31 & 12 & 29 & 56 & 22 & 35 & 2 & 13 & 57 \\
 45 & 46 & 47 & 48 & 32 & 21 & 53 & 11 & 49 & 8 & 26 & 17 & 19 & 25 & 6 & 50 & 10 & 55 & 24 & 30 \\
\hline
 49 & 50 & 51 & 52 & 19 & 30 & 39 & 4 & 10 & 48 & 27 & 16 & 45 & 11 & 15 & 28 & 32 & 17 & 3 & 40 \\
 53 & 54 & 55 & 56 & 26 & 12 & 46 & 13 & 6 & 35 & 21 & 43 & 24 & 41 & 8 & 34 & 47 & 14 & 25 & 9 \\
 57 & 58 & 59 & 60 & 44 & 33 & 5 & 23 & 1 & 20 & 38 & 29 & 31 & 37 & 18 & 2 & 22 & 7 & 36 & 42 \\
\hline %inserts single line
\end{tabular} }
\end{center}
\caption[Bases of the 600-cell]{The 75 Bases of the 600-cell, using the ray-indexes of Table \ref{RayTable60_75}.  Each group of 4 rays in a row is a basis.  Each group of 3 bases in a block is a 24-cell.  Each row and column of blocks is a decomposition of the 60 rays into 5 disjoint sets of 12 (the 24-cells).  The bases are organized so that a single period-5 rotation generates all of the blocks in each row, while another period-5 rotation generates the blocks in each column.}\label{BasisTable60_75} % is used to refer this table in the text
\end{table}

We conducted a large, but not exhaustive, computer search for parity proofs within this $60-75$ set \cite{WA_600cell_2}, and found proofs ranging from $26-13$ all the way up to $60-41$.  Based on the number of proofs we were able to enumerate, we estimate that there are roughly $10^8$ critical KS parity proofs within this set.  Another algorithm was also used to explore this system which finds critical KS sets of all types, not only parity proofs \cite{MP_600cell}.  Based on the results of this search we estimate that the total number of critical KS sets within the $60-75$ could be as large as $10^{12}$.  To see examples of every type of parity set that we have enumerated, and also how they are related to various geometric aspects of this figure, visit our Quantum Coloring website \cite{600cellWebsite}.

While the $26-13$ parity proof is the most compact proof within this set, the 26 rays are not vertex transitive.  There are two other types of proofs contained here which are of particular interest because they are vertex transitive and have other interesting geometric properties.  Both of these proofs have symbol $30_2 - 15_4$, and the 60 rays can be divided into one proof of each type in 120 different ways (which is to say that the proofs come in 120 complementary pairs).  Coxeter \cite{Coxeter} gave a particularly nice decomposition of the 120 vertices of the 600-cell into 4 disjoint groups of 30 vertices, with each group generated by a different period-30 rotation.  He called this the triacontagonal projection of the 600-cell.  It turns out that 2 of these triacontagons taken together correspond to one type of $30-15$ proof, while the other 2 triacontagons correspond to the complementary $30-15$ proof.  These sets can also be derived using arguments that follow from the Reyes Configuration of the 25 24-cells in this set, which is shown in our paper \cite{WA_600cell_1} and on the website.\newline

Finally, and somewhat tangentially, let us discuss the geometric solids corresponding to these $30-15$ proofs when they are projected back into $\mathfrak{R}^4$.  Each ray gives rise to a pair of antiparallel vectors, so that each set of 30 rays corresponds to 60 vectors.  If we define Edges between the nearest neighbors in such a set of 60 vertices, we find that each set contains 180 Edges.  Here is where the differences between the two 30-15 sets will become obvious, and we will refer to them as sets A and B respectively.

For set A, the Edges can be used to form 180 equilateral triangle Faces, which break up into 2 disconnected sets of 90 Faces each.  The 90 Faces in each set form 30 tetrahedral Cells, which form a single closed ring of shared Faces.  That is, each of the 30 Cells shares one face with each of its two neighbor Cells in the ring, and its other two faces are external.  Taken together then, the 30 tetrahedra actually form a closed Torus with 90 Edges, and 60 external Faces.  The other 90 Faces in this set form an identical but disjoint Torus.  It it noteworthy that these Tori meet most of the criteria to be closed regular polyhedra in $\mathfrak{R}^3$, though they live in $\mathfrak{R}^4$.  In some sense, the set can be seen as composed of these 2 identical Toric Cells, but these cells are rank-4 facets instead of rank 3-facets.

For the set B, the Edges form 120 equilateral triangle Faces, which do not form any tetrahedral cells.  Taken together these Faces form a single closed Torus, which again meets most criteria to be a regular polyhedral Cell in $\mathfrak{R}^3$, excepting that it is a rank-4 facet.

As a last note on Tori, the $18-9$ proof from within the Peres $24-24$ set also has a sort of toric structure when projected back into $\mathfrak{R}^4$.  This set of 36 real vertices form 72 Edges with nearest neighbors, and these in turn form 36 square Faces.  The Faces in turn form a single closed Torus in analogy to the $30-15$B set, with one important distinction.  Here, the `square' Faces are actually rank-4 objects, and so this set is even less like a regular polytope than the others.

The Kochen-Specker Diagram of a set is a graph with a vertex for each ray, and a line connecting pairs of orthogonal rays.  This tangent on geometric structures illustrates the subtle point that two sets with the same Kochen-Specker Diagram, like the $30-15$A and $30-15$B, may still be geometrically distinct (unitarily inequivalent), and require different experimental setups to examine.

\pagebreak

\subsection{Pentagon Inequalities}

Klyachko et al \cite{Klyachko} introduced a novel state-dependent proof that rules out NCHVTs in a Hilbert Space of $d=3$, and makes use of only pairwise correlations between projectors.  The proof uses a set of 5 rays which form an orthogonal ring, in which each ray is orthogonal to its 2 neighbors.  We define the observable $\Sigma$ as the sum of all 5 projectors in the set:
\begin{equation}
\Sigma = |0\rangle\langle 0| + |1\rangle\langle 1| + |2\rangle\langle 2| + |3\rangle\langle 3| + |4\rangle\langle 4|
\end{equation}
And we can express the orthogonality relations within this set as
\begin{equation}
\langle i | i \pm 1 \rangle = 0
\end{equation}
for $i=0,1,2,3,4$, with addition modulo 5.
It is trivial to see that in any NCHVT, no more than 2 of the 5 projectors can be assigned the truth-value 1, because of the rule that no two orthogonal rays can both be assigned the truth-value 1.  This gives us what we call the Pentagon Inequality
\begin{equation}
\langle\Sigma_{NC}\rangle \leq 2 \label{PentagonIneq}
\end{equation}
For the set of real rays chosen by Klyachko we obtain a $\Sigma$ whose largest eigenvalue is $\sqrt(5) = 2.236$, and thus by preparing the corresponding eigenstate, we can obtain
\begin{equation}
\langle\Sigma_{Q}\rangle = 2.236,
\end{equation}
which clearly violates the Pentagon Inequality of Eq. \ref{PentagonIneq}, ruling out NCHVTs.

Sets of rays that form orthogonal pentagons of this type can exist in any dimension $d \geq 3$, though it can be shown \cite{Cabello_Pentagons} that no $\Sigma$ in any dimension can have an eigenvalue greater than $\sqrt(5)$.  Nevertheless, the sets of rays we have discussed in this chapter do contain pentagons in $d=4$ which allow for violation of the Pentagon Inequality.  We will review them briefly here.

Using the 60 rays of the 2-qubit Pauli group one can build 17,280 distinct observables $\Sigma$ with largest eigenvalue greater than 2, which break down into two types.  We call sets of 5 rays that form pentagons of this type {\it Conflict Pentagons}.  In this set, there are 5,760 Conflict Pentagons with largest eigenvalue 2.172, and 11,520 weaker Conflict Pentagons with largest eigenvalue 2.085.

Using the 60 rays of the 600-cell one can build 18,000 distinct Conflict Pentagons which break down into 3 successively weaker types: 3,600 with largest eigenvalue 2.178, 7,200 with largest eigenvalue 2.114, and 3,600 with largest eigenvalue 2.0850.

Each Conflict Pentagon Inequality is maximally violated by the eigenstate corresponding to its largest eigenvalue.  It is also violated, though not maximally, by all states in a continuous region of Hilbert space surrounding that eigenstate, which we call a {\it Conflict Cap}.  The boundary of the Conflict Cap for a given Pentagon is the set of states for which $\langle\Sigma\rangle=2$.  A natural question arises: does the union of all of the Conflict Caps of all of the Pentagons in one of our KS sets cover all of Hilbert Space?  If so, this yields a state-independent proof that rules out NCHVTs, while making use of only pairwise measurements of projectors.

For the 60 rays of the 2-qubit Pauli group, the answer is no, which can be seen because the Max$(\langle\Sigma_S\rangle) = 2$ for all 60 of the rays in this set, with $S$ the set of all 17,280 Conflict Pentagons.

For the 60 Rays of the 600-cell, the Conflict Caps appear to cover the entire real portion of Hilbert space.  We find that $V =$ Max$(\langle \vec{r} | \Sigma_S | \vec{r}\rangle) \geq 2.059$ for all real states $\vec{r}$, and with $S$ the set of all 18,000 Conflict Pentagons.  This result was obtained numerically by parameterizing $\vec{r}$ as,
\begin{equation}
\vec{r} = (\cos\phi\sin\theta_1\sin\theta_2, \sin\phi\sin\theta_1\sin\theta_2, \cos\theta_1\sin\theta_2, \cos\theta_2)
\end{equation}
with $0 \leq \phi < 2 \pi$, $0 \leq \theta_1,\theta_2 < \pi$, and calculating $V$ over a very fine mesh of these parameters.

For further details about the pentagons and other $n$-gons within the 600-cell, and information about more compact subsets of rays that provide a conflict-covering of real Hilbert space, refer to our unpublished note \cite{WA_Pentagons}.\newline

Now we move on to make a subtle, but in some sense obvious, new point about Pentagons using two qubits, and this is simply that in order for the Pentagon Inequality to be violated, there must be entanglement between the two qubits.  This is noteworthy simply because an identical Pentagon Inequality in the space of a single 4-level particle could be violated without any type of entanglement, just as in the case of the 3-level system discussed by Klyachko.  If we want this 4-dimensional Hilbert space to be the space of 2 qubits, we can only prove contextuality if the qubits are entangled, as we now demonstrate.

\begin{figure}[ht]
\centering
\subfloat[][]{
\includegraphics[width=1.5in]{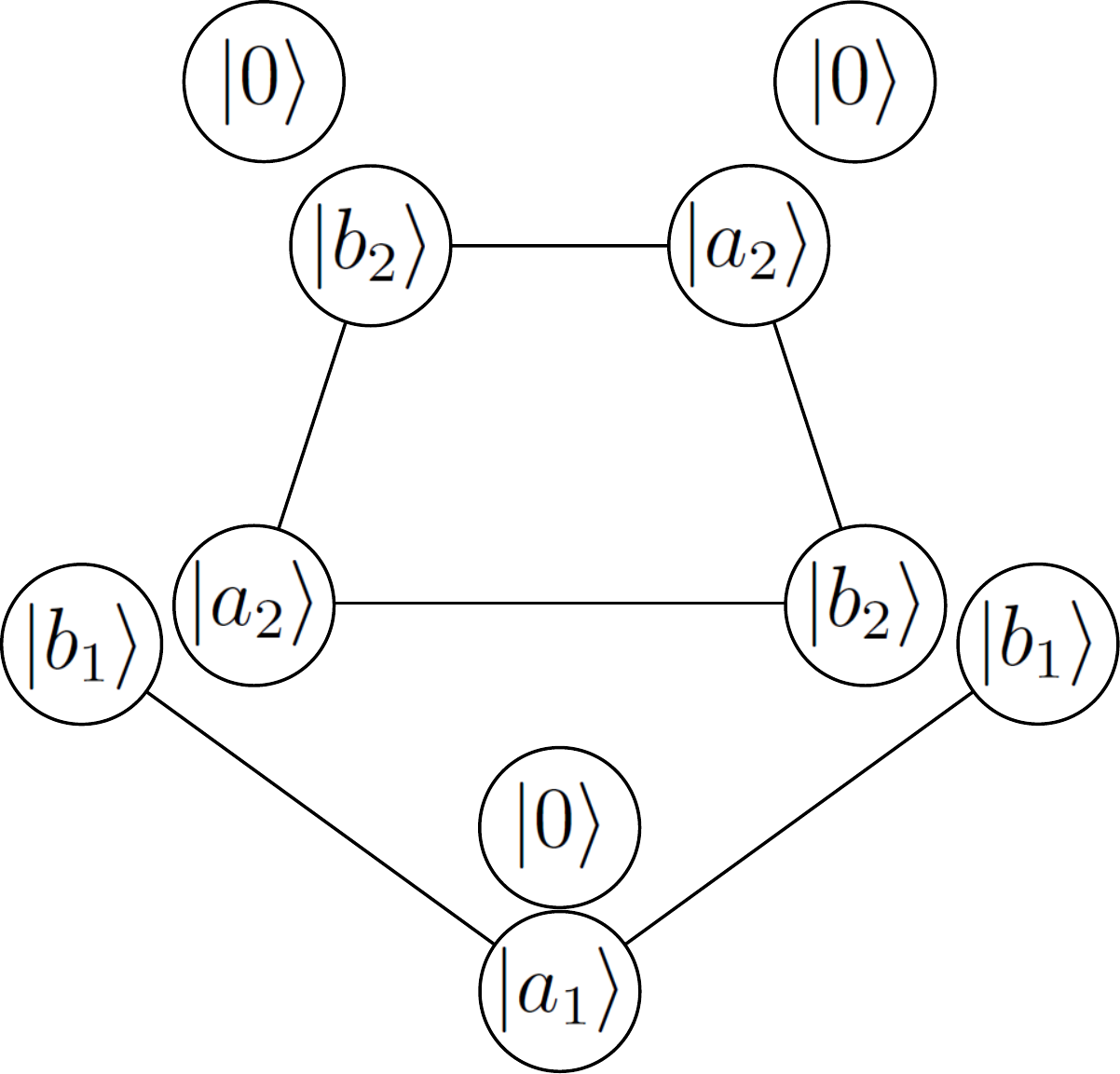}\label{PP1}}
\qquad
\subfloat[][]{
\includegraphics[width=1.5in]{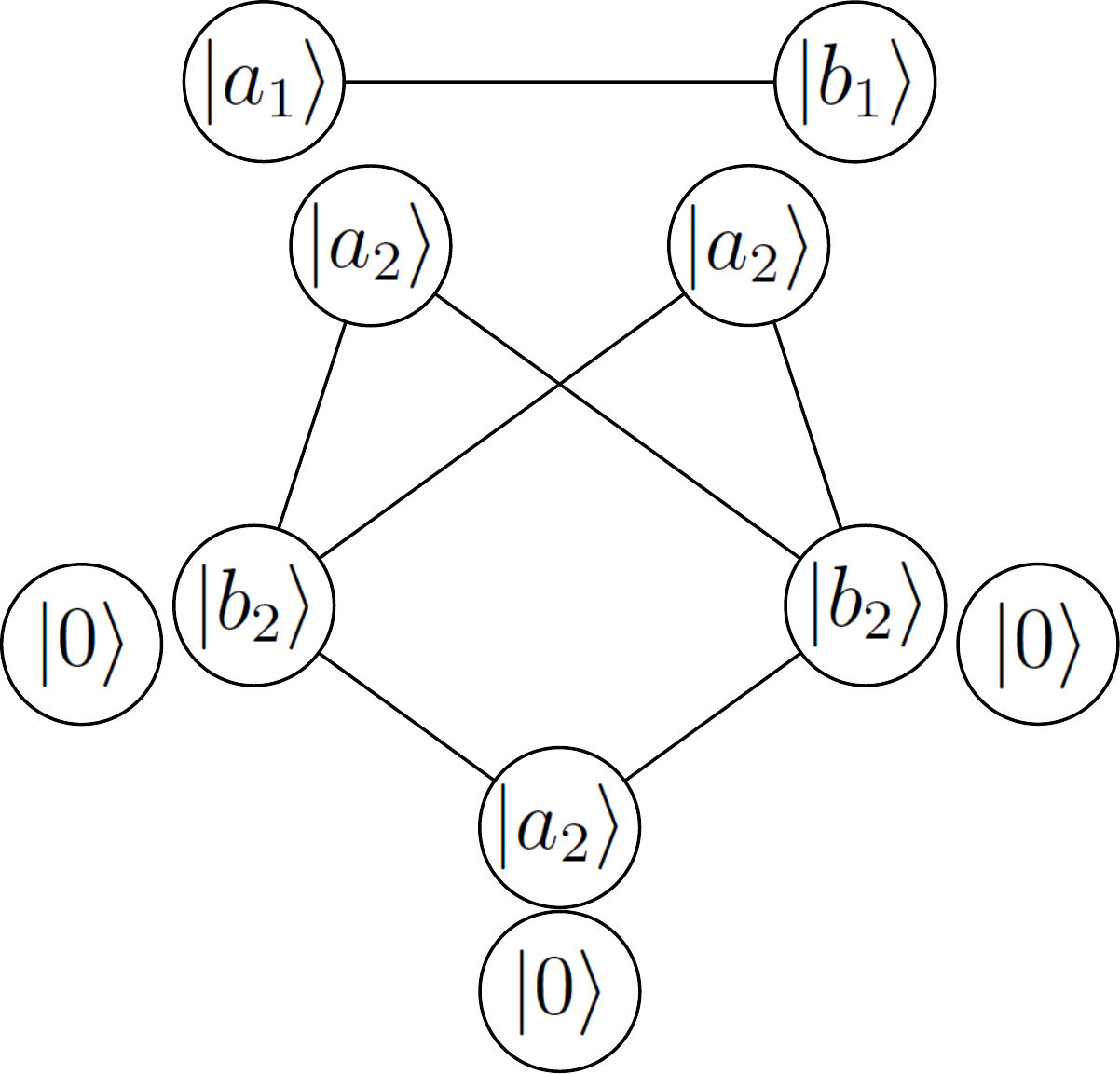}\label{PP2}}
\qquad
\subfloat[][]{
\includegraphics[width=1.5in]{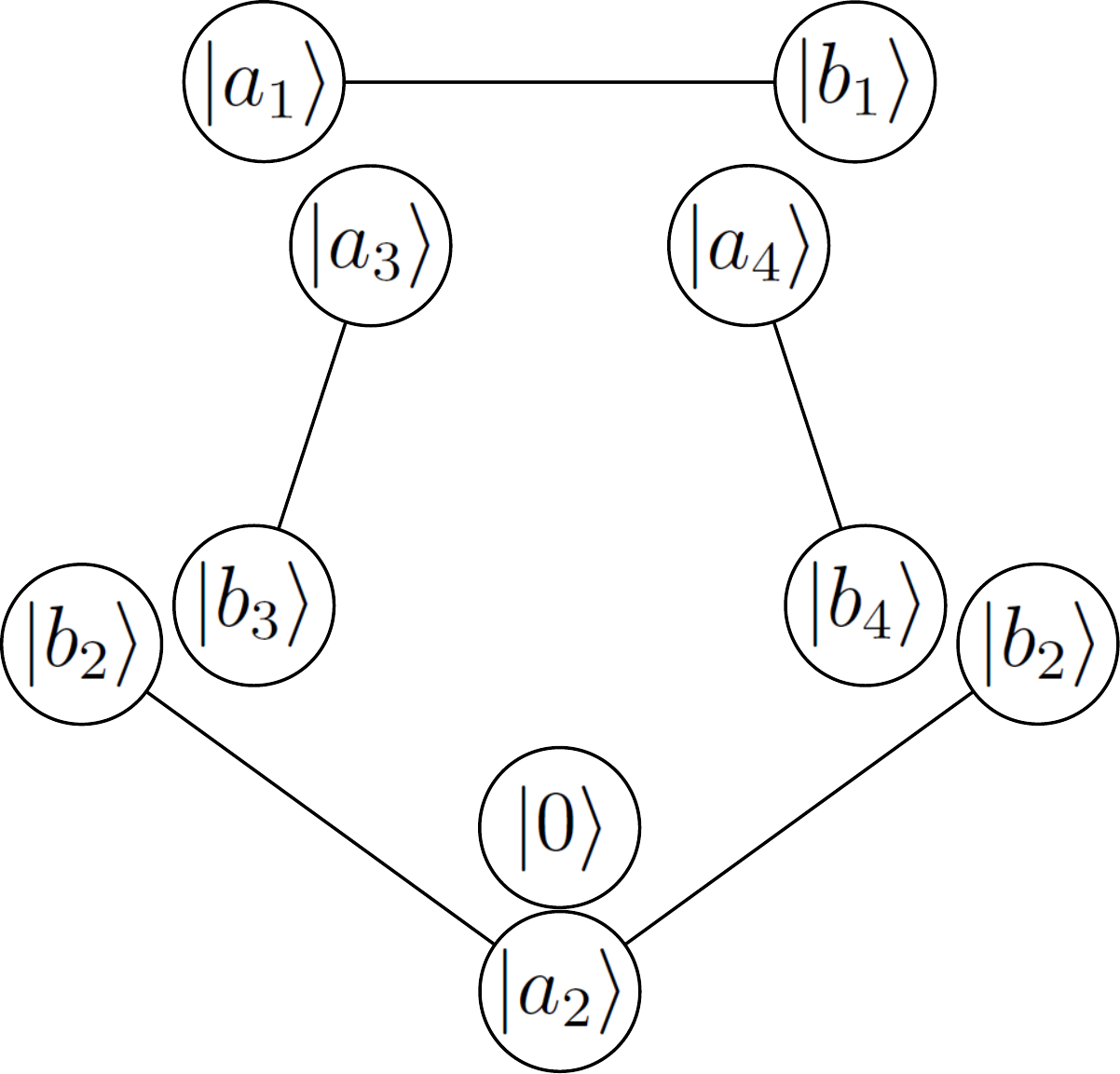}\label{PP3}}
\qquad
\caption[2-qubit Product Pentagon Diagrams]{The 3 possible Product Pentagons.  The 5 outer states are assigned to qubit 1 and the 5 inner states are assigned to qubit 2.  The two states at each corner of the Pentagon form a product state of the two qubits. Orthogonal states are connected by lines.}\label{ProductPentagons}
\end{figure}

In order to show that entanglement is required for a violation of a Pentagon Inequality, we construct the most general Pentagon using the product states of 2 individual qubits.  With no loss of generality, we will look to see if any Product Pentagon Inequality can be violated by the state $|\Psi_0\rangle = |0\rangle \otimes |0\rangle$.  For a Pentagon Operator of the form,
\begin{equation}
\Sigma = \sum_{i=0}^4 |\psi_i\rangle \langle\psi_i|,
\end{equation}
The expectation value in the above state is then simply
\begin{equation}
\langle\Sigma\rangle = \sum_{i=0}^4 |\langle\psi_i|\Psi_0\rangle|^2.
\end{equation}
From here we can see that the phases of the $|\psi_i\rangle$ do not appear in the expectation value, and so with no loss of generality we can parameterize an arbitrary pair of orthogonal single-qubit states as $|a_j\rangle = \cos{\theta_j}|0\rangle + \sin{\theta_j}|1\rangle$ and $|b_j\rangle = \sin{\theta_j}|0\rangle - \cos{\theta_j}|1\rangle$.  We will build the three possible Product Pentagons using orthogonal pairs like this.  Since our goal is to find the maximum of the expectation value, we will simply assign $|0\rangle$ to any single-qubit state in the Pentagon that does not belong to one of the orthogonal pairings.  The three Pentagons and the states we assign to them are shown in Figure \ref{ProductPentagons}.

For the Pentagons in Figures \ref{PP1} and \ref{PP2}, the expectation value does not depend on the values of the parameters, and is always exactly 2.  This is no surprise, since the extra orthogonality on the diagonals of these Pentagon graphs causes their Independence numbers and Lov\'{a}sz numbers to be equal, meaning their Pentagon Inequalities cannot be violated \cite{sadiq2011cabello}.

The Pentagon in Figure \ref{PP3} does have a higher Lov\'{a}sz number than its Independence number, and the expectation value does depend on the 4 parameters, but a simple calculation shows that its maximum is 2, and so as we expected, none of the 2-qubit Product Pentagon Inequalities can be violated, Q.E.D.

\pagebreak

%
%\end{document}

%\documentclass[12pt]{article}
%\usepackage{graphicx}
%\usepackage{epsfig}
%\usepackage{amsfonts}
%\usepackage[lofdepth,lotdepth]{subfig}
%
%\begin{document}

\section{3 qubits}\label{sec:q3}

In this chapter we discuss a number of interesting structures within the $d=8$ Hilbert space of 3 qubits.  To begin we will discuss additional features of IDs, and introduce the complete family of IDs that exist in the 3-qubit Pauli group.  Next we will discuss how these IDs can be used to form Kernels and discuss the structure of Kernels.  We will then discuss the Mermin Star and the set of 40 rays it generates.  After this we will introduce the 3-qubit Kite, which is part of a family of Kite proofs that extends from $2$ to $N$ qubits.  Finally we will discuss the 3-qubit Wheel, which has a structure isomorphic to the Mermin Square, and the related families of Wheel and Whorl proofs which also extend from 3 to $N$ qubits.

\subsection{The IDs of the 3-qubit Pauli Group}\label{sec:q3IDs}

An {\it Identity Product} (ID) is a set of mutually commuting observables of the $N$-qubit Pauli group whose overall product is $\pm I$.  We will represent an ID as an array of Pauli observables $(Z,X,Y)$, with each row showing a different observable, and each column showing a different qubit.

There are just 3 unique types of IDs that exist within the 3-qubit Pauli group.  Two of them are new for 3 qubits, and the other is a trivial extension of the 2-qubit ID.  At this point, it is important to stress what we mean here by a {\it unique} type.  Since an ID is a mutually commuting set, the ordering of the observables is irrelevant.  Furthermore, the order we choose to index the qubits is arbitrary, and so the reordering of both the rows (observables) and columns (qubits) in an ID does not alter its underlying structure.

The next issue concerns the columns of an ID, which we call {\it Single-Qubit-Products} (SQPs).  The SQPs in an ID can come in only two varieties: those containing each of the single-qubit Pauli observables $(Z,X,Y)$ an even number of times, whose product is $\pm I$ (the single-qubit Identity), and those containing each of the single-qubit Pauli observables an odd number of times, whose product is $\pm i I$.  We call these {\it Even} and {\it Odd} SQPs respectively, and it should be clear that all IDs contain an even number of Odd SQPs, since their overall product must be real.  Furthermore, if a Negative ID contains no Odd SQPs, then it is automatically a single-ID Kernel.  If a Positive ID contains no Odd SQPs, it cannot be used in any Kernel, and so we call it {\it Null} and exclude it from our definition of IDs.  Among the even SQPs, there are a subset we will also call {\it Trivial} because they contain either none, or only one of the single-qubit Pauli Operators (which appears an even number of times), meaning that they have product $+I$ and all of their elements mutually commute.

Now, each of the nontrivial SQPs in an ID actually has 6 possible permutations:  For Even SQPs with only 2 different single-qubit Pauli observables, the 6 permutations are the assignments $(ZX$, $XZ$, $ZY$, $YZ$, $XY$, $YX)$ to those observables.  For SQPs with all 3 different single-qubit Pauli observables, the 6 permutations are the assignments $(ZXY$, $XYZ$, $YZX$, $ZYX$, $YXZ$, $XZY)$.  Permutations of this second type may also change the overall sign of an Odd SQP.  Each SQP in an ID can be permuted independently in this way without altering the underlying structure, since $(Z,X,Y)$ all mutually anticommute.

Finally, in looking for {\it unique} ID structures, we can ignore the order of rows and columns, and we need only consider SQPs with their observables in the orders $ZX$ or $ZXY$.  With all of that understood, we can now see that the only unique ID structure for 2 qubits is the upper one in Table \ref{q2kernel}.  Table \ref{q3ID4s} shows the 2 unique types of 3-qubit IDs.
\begin{table}[ht]
\centering
\subfloat[][This ID has no Odd SQPs, and is therefore also a single-ID Kernel.  This Kernel generates the Mermin Star.]{
\begin{tabular}{ccccc}
& $Z$ & $Z$ & $Z$ & \\
& $Z$ & $X$ & $X$ & \\
& $X$ & $Z$ & $X$ & \\
& $X$ & $X$ & $Z$ & \\
\end{tabular}\label{GHZKernel}}
\qquad
\subfloat[][This ID has two Odd SQPs, and can be combined with other IDs to form numerous Kernels.]{
\begin{tabular}{ccccc}
& $Z$ & $I$ & $Z$ & \\
& $I$ & $Z$ & $Z$ & \\
& $X$ & $X$ & $X$ & \\
& $Y$ & $Y$ & $X$ & \\
\end{tabular}\label{q3PartialID}}
\qquad
\caption[3-qubit critical IDs]{The two types of unique IDs for 3 qubits.  These IDs are critical, in the sense that none of the qubits and/or observables can be deleted such that the remaining set is still an ID.}\label{q3ID4s}
\end{table}

We can also obtain 3-qubit IDs by taking the 2-qubit ID and adding to it a trivial SQP.  Again, trivial SQPs can contain either all $I$s, or they contain some mix of $I$s with an even number of one of the three single-qubit Pauli observables $(Z,X,Y)$.  When choosing a trivial SQP, we can choose any even number of any one of the three Pauli observables, and any order, since all of its elements mutually commute.  Table \ref{q3ID3s} shows some 3-qubit IDs of this type.
\begin{table}[ht]
\centering
\subfloat[][Here the most trivial SQP is added to the 2-qubit ID in order to form a 3-qubit ID.]{
\begin{tabular}{ccccc}
& $Z$ & $Z$ & $I$ & \\
& $X$ & $X$ & $I$ & \\
& $Y$ & $Y$ & $I$ & \\
\end{tabular}\label{q3TrivialID}}
\qquad
\subfloat[][Here another of 10 possible 3-element trivial SQPs  is added to the 2-qubit ID in order to form a 3-qubit ID.]{
\begin{tabular}{ccccc}
& $Z$ & $Z$ & $Z$ & \\
& $X$ & $X$ & $Z$ & \\
& $Y$ & $Y$ & $I$ & \\
\end{tabular}}
\qquad
\caption[3-qubit noncritical IDs]{The 3-qubit IDs shown here are not critical, in the sense that we could simply delete the trivial SQP and the remaining set would still be an ID.  However, these noncritical IDs can be combined to build Kernels that are critical, in the sense that no combination of qubits and/or IDs can be deleted such that the remaining set is still a Kernel.}\label{q3ID3s}
\end{table}

As a last detail, we say that an ID is critical if none of the qubits and/or observables can be deleted such that the remaining set is still an ID.  Uniqueness and criticality of these sets have been verified by exhaustive computer searches.\newline

As we consider more qubits, the number of IDs in the $N$-qubit Pauli Group grows quite rapidly.  In the $N$-qubit Pauli group, one can always find sets of $2^N-1$ mutually commuting observables - stabilizer group for joint eigenbasis of the ID, and these complete sets are always IDs (many including numerous trivial SQPs), but only for $N=2$ are these IDs critical.  For $N=3$, we obtain sets of 7 mutually commuting observables, but these sets can always be separated (in 7 ways) into the product of a 3-observable ID and 4-observable ID.  To help handle the proliferation of different types of IDs, we now introduce the expanded symbol ID$M^N$ for an $N$-qubit ID containing $M$ observables.  We may omit the $M$ and/or $N$ depending on the context.   The symbols for some of the sets we have been considering are, ID$4^3$ (Table \ref{q3ID4s}), ID$3^3$ (Table \ref{q3ID3s}), and ID$3^2$ (both IDs in Table \ref{q2kernel}).  We can also describe the sets of 7 mutually commuting observables as noncritical ID$7^3$s, which can always be separated into the product of an ID$3^3$ and ID$4^3$.

Furthermore, the ID$3^3$ introduces a new wrinkle regarding eigenbases that was not present for $N=2$, but persists for all larger $N$.  For a given ID$M^N$, there are $2^{M-1}$ different ways to choose the eigenvalues of the $M$ observables, meaning that its joint eigenbasis will contain $2^{M-1}$ rays.  For some values of $N$ and $M$, the rays actually describe projectors with rank higher than one, meaning that each ray spans more than one dimension of Hilbert space, and contains internal degrees of freedom.  In general, the rank of the projectors in the eigenbasis of an ID$M^N$ is $2^{N-M+1}$.  So for our ID$3^3$s we will have eigenbases with 4 rank-2 rays, and for the ID$4^3$s we will have eigenbases with 8 of the usual rank-1 rays.

As we will see later, this remarkable feature of IDs will allow for surprisingly compact proofs of the KS theorem for larger numbers of qubits.

\pagebreak

\subsection{3-qubit Kernels}\label{sec:q3kernels}

Now that we have introduced the available ID$^3$s, let us consider how they form Kernels.  Recall that a Kernel is defined as a set of IDs, an odd number of which are Negative IDs, with each single-qubit Pauli observable appearing an even number of times.  We will say that a Kernel is critical if no combination of qubits and/or IDs can be deleted such that the remaining set is still a Kernel.

The structure of a Kernel depends almost entirely on the Odd SQPs within its IDs.  A negative ID with no Odd SQPs is also automatically a {\it single-ID Kernel}, as can be seen from the above definition of Kernels.  In general, we will call IDs without any Odd SQPs {\it Whole} IDs, and those with Odd SQPs {\it Partial} IDs.

The single-ID$4^3$s Kernel of Table \ref{GHZKernel} can be used to prove the GHZ theorem using only 3 qubits.  A source repeatedly produces a particular joint eigenstate of the ID, and sends one particle to Alice, another to Bob, and the last to Charlie.  Alice, Bob, and Charlie each randomly choose one of the two measurement bases, $Z$ or $X$, and measure their qubit, such that the random choice and measurement for all 3 qubits are spacelike separated.  In any run where Alice, Bob, and Charlie randomly measure one of the 3-qubit observables of the ID, the product of their three individual results is correlated, and must equal the eigenvalue of that 3-qubit observable in the prepared state.  The product of all 4 such eigenvalues is -1 in any prepared eigenstate, because the ID is negative.  Because the measurements are chosen randomly, a local hidden variables theory requires that a truth-value be preassigned to all single-qubit observables that might be measured.  Furthermore, locality also guarantees that Alice's measurement result cannot be affected by Bob's or Charlie's choice of measurement basis, meaning the single-qubit observables in the ID must be assigned noncontextual truth-values.  If we define $B$ as the product of the IDs, then because each single-qubit observable appears twice in the ID, the truth-values assigned to each one will be squared, and so the overall product will always be $B_L = +1$.  But we know already that the experimental value is $B_Q = -1$ for any prepared eigenstate, which rules out local hidden variable theories, and proves the GHZ theorem.  All single-ID$^N$ Kernels can be used to give an analogous proof using $N$ qubits and $N$ spacelike separated parties.

Table \ref{GHZKernel} is only one unique Whole ID$4^3$ (single-ID Kernel) that exists.  This Kernel generates the Mermin Star of Figure \ref{MerminStar} following the method described in Chapter \ref{sec:MerminSquare}.  We will return to this in Chapter \ref{sec:MerminStar}.  It is noteworthy that single-ID Kernels only exist for $N\geq3$.

Now we move on to the subject of Kernels composed of more than one different ID, which we call {\it Composite Kernels}.  Composite Kernels are always formed using some set of Partial IDs.  To do this, we must simply choose the order of the SQPs within each ID so that every Odd SQP in a given ID is paired with an Odd SQP in another ID in the set.  With the Odd SQPs paired off in this way, each single-qubit Pauli Observable now appears an even number of times for each qubit, which is one of the requirements of a Kernel.  The other requirement is very easy to satisfy, since we can change the sign of any Partial ID by permuting $(Z,X,Y)$ within one of its Odd SQPs.  We fix the orders such that an odd number of the IDs are negative, and then we have a complete Composite Kernel.

Before continuing, we will add one final feature to complete our ID symbol.  We will denote an $N$-qubit ID with $M$ observables, and $O$ Odd SQPs as an ID$M^N_O$.  For example the ID of Table \ref{GHZKernel} has symbol ID$4^3_0$, while the ID of Table \ref{q3PartialID} has symbol ID$4^3_2$.  In constructing Composite Kernels we really only need to concern ourselves with the value of $O$, and as we will see, once a {\it Composite Kernel Structure} (CKS) has been selected, we can actually assign to each of its elements {\it any} ID with the appropriate value of $O$.

To build the aforementioned Composite Kernel Structures, we will first introduce a compact notation for IDs that we call an {\it Oddness Profile}, in which we collapse each Odd SQP to an `O', each Even SQP to an `E' and each trivial SQP to an `I'.  To be clear, the Oddness Profile of the ID in Table \ref{GHZKernel} is simply `EEE', for Table \ref{q3PartialID} it is `OOE', while for the two IDs of Table \ref{q3ID3s} it is `OOI'.  We say that a Composite Kernel is critical if no deletion of IDs and/or qubits can result in a smaller Kernel.  For 2 qubits, there is only one critical CKS, as shown in Table \ref{CompKernel2}.  The 6 critical Composite Kernels for 3 qubits are shown in Table \ref{CompKernels3}.  In both cases, these are unique up to the overall order of the qubits.
\begin{table}[ht]
\centering
\begin{tabular}{cc}
  O & O \\
  O & O \\
\end{tabular}\caption[2-qubit Composite Kernel Structure]{This is the only critical Composite Kernel Structure for 2 qubits.  Each row of the structure represents a different ID, and each column represents a different qubit.}\label{CompKernel2}
\end{table}
\begin{table}[ht]
\centering
\subfloat[][A 2-ID Kernel]{
\begin{tabular}{ccccc}
& O & O & E & \\
& O & O & I & \\
\end{tabular}\label{CKS2_3}}
\qquad
\subfloat[][A 2-ID Kernel]{
\begin{tabular}{ccccc}
& O & O & E & \\
& O & O & E & \\
\end{tabular}\label{Kite3Struct}}
\qquad
\subfloat[][A 3-ID Kernel]{
\begin{tabular}{ccccc}
& O & O & I & \\
& I & O & O & \\
& O & I & O & \\
\end{tabular}\label{Wheel3Struct}}
\qquad
\subfloat[][A 3-ID Kernel]{
\begin{tabular}{ccccc}
& O & O & E & \\
& I & O & O & \\
& O & I & O & \\
\end{tabular}}
\qquad
\subfloat[][A 3-ID Kernel]{
\begin{tabular}{ccccc}
& O & O & E & \\
& E & O & O & \\
& O & I & O & \\
\end{tabular}}
\qquad
\subfloat[][A 3-ID Kernel]{
\begin{tabular}{ccccc}
& O & O & E & \\
& E & O & O & \\
& O & E & O & \\
\end{tabular}}
\qquad
\caption[3-qubit Composite Kernels]{The 6 critical Composite Kernels for 3 qubits.  Each row of the structure is the Oddness Profile of a different ID, and each column is a different qubit.  {\it Any} set of IDs with the correct Oddness Profiles can be assigned to the rows of a Composite Kernel Structure to give a Kernel.}\label{CompKernels3}
\end{table}

To build a Composite Kernel from one of the elements of Table \ref{CompKernels3}, we can assign any Partial ID with the correct Oddness Profile to each of its rows.  Note that the qubits in a given ID may need to be reordered to match the Oddness Profile.  Using the IDs from Tables \ref{q3ID4s} and \ref{q3ID3s}, and allowing all permutations of the individual SQPs, as well as all 10 choices for the trivial qubit in Table \ref{q3ID3s}, we can see the number of possible Composite Kernels for 3-qubits is quite large.  We will consider only the two simplest cases here.

For the first case, we will take the Composite Kernel Structure of Table \ref{Kite3Struct}, and assign to it two different permutations of the ID in Table \ref{q3PartialID}.  As shown specifically in Table \ref{Kite3Kernel}, we will use the given Negative ID, and also the Positive ID obtained by transposing the $Z$ and $X$ in the second SQP of the original.
\begin{table}[ht]
\centering
\begin{tabular}{|ccc|}
\hline
$Z$ & $I$ & $Z$ \\
$I$ & $Z$ & $Z$ \\
$X$ & $X$ & $X$ \\
$Y$ & $Y$ & $X$ \\
\hline
$Z$ & $I$ & $Z$ \\
$I$ & $X$ & $Z$ \\
$X$ & $Z$ & $X$ \\
$Y$ & $Y$ & $X$ \\
\hline
\end{tabular}\caption[3-qubit Kite Kernel]{This is the Kernel of the 3-qubit Kite, an Observable-based KS proof which is examined in further detail in Chapter \ref{sec:Kite3}.  Note that by permuting just one Odd SQP within a Partial ID, we are able to form a 2-ID Kernel.  It should also be easy to see that this will work for {\it any} Partial ID of any size, and in fact the Kernel of Figure \ref{q2kernel} is obtained in just the same way.}\label{Kite3Kernel}
\end{table}

For the second case, we will take the Composite Kernel Structure of Table \ref{Wheel3Struct}, and assign to it three different permutations of the ID of Table \ref{q3TrivialID}.  As shown specifically in Table \ref{Wheel3Kernel}, we use the given Negative ID, and also two more Negative IDs obtained by permuting the qubits of the original.
\begin{table}[ht]
\centering
\begin{tabular}{|ccc|}
\hline
$Z$ & $Z$ & $I$ \\
$X$ & $X$ & $I$ \\
$Y$ & $Y$ & $I$ \\
\hline
$I$ & $Z$ & $Z$ \\
$I$ & $X$ & $X$ \\
$I$ & $Y$ & $Y$ \\
\hline
$Z$ & $I$ & $Z$ \\
$X$ & $I$ & $X$ \\
$Y$ & $I$ & $Y$ \\
\hline
\end{tabular}\caption[3-qubit Wheel Kernel]{This is the Kernel of the 3-qubit Wheel, an Observable-based KS proof which is examined in further detail in Chapter \ref{sec:Wheel3}.  Note that by permuting the qubits of a Partial ID with two Odd SQPs, we are able to create a closed loop of three pairs of Odd SQPs, forming a 3-ID Kernel.  It should also be easy to see that this will work for {\it any} Partial ID with two Odd SQPs, regardless of its size, and that the size of the loop is only limited by the number of qubits.  It should also be noted that while none of the three IDs of this Kernel are individually critical (as defined for IDs), the Kernel they form is critical (as defined for Kernels).}\label{Wheel3Kernel}
\end{table}

It should be easy to see how mixing and matching the various 3-qubit IDs can create many distinct Kernels.  For more examples of 3-qubit Kernels and the Observable-based KS proofs that they generate, refer to our paper \cite{WA_3qubits}.

\pagebreak

\subsection{The Mermin Star}

\label{sec:MerminStar}

\begin{figure}[ht]
  \centering
  \includegraphics[width=3in]{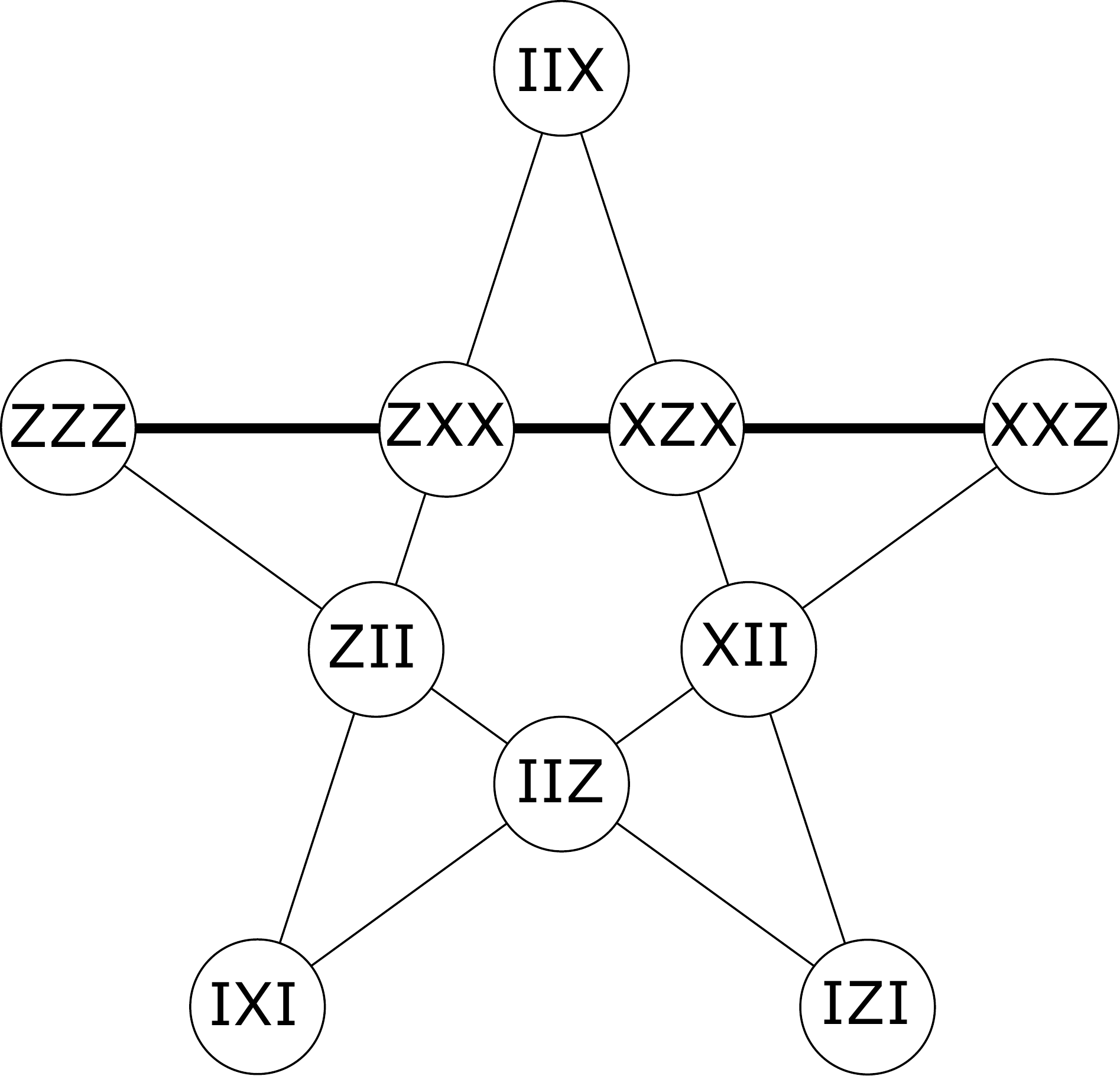}\\
  \caption[3-qubit Mermin Star]{The Mermin Star, a $10_2-5_4$ Observable-based KS proof for 3 qubits.  Each straight line connects the 4 observables of an ID.  The thin lines denote Positive IDs, while the thick line denotes a Negative ID (which in this case is also the Kernel).}\label{MerminStar}
\end{figure}

The Mermin Star \cite{Mermin_SquareStar}, shown in Figure \ref{MerminStar}, is an Observable-based KS proof introduced by David Mermin and related to the work of Greenberger, Horne, Shimony, and Zeilinger \cite{GHSZ}.

The Mermin Star can be generated directly from the 3-qubit Kernel of Table \ref{GHZKernel}, using the method described in Chapter \ref{sec:MerminSquare}.  To be explicit, we will show the full process for this case here.  For each observable of this ID$4^3_0$, we will generate a new Positive ID by supplementing that observable with its own single-qubit decomposition.  For example, for the first observable of Table \ref{GHZKernel}, $ZZZ$, we supplement with $ZII$, $IZI$, and $IIZ$.  This process guarantees that all of the IDs taken together will then form an Observable-based KS proof.  Table \ref{MerminStarIDs} shows the same 5 IDs of Figure \ref{MerminStar}, and how they are generated from the Kernel.

\begin{table}[ht]
\centering
\subfloat[][]{
\begin{tabular}{ccccc}
$Z$ & $Z$ & $Z$ \\
$Z$ & $X$ & $X$ \\
$X$ & $Z$ & $X$ \\
$X$ & $X$ & $Z$ \\
\end{tabular}\label{IDK}}
\qquad
\subfloat[][]{
\begin{tabular}{ccccc}
$Z$ & $Z$ & $Z$ \\
$Z$ & $I$ & $I$ \\
$I$ & $Z$ & $I$ \\
$I$ & $I$ & $Z$ \\
\end{tabular}}
\qquad
\subfloat[][]{
\begin{tabular}{ccccc}
$Z$ & $X$ & $X$ \\
$Z$ & $I$ & $I$ \\
$I$ & $X$ & $I$ \\
$I$ & $I$ & $X$ \\
\end{tabular}}
\qquad
\subfloat[][]{
\begin{tabular}{ccccc}
$X$ & $Z$ & $X$ \\
$X$ & $I$ & $I$ \\
$I$ & $Z$ & $I$ \\
$I$ & $I$ & $X$ \\
\end{tabular}}
\qquad
\subfloat[][]{
\begin{tabular}{ccccc}
$X$ & $X$ & $Z$ \\
$X$ & $I$ & $I$ \\
$I$ & $X$ & $I$ \\
$I$ & $I$ & $Z$ \\
\end{tabular}}
\caption[The 5 IDs of the Mermin Star]{The 5 IDs of the Mermin Star (Figure \ref{MerminStar}).  The Kernel is shown in \subref{IDK}.  The other 4 Positive IDs show how the Kernel generates the set.  In each one, the top observable comes from the Kernel, and the observables below are its single-qubit decomposition.}\label{MerminStarIDs}
\end{table}

Let us now consider the set of rays that are generated by the Mermin Star.  Each ID$4^3$ has an eigenbasis composed of 8 rank-1 rays, yielding a total of 40 rays, which was first examined in detail by Peres and Kernaghan \cite{kernaghan_peres}.  Since each of the 10 observables is shared by two IDs, there will be 10 complementary pairs of hybrid bases, for a total of 25 bases in the set, which we now see has symbol $40_5-25_8$.  This set contains a total of $2^{10}=1,024$ parity proofs, which can be obtained using the 10 complementary pairs of hybrid bases, just as in Chapter \ref{sec:MerminSquare} and \ref{sec:q2Whorl}.  The breakdown of the different types of parity proofs is shown in Table \ref{StarProofs}.  It should be noted that this set does not contain complementary pairs of proofs, as did those of Chapter \ref{sec:MerminSquare} and \ref{sec:q2Whorl}.

\begin{table}[h]
\begin{center} {
\begin{tabular}{|ccc|}
\hline
Compact Symbol & Expanded Symbol & \# of Proofs \\
\hline
$36-11$ & $28_2 8_4 -11_8$ & 320 \\
$38-13$ & $24_2 14_4 -13_8$ & 640 \\
$40-15$ & $20_2 20_4 -15_8$ & 64 \\
\hline
\end{tabular} }
\end{center}
\caption[Critical Parity Proofs of the Mermin Star]{The 1024 Critical Parity Proofs of the KS Theorem within the $40_4-25_8$ set of Peres and Kernaghan, generated by the Mermin Star.}
\label{StarProofs}
\end{table}

For more details about the proofs in this set, refer to our paper \cite{WA_3qubits}.

\pagebreak

\subsection{The 3-qubit Kite}\label{sec:Kite3}

\begin{figure}[ht]
  \centering
  \includegraphics[width=3.5in]{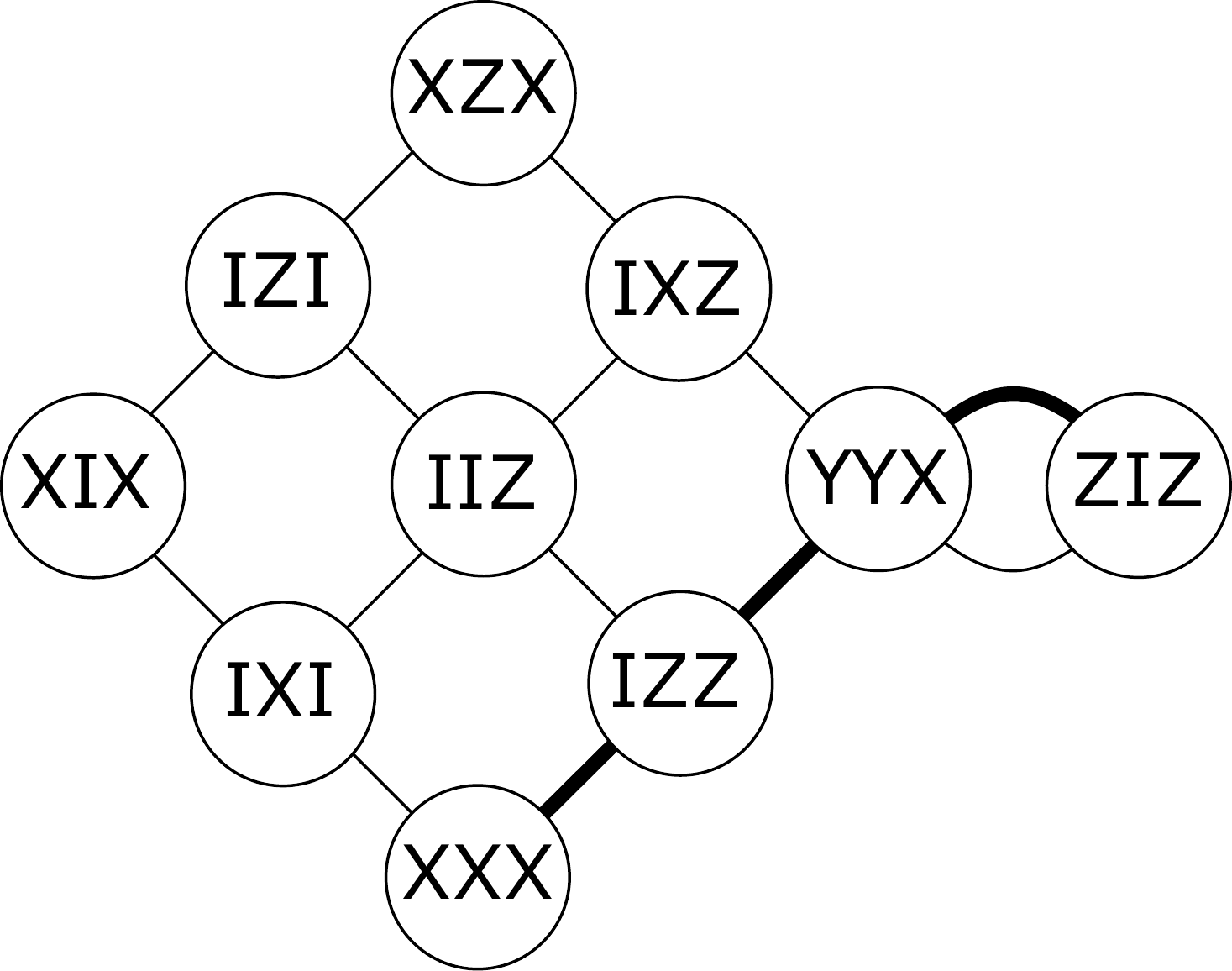}\\
  \caption[3-qubit Kite]{The 3-qubit Kite, a $10_2-2_4 4_3$ Observable-based KS proof for 3 qubits.  Each line or arc connects the observables of an ID.  The thin lines denote Positive IDs, while the thick line denotes a Negative ID. The 2-IDs of the Kernel form the `tail' of the Kite.  Most any Partial ID$M^N$ can be used to form a Kernel like the one in Table \ref{Kite3Kernel}, which can in turn be used to form a Kite proof in just the same way, with the length of the tail determined by $M$.}\label{q3Kite}
\end{figure}

Here we consider the 3-qubit Kite \cite{WA_3qubits}, as shown in Figure \ref{q3Kite}, an Observable-based KS proofs that can be generated from the Kernel of Table \ref{Kite3Kernel}.

To get this proof, we will make a slight modification to the process we have used to generate an Observable-based KS proof from a Kernel.  Again, we take each observable that appears an odd number of times in the Kernel, and supplement it with its own decomposition in order to form a new positive ID.  In this case however, we need not necessarily go all the way down to the single-qubit decomposition.  Any portion of the observable that is shared by another observable can be kept together in the decomposition.  As an example of this consider that this Kernel contains the observables $XXX$ and $XZX$, which means that the decomposition of the first is $XIX$ and $IXI$, while the decomposition of the second is $XIX$ and $IZI$.  This simplification still guarantees that the generated set of IDs contain each observable an even number of times, resulting in the generation of a more compact Observable-based KS proof.  Again, the specific process is shown in Table \ref{q3KiteIDs}.
\begin{table}[ht]
\centering
\qquad
\subfloat[][]{
\begin{tabular}{ccccc}
$Z$ & $I$ & $Z$ \\
$I$ & $Z$ & $Z$ \\
$X$ & $X$ & $X$ \\
$Y$ & $Y$ & $X$ \\
\end{tabular}\label{IDK1}}
\qquad
\subfloat[][]{
\begin{tabular}{ccccc}
$Z$ & $I$ & $Z$ \\
$I$ & $X$ & $Z$ \\
$X$ & $Z$ & $X$ \\
$Y$ & $Y$ & $X$ \\
\end{tabular}\label{IDK2}}
\qquad\qquad\qquad\qquad\qquad\qquad\qquad\qquad
\subfloat[][]{
\begin{tabular}{ccccc}
$I$ & $Z$ & $Z$ \\
$I$ & $Z$ & $I$ \\
$I$ & $I$ & $Z$ \\
\end{tabular}}
\qquad
\subfloat[][]{
\begin{tabular}{ccccc}
$X$ & $X$ & $X$ \\
$X$ & $I$ & $X$ \\
$I$ & $X$ & $I$ \\
\end{tabular}}
\qquad
\subfloat[][]{
\begin{tabular}{ccccc}
$I$ & $X$ & $Z$ \\
$I$ & $X$ & $I$ \\
$I$ & $I$ & $Z$ \\
\end{tabular}}
\qquad
\subfloat[][]{
\begin{tabular}{ccccc}
$X$ & $Z$ & $X$ \\
$X$ & $I$ & $X$ \\
$I$ & $Z$ & $I$ \\
\end{tabular}}
\caption[The 6 IDs of the 3-qubit Kite]{The 6 IDs of the 3-qubit Kite (Figure \ref{q3Kite}).  The 2 ID4s of the Kernel are shown in \subref{IDK1} and \subref{IDK2}.  The other 4 Positive ID3s show how the Kernel generates the set.  In each one, the top observable comes from the Kernel, and the observables below are its decomposition.}\label{q3KiteIDs}
\end{table}

Now let us consider the set of rays generated by the 3-qubit Kite.  The structure of this proof introduces some new features that we have not seen before.  If two IDs share exactly one observable in common, then the rays of their two eigenbases mix to form two complementary hybrid bases.  However, it can be shown more generally that if two IDs share $s$ observables in common, then the rays of their two eigenbases mix to form $h = 2^{2^s} - 2$ hybrid bases, which still break down into complementary pairs (see Chapter \ref{sec:RBSets} for more details).

So, to begin with, we have 2 ID$4^3$ each of which have an eigenbasis of 8 rank-1 rays, and 4 ID$3^3$s, each of which have an eigenbasis of 4 rank-2 rays, for a total of 32 rays.  To find all of the hybrid bases, we consider every pairing of intersecting IDs.  We see that there are 4 pairs of ID$3$s that share one observable in common, giving us 8 hybrid bases with rays of rank 2, 4 pairings of an ID$4$ with an ID$3$, giving us 8 hybrid bases with rays of mixed rank, and 1 pair of ID$4$s which share two observables in common, giving us $h = 2^{2^2} - 2= 14$ hybrid bases with rays of rank 1, giving us 32 bases in total.

Before giving the expanded symbol for this $32-36$ set, we will add one final feature to our symbols for sets of projectors and bases.  We will now use the symbol $R^r_m - B_n$ for a set of $R$ projectors of rank $r$ that appear in $m$ of the $B$ bases, each of which has $n$ elements.  Again, if a set contains projectors of differing $r$ and/or $m$, or bases with differing $n$, the symbol on either side of the dash can be repeated.  We can use the expanded symbol for this set to illustrate this.  This set has symbol $16^1_{10} 16^2_4 - 16_8 8_6 12_4$, meaning it has 16 rank-1 projectors that each appear in 10 bases, 16 rank-2 projectors that each appear in 4 bases, 16 8-element bases, 8 6-element bases, and 12 4-element bases.  We will still refer generally to projectors of all ranks as rays.

The presence of a pair of IDs with $s>1$ in the Observable-based KS proof also seems to break the simple pattern we have used in other cases to obtain $2^H$ parity proofs from a set with $H$ complementary pairs of hybrid bases.  An exhaustive computer search reveals that there are 33,152 distinct parity proofs of the KS theorem within this set, with the smallest being a $24-9$ set with expanded symbol $12^1_2 12^2_2 - 1_8 4_6 4_4$, and the largest a $32-17$, with expanded symbol $8^1_2 4^1_4 4^1_6 12^2_2 4^2_4 - 5_8 4_6 8_4$.

The Kite family of Observable-based KS proofs has members for all number of qubits $N\geq2$, which we will discuss in Chapter \ref{sec:Kites}.

\pagebreak

\subsection{The 3-qubit Wheel}\label{sec:Wheel3}

\begin{figure}[ht]
  \centering
  \includegraphics[width=3in]{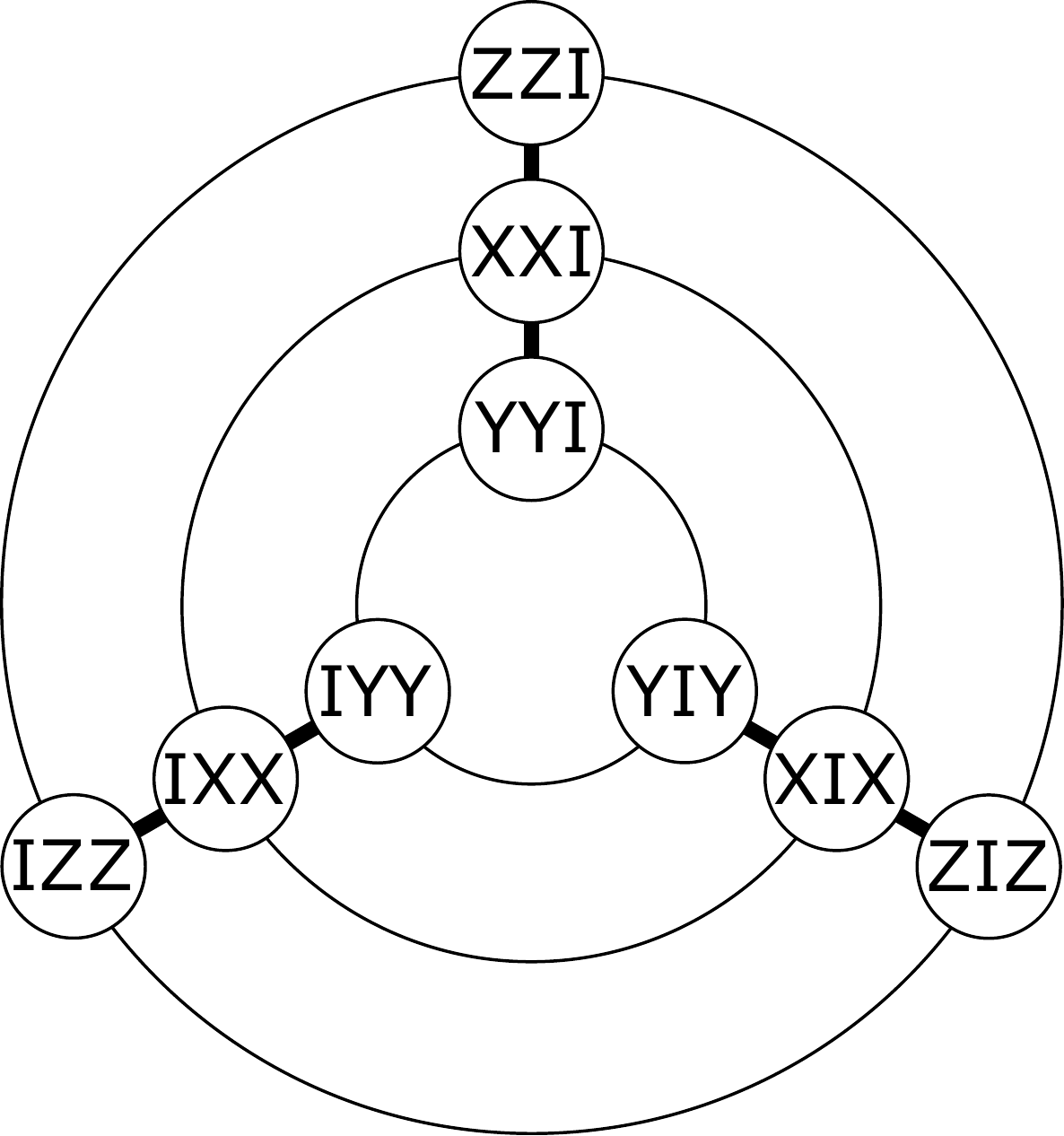}\\
  \caption[3-qubit Wheel]{The 3-qubit Wheel, a $9_2 -6_3$ Observable-based KS proof for 3 qubits, isomorphic to the Mermin Square.  Each line or large circle connects the observables of an ID.  The thin lines denote Positive IDs, while the thick lines denote Negative IDs.  The 3 ID$3^3_2$ of the Kernel form the `spokes' of the Wheel.}\label{Wheel3}
\end{figure}

Now we move on to discuss the 3-qubit Wheel \cite{WA_3qubits} of Figure \ref{Wheel3}, an Observable-based KS proof generated from the Kernel of Table \ref{Wheel3Kernel}.

In this remarkable case, we do not need to introduce any additional observables in order to form the new IDs of an Observable-based KS proof.  Instead we will form 3 new Positive ID3s simply by regrouping the observables of the 3 original Negative ID3s in different ways.  The specific process is shown in Table \ref{Wheel3IDs}.  It is important to note that this diagram is isomorphic to the Mermin square, and we could just as easily draw it as a square.  The wheel form is used instead because it is more convenient for the complete family of $N$-qubit Wheels ($N$ odd), and the related family of $N$-qubit Whorls ($N$ even), which we will discuss in Chapter \ref{sec:Wheel_Whorl}.

\begin{table}[ht]
\centering
\subfloat[][]{
\begin{tabular}{ccccc}
$Z$ & $Z$ & $I$ \\
$X$ & $X$ & $I$ \\
$Y$ & $Y$ & $I$ \\
\end{tabular}\label{IDW1}}
\qquad
\subfloat[][]{
\begin{tabular}{ccccc}
$I$ & $Z$ & $Z$ \\
$I$ & $X$ & $X$ \\
$I$ & $Y$ & $Y$ \\
\end{tabular}\label{IDW2}}
\qquad
\subfloat[][]{
\begin{tabular}{ccccc}
$Z$ & $I$ & $Z$ \\
$X$ & $I$ & $X$ \\
$Y$ & $I$ & $Y$ \\
\end{tabular}\label{IDW3}}
\qquad\qquad\qquad\qquad\qquad
\subfloat[][]{
\begin{tabular}{ccccc}
$Z$ & $Z$ & $I$ \\
$I$ & $Z$ & $Z$ \\
$Z$ & $I$ & $Z$ \\
\end{tabular}}
\qquad
\subfloat[][]{
\begin{tabular}{ccccc}
$X$ & $X$ & $I$ \\
$I$ & $X$ & $X$ \\
$X$ & $I$ & $X$ \\
\end{tabular}}
\qquad
\subfloat[][]{
\begin{tabular}{ccccc}
$Y$ & $Y$ & $I$ \\
$I$ & $Y$ & $Y$ \\
$Y$ & $I$ & $Y$ \\
\end{tabular}}
\qquad
\caption[The 6 IDs of the 3-qubit Wheel]{The 6 IDs of the 3-qubit Wheel (Figure \ref{Wheel3}).  The 3 Negative ID3s of the Kernel are shown in \subref{IDW1}, \subref{IDW2}, and \subref{IDW3}.  The other 3 Positive ID3s show how the Kernel generates the set.  In each case, the one observable from each of the 3 Kernel IDs is used to form a new ID.}\label{Wheel3IDs}
\end{table}

This proof reveals another important feature of Observable-based KS proofs having to do with the fact that it is isomorphic to the Mermin Square.  By isomorphic, we mean that the observables and IDs of the two proofs fit together in exactly the same way, and thus that their diagrams have exactly the same structure.  Remarkably, if two Observable-based KS proofs have isomorphic diagrams, then they also generate isomorphic sets of rays, meaning that the rays and bases of both sets also fit together in exactly the same way.  The ranks of the associated rays may be different, since the ID$M^N$s in two isomorphic Observable-based KS proofs need only have the same set of $M$ values, but need not have the same number of qubits $N$.  This is exactly the case for the 3-qubit Wheel and Mermin Square, and so we can simply refer back to the $24^1_4-24_4$ of Chapter \ref{sec:MerminSquare} to see the structure of the $24^2_4-24_4$ set generated by the 3-qubit Wheel.  As the symbols indicate, the only difference between the two sets is that the 24 rays generated by the 3-qubit Wheel are of rank 2 and occupy a Hilbert space of $d=8$.  Another important thing to note here is that two Observable-based KS proofs (and their corresponding $R-B$ sets) can still be isomorphic if the numbers of negative IDs they each contain are not the same, provided they are both odd - because the sign of the IDs does not affect the overall orthogonality relations between the eigenbases.

It is worth noting that the $18^2_2 -9_4$ parity proof contained within this set is the most compact proof of the KS theorem yet discovered in $d=8$, and we doubt that a smaller proof exists.  This proof also encompasses a continuum of $36^1_2-9_8$ proofs that can be obtained by choosing particular values for the internal degrees of the freedom of the 18 rank-2 projectors.  A proof like this was given by Ruuge \cite{Ruuge},\cite{Ruuge_Err}.  The $12^1_2 12^2_2 - 1_8 4_6 4_4$ proof from Chapter \ref{sec:Kite3} is also a case like this, with the internal degrees of freedom fixed for 6 of the rays to give 12 rank-1 projectors.  This issue is examined in more detail in Chapter \ref{IDEntanglement}.

\begin{figure}[h!]
  \centering
  \includegraphics[width=3in]{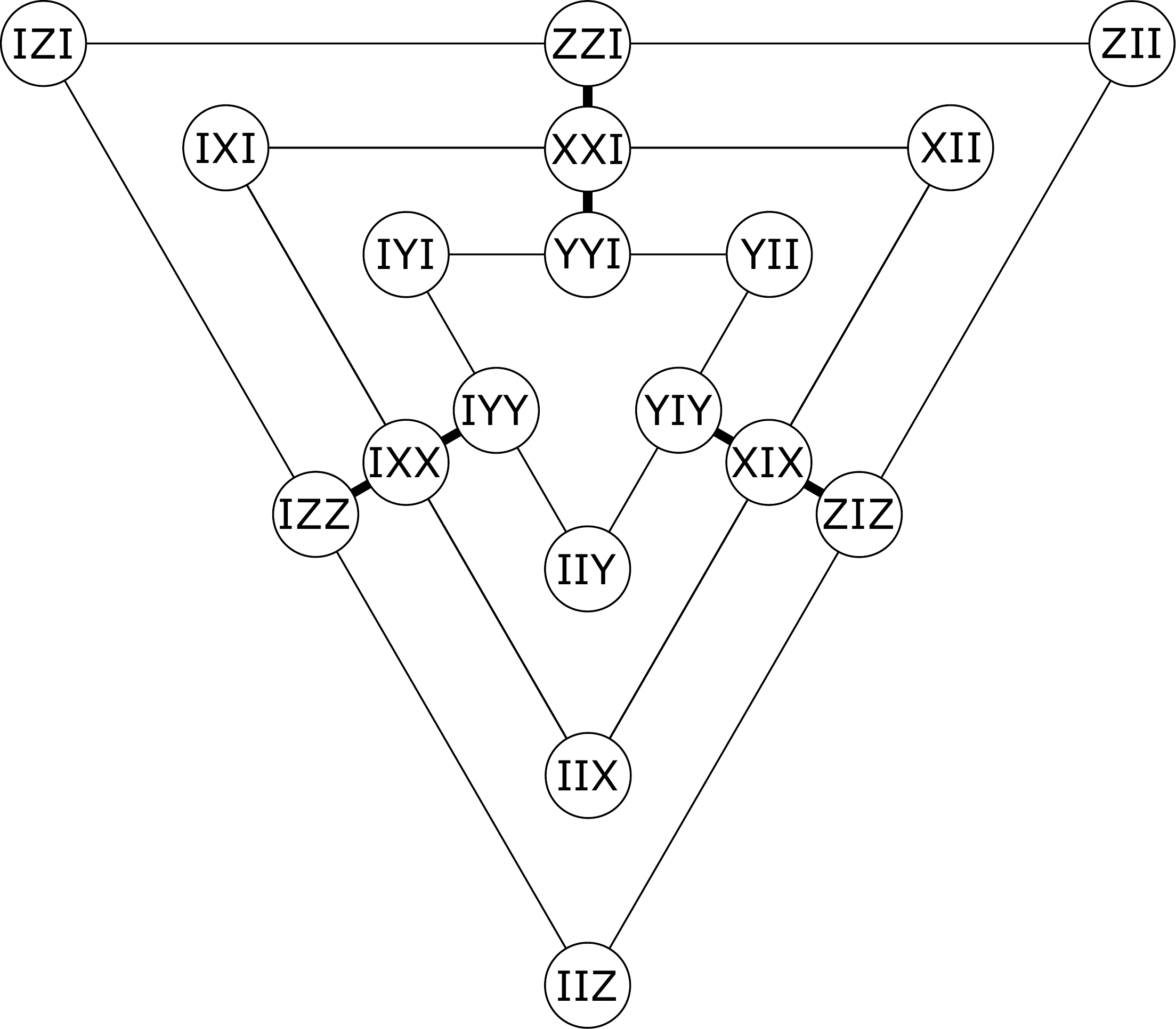}\\
  \caption[Expanded 3-qubit Wheel]{This is the expanded version of the Wheel diagram of Figure \ref{Wheel3}, obtained by replacing each circular ID3 with a triangle of 3 new ID3s with 3 new observables at the corners.}\label{Wheel3_Exp}
\end{figure}

The Wheel family of proofs has members for all odd $N\geq 3$, and is closely related to the Whorl family which has members for all even $N\geq 2$.  We will discuss both of these families in Chapter \ref{sec:Wheel_Whorl}. For now, let us consider some alternate proofs that can be generated from the same Kernel.  Each circular ID3 in Figure \ref{Wheel3} can be independently replaced by a triangle of 3 new ID3s, with 3 new observables at the corners.  The case where all 3 circles are replaced in this way is the $18_2 - 12_3$ proof shown in Figure \ref{Wheel3_Exp}, and discussed by Saniga et al \cite{Planat_Saniga2}.  Because each of the 18 observables is shared by exactly two IDs, there are exactly 18 complementary pairs of hybrid bases.  This proof then generates a $48^2_4-48_4$ set that contains $2^{18} = 262,144$ critical parity proofs, which can be obtained using the complementary pairs of hybrid bases as described in Chapter \ref{sec:MerminSquare}.  These proofs also come in complementary pairs, where the two proofs in a pair together use all 48 bases.
\begin{figure}[h!]
  \centering
  \includegraphics[width=2.5in]{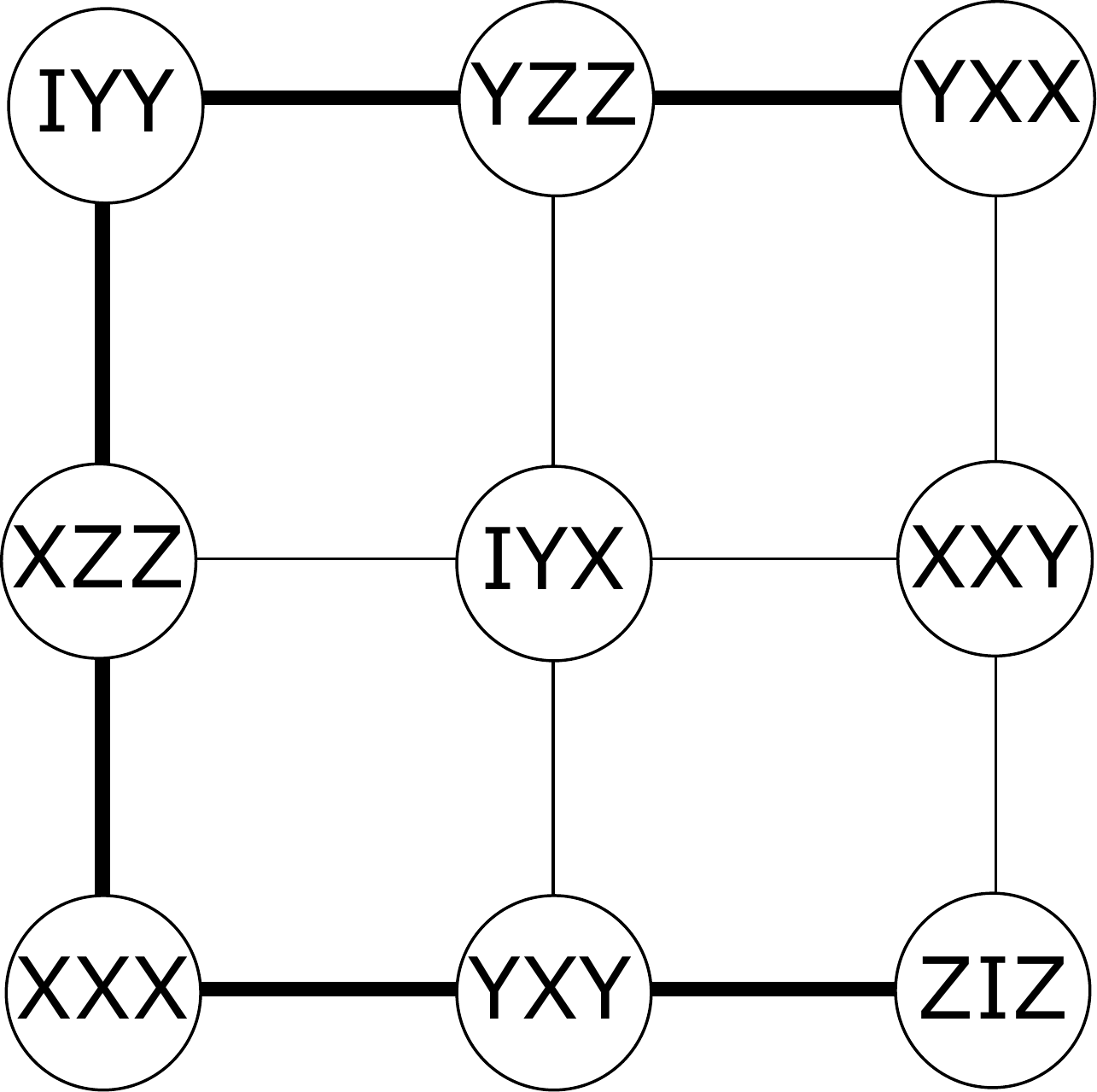}\\
  \caption[Special 3-qubit Square]{This critical 3-qubit Observable-based KS proof is isomorphic to both the Mermin Square and the 3-qubit Wheel.  Each line connects the observables of an ID.  The thin lines denote Positive IDs, while the thick lines denotes Negative IDs.  The 2-qubit Kernel that generates this set is composed of the first vertical ID, and the middle horizontal ID.  The Kernel is comprised of the second and third qubits, while the first qubit is trivial in both IDs.}\label{q3Square_K2}
\end{figure}

There is one final proof that we will discuss in this section, because it also introduces a new structural feature of the Observable-based KS proofs within the $N$-qubit Pauli group.  Figure \ref{q3Square_K2} shows another 3-qubit proof isomorphic to the Mermin Square and 3-qubit Wheel.  What is unique about this 3-qubit Observable-based KS proof, is that it is critical in the sense that no subset of qubits and/or IDs can be deleted such that the remaining set is still an Observable-based KS proof, but it is generated from a 2-qubit critical Kernel (a permutation of the Kernel in Figure \ref{q2kernel}).

So the relationship between the criticality of IDs, Kernels, and Observable-based KS proofs is quite intricate.  Noncritical IDs can generate critical Kernels and/or Observable-based KS sets, and noncritical Kernels can generate critical Observable-based KS sets.  Criticality is still a very useful idea, because it allows us to find and catalog the minimal elemental structures that exist within the $N$-qubit Pauli group, but we will have to carefully catalog each of these three structures separately.  In reality, the number of possible Kernels and Observable-based KS proofs grows so prolifically as $N$ increases that we will make no attempt to catalog all of them.  However, the number of unique critical IDs, at least for some small values of $N$ can actually be explored in complete detail, as we will see in Chapter \ref{sec:IDs}.  It turns out that we can also build the complete set of possible critical Composite Kernel Structures for small values of $N$, which we discuss in Chapter \ref{sec:Kernels}.

\pagebreak

%
%\end{document}

%\documentclass[12pt]{article}
%\usepackage{graphicx}
%\usepackage{epsfig}
%\usepackage{amsfonts}
%\usepackage[lofdepth,lotdepth]{subfig}
%
%\begin{document}

\section{4 qubits}\label{sec:q4}

In this chapter we will discuss several interesting structures within the $d=16$ Hilbert space of 4 qubits.  The situation for 5 qubits becomes much more involved, and so this is the largest number of qubits we will devote a full chapter to.  To begin we will introduce the complete set of critical ID$^4$s of the 4-qubit Pauli group.  Next we will introduce the complete set of critical Composite Kernel structures for 4 qubits.  Then we will move on to discuss a few specific proofs, beginning with the 4-qubit Star, which is the first member of a family of Observable-based KS proofs that exist for all even $N\geq4$.  Finally, we will examine the 4-qubit Whorl, and another new proof we call the 4-qubit Windmill.

\subsection{The IDs of the 4-qubit Pauli Group}\label{sec:q4IDs}

In this section we will introduce the complete set of critical ID$^4$s.  Unlike the previous chapter, we will show only the critical IDs, but clearly we can form additional IDs by adding trivial qubits to the ID$^N$s for $N<4$, just as was done to obtain the IDs of Table \ref{q3ID3s}.  The complete families have been obtained by an exhaustive computer search.  The full search actually returns 4 different critical ID$4^4$s, and 68 different critical ID$5^4$s, but it appears that just 2 of the ID$4^4$s and 7 of the ID$5^4$s are unique.  We say `appears' here, because it is possible that we have missed some subtle distinction between two of the 68 ID$5^4$s that actually makes them both unique.  The full search results are undoubtedly complete, but it is possible they contain additional unique IDs beyond the 9 we list in Table \ref{q4IDs}.

Again, the permutations of the Partial IDs given here can be assigned to Composite Kernel Structures in an enormous variety of combinations to form Kernels and Observable-based KS proofs.  We will examine only a few interesting cases in detail, later in this chapter.
\begin{table}[ht]
\centering
\qquad
\subfloat[][ID$4^4_2$]{
\begin{tabular}{cccc}
$Z$ & $Z$ & $Z$ & $Z$ \\
$X$ & $X$ & $X$ & $X$ \\
$Y$ & $I$ & $Z$ & $X$ \\
$I$ & $Y$ & $X$ & $Z$ \\
\end{tabular}\label{ID4_q4_1}}
\qquad
\subfloat[][ID$4^4_4$]{
\begin{tabular}{cccc}
$Z$ & $Z$ & $Z$ & $I$ \\
$X$ & $X$ & $I$ & $Z$ \\
$Y$ & $I$ & $X$ & $X$ \\
$I$ & $Y$ & $Y$ & $Y$ \\
\end{tabular}\label{ID4_q4_2}}
\qquad
\subfloat[][ID$5^4_0$]{
\begin{tabular}{cccc}
$Z$ & $Z$ & $Z$ & $Z$ \\
$Z$ & $Z$ & $X$ & $X$ \\
$X$ & $X$ & $I$ & $I$ \\
$X$ & $I$ & $Z$ & $X$ \\
$I$ & $X$ & $X$ & $Z$ \\
\end{tabular}\label{ID5_q4_1}}
\qquad
\subfloat[][ID$5^4_2$]{
\begin{tabular}{cccc}
$Z$ & $Z$ & $Z$ & $Z$ \\
$X$ & $X$ & $Z$ & $Z$ \\
$Y$ & $I$ & $X$ & $I$ \\
$I$ & $Y$ & $I$ & $X$ \\
$I$ & $I$ & $X$ & $X$ \\
\end{tabular}\label{ID5_q4_2}}
\qquad
\subfloat[][ID$5^4_2$]{
\begin{tabular}{cccc}
$Z$ & $Z$ & $Z$ & $Z$ \\
$X$ & $I$ & $X$ & $I$ \\
$Y$ & $I$ & $Z$ & $X$ \\
$I$ & $X$ & $X$ & $Z$ \\
$I$ & $Y$ & $I$ & $X$ \\
\end{tabular}\label{ID5_q4_3}}
\qquad
\subfloat[][ID$5^4_2$]{
\begin{tabular}{cccc}
$Z$ & $Z$ & $Z$ & $I$ \\
$X$ & $X$ & $I$ & $Z$ \\
$Y$ & $I$ & $X$ & $X$ \\
$I$ & $Y$ & $X$ & $X$ \\
$I$ & $I$ & $Z$ & $Z$ \\
\end{tabular}\label{ID5_q4_4}}
\qquad
\subfloat[][ID$5^4_2$]{
\begin{tabular}{cccc}
$Z$ & $Z$ & $Z$ & $I$ \\
$X$ & $X$ & $Z$ & $Z$ \\
$Y$ & $Z$ & $X$ & $X$ \\
$I$ & $Z$ & $I$ & $X$ \\
$I$ & $Y$ & $X$ & $Z$ \\
\end{tabular}\label{ID5_q4_5}}
\qquad
\subfloat[][ID$5^4_2$]{
\begin{tabular}{cccc}
$Z$ & $Z$ & $Z$ & $I$ \\
$X$ & $X$ & $I$ & $Z$ \\
$Y$ & $Z$ & $X$ & $Z$ \\
$I$ & $Z$ & $Z$ & $X$ \\
$I$ & $Y$ & $X$ & $X$ \\
\end{tabular}\label{ID5_q4_6}}
\qquad
\subfloat[][ID$5^4_2$]{
\begin{tabular}{cccc}
$Z$ & $Z$ & $Z$ & $I$ \\
$Z$ & $X$ & $X$ & $Z$ \\
$Z$ & $Y$ & $X$ & $X$ \\
$X$ & $X$ & $Z$ & $Z$ \\
$Y$ & $X$ & $I$ & $X$ \\
\end{tabular}\label{ID5_q4_7}}
\qquad
\caption[The 9 critical 4-qubit IDs]{The 9 critical ID$^4$s of the 4-qubit Pauli group.  The Whole ID of \subref{ID5_q4_1} is definitely unique, and is the only single-ID Kernel for 4 qubits.}\label{q4IDs}
\end{table}

\pagebreak

\subsection{Composite Kernel Structures for 4 qubits}

In this section we will expand on the definition and description of the Critical Kernel Structures (CKSs) introduced in Chapter \ref{sec:q3kernels}.  Table \ref{CompKernels3} shows the list of Composite Kernels that can be built using the IDs available for 3 qubits.  Note that the Composite Kernels of Tables \ref{CKS2_3} and \ref{Kite3Struct} are actually built using the 2-qubit CKS of Table \ref{CompKernel2}, but these 3-qubit Kernels are still critical because the ID$4^3_2$ of Table \ref{q3PartialID} is critical.  The other 4 Composite Kernels of Table \ref{CompKernels3} are actually built on the CKS of Table \ref{CKS3}, as we will explain.
\begin{table}[ht]
\centering
\qquad
\subfloat[][]{
\begin{tabular}{cc}
O & O \\
O & O \\
\end{tabular}\label{CKS2}}
\qquad
\subfloat[][]{
\begin{tabular}{ccc}
O & O & I \\
I & O & O \\
O & I & O \\
\end{tabular}\label{CKS3}}
\qquad
\subfloat[][]{
\begin{tabular}{cccc}
O & O & O & O \\
O & O & O & O \\
\end{tabular}\label{CKS4_2}}
\qquad
\subfloat[][]{
\begin{tabular}{cccc}
O & O & O & O \\
O & O & I & I \\
I & I & O & O \\
\end{tabular}\label{CKS4_3}}
\qquad
\subfloat[][]{
\begin{tabular}{cccc}
O & O & O & O \\
O & O & I & I \\
O & I & O & I \\
O & I & I & O \\
\end{tabular}\label{CKS4_4_1}}
\qquad
\subfloat[][]{
\begin{tabular}{cccc}
O & O & I & I \\
I & O & O & I \\
I & I & O & O \\
O & I & I & O \\
\end{tabular}\label{CKS4_4_2}}
\qquad
\caption{The 6 Critical Composite Kernel Structures for $N=2,3,4$ qubits.}\label{CompStructs234}
\end{table}

The only relevant property of an ID where CKSs are concerned is its Oddness.  In fact, we no longer need to distinguish between `E' and `I' in our Oddness Profiles either, since these can be arbitrarily determined by the choice of IDs we assign to the structure, as seen in Table \ref{CompKernels3}.  In constructing CKSs for a given number $N$ of qubits, we will only consider cases where the Odd SQPs of all IDs in the set are assigned to all $N$ qubits.  A Composite Kernel Structure for $N$ qubits can then be composed of any mix of Partial IDs with Oddnesses from the set of even integers from 2 to N.  Finally, we will say that a CKS is critical if no subset of IDs and/or Odd qubits can be deleted such that the remaining set is a CKS - where an Odd SQP in any ID in the set means the corresponding qubit is Odd.

The critical CKSs for 2, 3, and 4 Odd qubits are shown in Table \ref{CompStructs234}, where again each row is a different ID, and each column a different qubit.  It is important to note that these are only the Odd qubits, since as we have seen in Tables \ref{CKS2_3} and \ref{Kite3Struct}, such a structure can easily be used to generate a critical Kernel with more qubits.  Likewise, the `I's can be trivially replaced with `E's when assigning IDs to a CKS.  These complete listings of CKSs were generated by an exhaustive computer search.

We explain a few specific cases to make certain the meaning of these tables is clear.  In Table \ref{CKS4_2}, we assign any two different IDs of Oddness 4.  In Table \ref{CKS4_3}, we assign an ID of Oddness 4, and 2 IDs of Oddness 2.  Finally, in Table \ref{CKS4_4_2}, we assign any 4 different IDs of Oddness 2.  The number of observables and/or qubits in each ID plays no role in these assignments, so long as their Odd SQPs are matched up as shown.  We will discuss several interesting consequences of this in Chapter \ref{sec:Kernels}.

We already knew that the number of permutations of IDs within the 4-qubit Pauli group is quite large, meaning that the number of ways that they can be assigned to these Composite Kernel Structures is truly astounding, never mind how many Observable-based KS proofs they might lead to.  Again, we will only consider a few special cases in this Chapter. 

\pagebreak

\subsection{The 4-qubit Star}\label{sec:q4Star}

\begin{figure}[ht]
  \centering
  \includegraphics[width=3in]{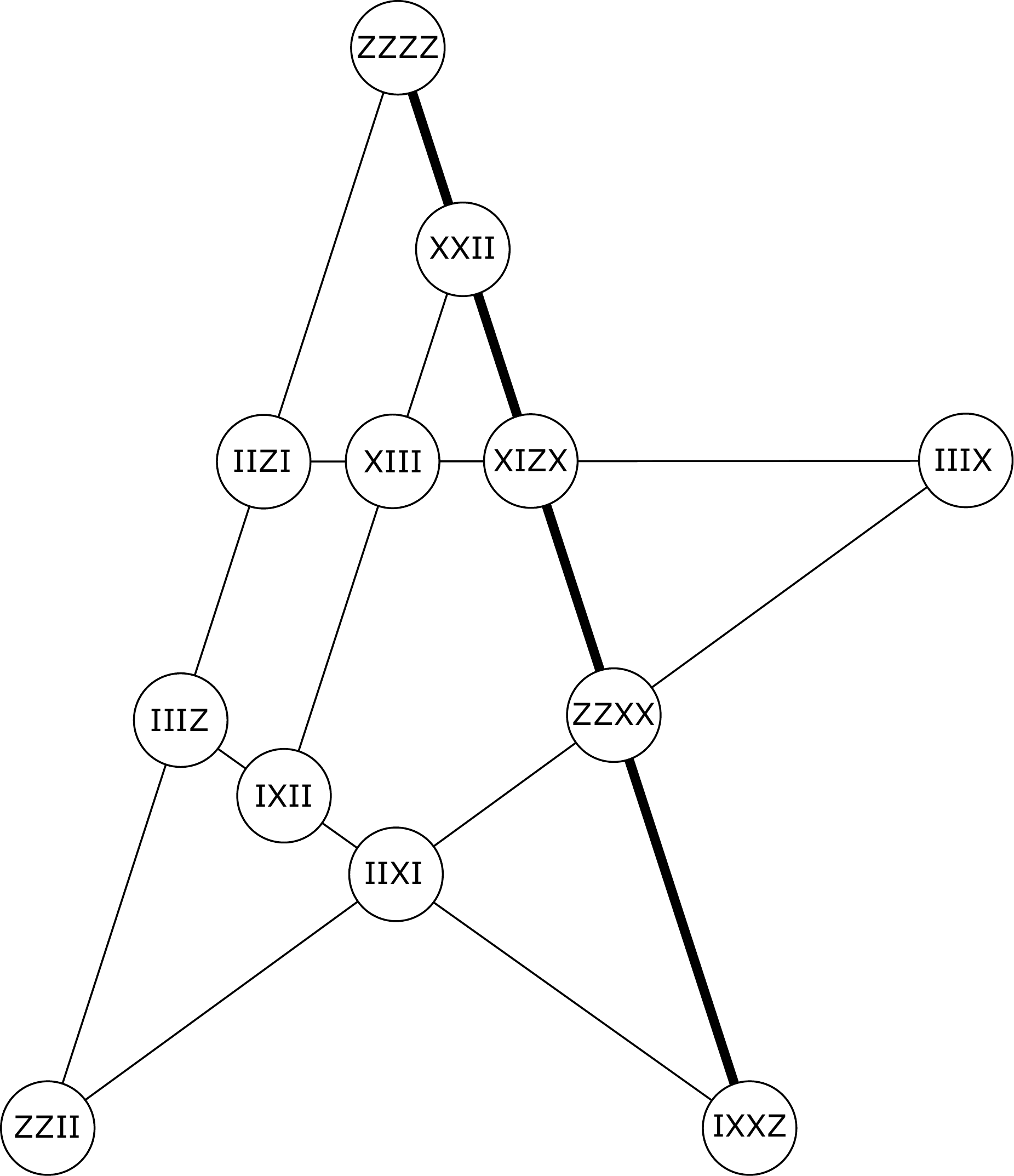}\\
  \caption[4-qubit Star]{The 4-qubit Star, a $12_2 - 1_5 4_4 1_3$ Observable-based KS proof.  Each line connects all of the observables in an ID.  Thin lines denote Positive IDs, while the thick line denotes the Negative ID$5^4_0$ of the Kernel.}\label{q4Star}
\end{figure}

In this section we discuss the 4-qubit Star shown in Figure \ref{q4Star}, an Observable-based KS proof generated by a permutation of the single-ID$5^4_0$ Kernel of Table \ref{ID5_q4_1}.  This proof is generated from the Kernel by the same process we used in \ref{sec:Kite3}, as should be clear from the diagram.

Like the Kernel that generates the Mermin Star of Chapter \ref{sec:MerminStar}, the ID$5^4_0$ that generates this proof can be used to prove the GHZ theorem using its own eigenstates \cite{WA_4qubits}.  The same is true of all single-ID Kernels.

For this set, each of the 12 observables is shared by exactly two IDs, and thus the rays of the 6 eigenbases mix to form 12 complementary pairs of hybrid bases.  The complete $52-30$ set has expanded symbol $16^1_6 32^2_5 4^4_4 - 1_{16} 8_{12} 2_{10} 14_8 4_6 1_4$.  In the same way we did for the other sets in which no two IDs share more than one observable in common (see Chapter \ref{sec:MerminSquare}), we can use the 12 pairs of complementary hybrid bases to generate the $2^{12} = 4,096$ critical parity proofs of this set.  The smallest critical parity proof in this set is a $47-13$, and the largest $51-17$.

There is one more interesting detail of the particular ID$5^4_0$ Kernel used in Figure \ref{q4Star}, which is the type of entanglement possessed by the states of its eigenbasis.  As discussed in our paper \cite{WA_4qubits}, these states exhibit a remarkable type of robust entanglement that might have many interesting applications.  They belong to a family of fully entangled states that we call Web States, that spans all numbers of qubits $N\geq 2$.  The property of these states that earns them their name is that if any qubit from an $N$-qubit Web State is measured in the product basis, the remaining qubits are always left in an $(N-1)$-qubit Web State.  This entanglement persists in this way as successive qubits are measured, regardless of order, until the last two that remain are in a Bell state.  We will discuss the two distinct families of Web States within the $N$-qubit Pauli group elsewhere.

\pagebreak

\subsection{The 4-qubit Whorl and the 4-qubit Windmill}
\begin{figure}[ht]
  \centering
  \includegraphics[width=3.5in]{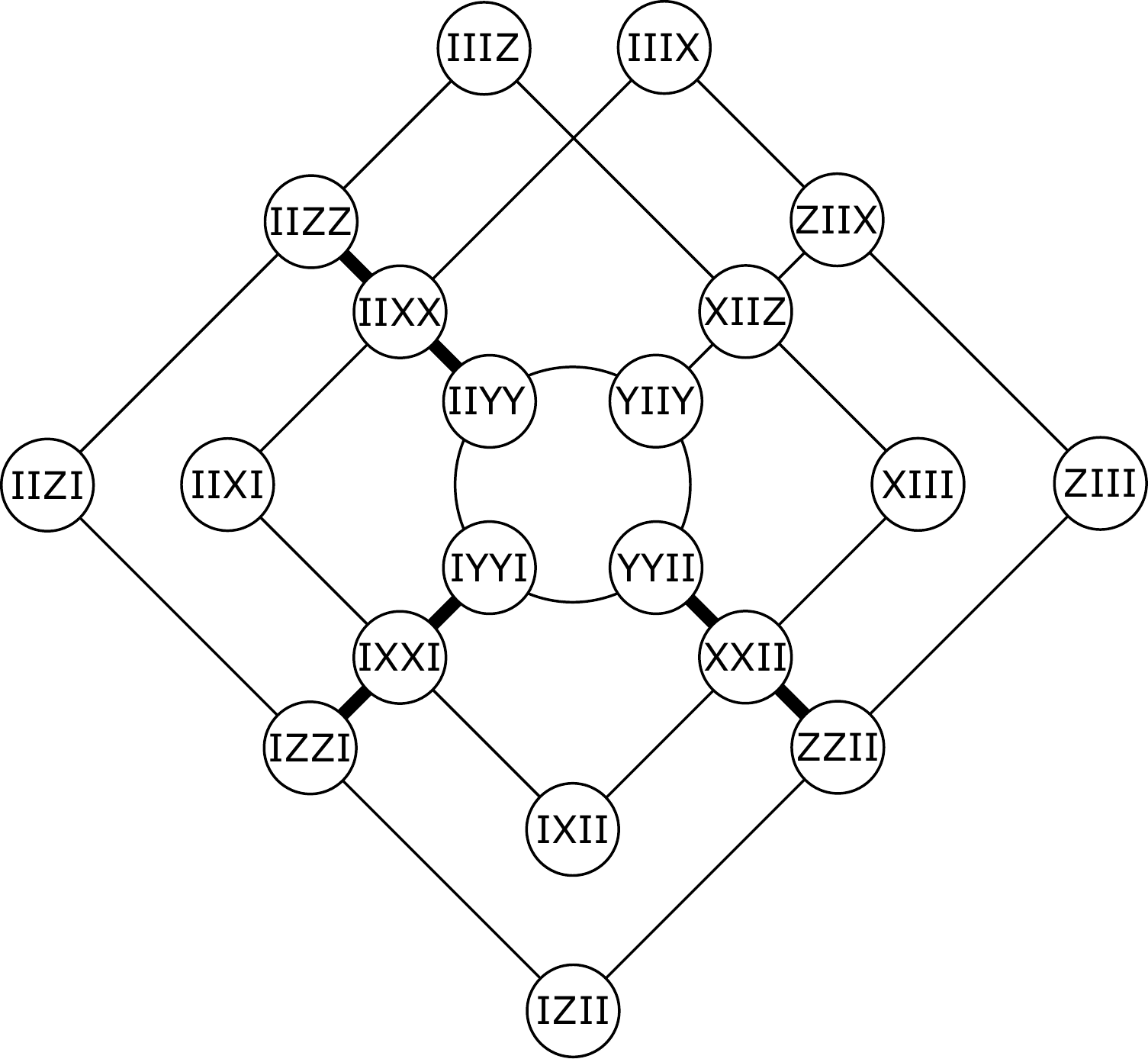}\\
  \caption[4-qubit Whorl]{The 4-qubit Whorl, a $20_2 - 1_4 12_3$ Observable-based KS proof.  Each line and large circle connects all of the observables in an ID.  Thin lines denote Positive IDs, while thick lines denote Negative IDs.  The 4 ID$3^4_2$s of the Kernel form the `spokes' of the Whorl.}\label{q4Whorl}
\end{figure}
\begin{figure}[ht]
  \centering
  \includegraphics[width=3in]{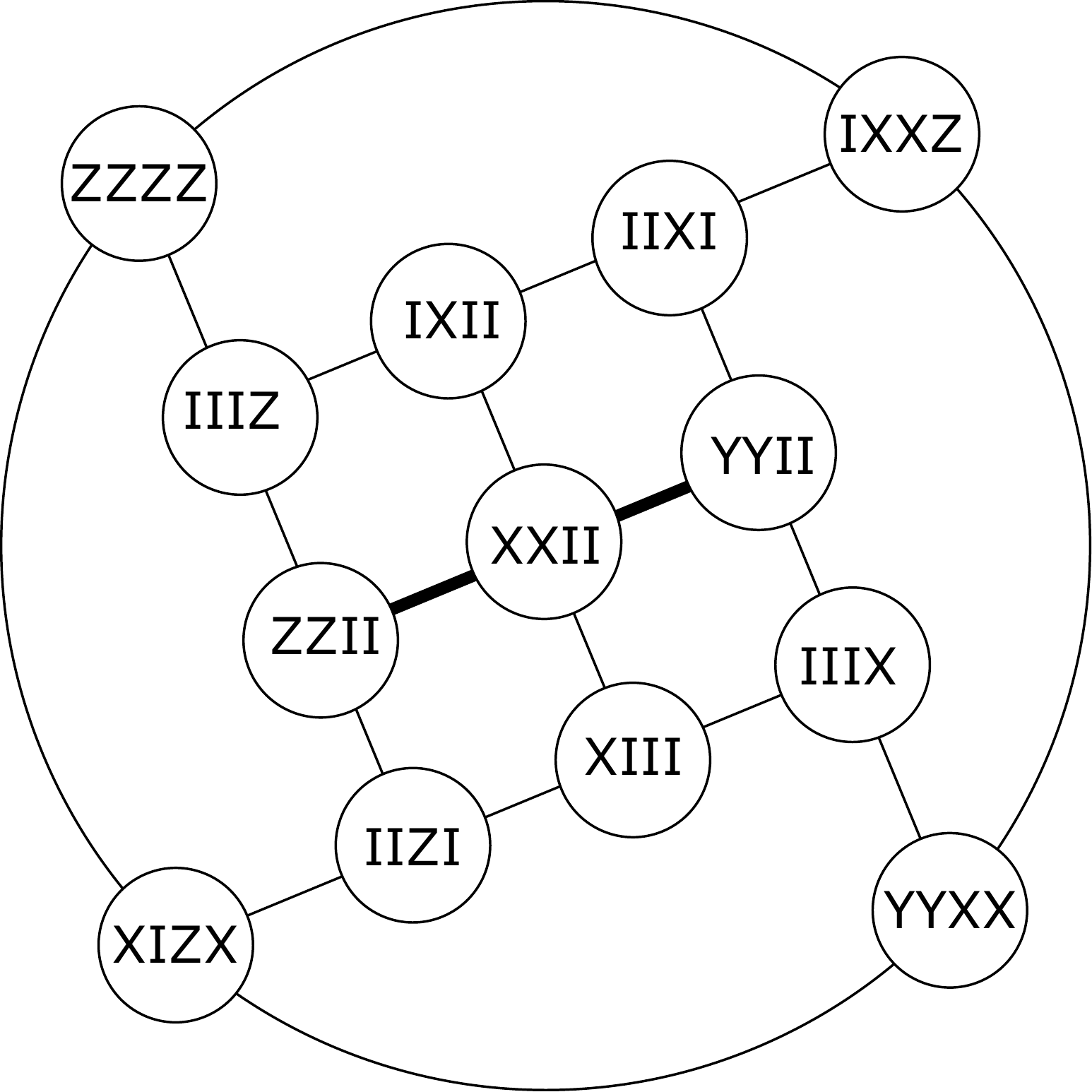}\\
  \caption[4-qubit Windmill]{The 4-qubit Windmill, a $13_2 - 5_4 2_3$ Observable-based KS proof.  The Kernel is composed of the Negative ID$3^4_2$ at the center and the circular ID$4^4_2$.}\label{q4Windmill}
\end{figure}
In this section we introduce the 4-qubit Whorl \cite{WA_4qubits} shown in Figure \ref{q4Whorl}.  The Kernel for this $20_2 - 1_4 12_3$ proof is constructed using the Composite Kernel Structure of Table \ref{CKS4_4_2}, and involves 4 permutations of the noncritical ID$3^4_2$ composed of the critical ID$3^2_2$ paired with two trivial SQPs, chosen such that an odd number of IDs (in this case three) are negative.  The Kernel shows up as the four ``spokes'' surrounding the central wheel in Figure \ref{q4Whorl}.  To generate the full Observable-based KS proof, we first add one ID composed of one observable from each of the IDs in the Kernel, which shows up as the central wheel in Figure \ref{q4Whorl}, and then generate all the remaining IDs using the usual qubit-decomposition method.  The Whorl is never as compact as its cousin the Wheel, because the even number of `spokes' in the Kernel requires the introduction of new observables to form the needed IDs.  We will discuss these two related families of proofs in Chapter \ref{sec:Wheel_Whorl}.

As in similar cases, since each of the 20 observables is shared by exactly two IDs, there are 20 complementary pairs of hybrid bases.  So the 4-qubit Whorl generates a $56-53$ set with expanded symbol $8^2_5 48^4_4 - 1_8 8_6 44_4$, which contains $2^{20} = 1,048,576$ critical parity proofs that can be obtained in the usual way (see Chapter \ref{sec:MerminSquare}).  The smallest critical parity proof in this set is a $40-21$, and the largest a $52-33$.

We can obtain an alternate proof by choosing to use the qubit decomposition method for the entire Kernel, rather than forming the central ID$4$ of Figure \ref{q4Whorl}.  This simply replaces the circular ID$4$ by a square composed of 4 ID$3$s, introducing 4 new observables for the corners.  This alternate $24_2 - 16_3$ proof has a simpler but less compact form.  This proof generates a $64^4_4 - 64_4$ set, which contains $2^{24} = 16,777,216$ critical parity proof that can be obtained in the usual way.  This is another case where the proofs break up into complementary pairs that together use all 64 bases.

In general, the family of $R_4 - B_4$ sets with $R=B$ that are generated by the families of expanded Whorl and Wheel proofs seem always to contain critical parity proofs that come in complementary pairs.

Now we move on to describe the 4-qubit Windmill of Figure \ref{q4Windmill}, a $13_2 - 5_4 2_3$ Observable-based KS proof based on the 2-qubit Composite Kernel Structure, and using the ID$3^2_2$ with two trivial qubits and the ID$4^4_2$ of Table \ref{ID4_q4_1}. For reasons that should now be familiar, this set generates 13 complementary pairs of hybrid bases that can be used to obtain the $2^{13} = 8,192$ critical parity proofs in the usual way.  The smallest proof is a $41-13$, and the largest a $47-19$.  The full $48-33$ has expanded symbol $8^4_4 40^2_5 - 21_8 8_6 4_4$.

\pagebreak

%\end{document}

%\documentclass[12pt]{article}
%\usepackage{graphicx}
%\usepackage{epsfig}
%\usepackage{amsfonts}
%\usepackage[lofdepth,lotdepth]{subfig}
%
%\begin{document}

\section{Structures of the $N$-qubit Pauli Group} \label{sec:Structures}

Throughout the preceding Chapters of this text, we have introduced a number of important types of structures of the $N$-qubit Pauli group.  In this chapter we will review them and formalize their definitions, and importantly the definition of what makes each structure critical.  To begin, we will review IDs and the SQPs that form them.  Next we will review Kernels and Composite Kernel Structures.  After this, we will discuss Observable-based KS proofs.  We will next consider the sets of rays generated by Observable-based KS proofs, and the critical KS sets they contain.  Finally we will discuss the continuous classes of entangled states that are generated by critical IDs.

\begin{figure}[h]
  \centering
  \includegraphics[width=6in]{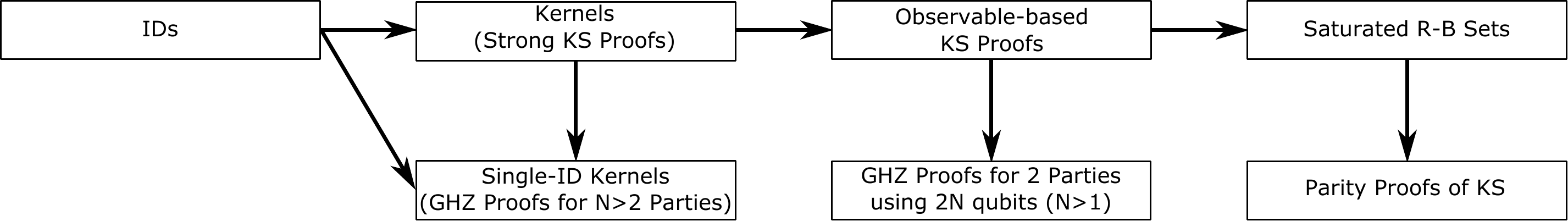}\\
  \caption[$N$-qubit Nonclassicality Flowchart]{This flow chart shows conceptually how each type of nonclassical structure is constructed from more elementary components.}\label{FlowChart}
\end{figure}

All of these structures are related in that the simpler structures are used to build or generate the successively more elaborate ones.  Figure \ref{FlowChart} shows the general flow of how the structures lead into one another, and how they are related to proofs of the GHZ and KS theorems.\newline

\subsection{Identity Products (IDs) and Single-Qubit Products (SQPs)} \label{sec:IDs}

In this section we will formalize the definition of an Identity Product (ID) of observables.  An ID$M^N_O$ is a set of $M$ mutually commuting observables from the $N$-qubit Pauli group, whose combined product is $\pm I$ (the $N$-qubit Identity), and Oddness $O$ (any portion of this symbol can be omitted depending on context).  We will refer to an ID as Positive or Negative, depending on the sign of its product.  An ID is depicted as an $M \times N$ table containing elements from the set $(Z, X, Y, I)$.  Each row of the ID shows a different $N$-qubit Pauli observable, and each column shows the SQP of a different qubit.

We will also refer to these columns as Single-Qubit Products (SQPs), and note that they only occur in two distinct types.  The first type, that we will call an Even SQP, contains an even number of each element of the set $(Z, X, Y)$, and has combined product $\pm I$ (the single-qubit Identity).  The second type, that we will call an Odd SQP, contains an odd number of each element of the set $(Z, X, Y)$, and has combined product $\pm i I$.  We define the Oddness, $O$, of an ID, as the number of Odd SQPs in that ID.  We call IDs with $O=0$ Whole IDs, and those with Oddness $>0$ Partial IDs.  It should be clear that since the product of all the observables of any ID is always real, the Oddness of all IDs is even.

We call Negative ID$_0$s Whole IDs, because they are also single-ID Kernels.  We call Positive ID$_0$s Null IDs, and exclude them from our enumeration of IDs.  This is done because these IDs are never needed to form critical Kernels, though they can be used in the process of generating an Observable-based KS proof from a critical Kernel.  They can also be used as a set of Trivial SQPs to form noncritical IDs for a Composite Kernel.
We call an ID$_O$ with $O>0$ a Partial ID.  The sign of a Partial ID is flexible because it can be changed by permutations of any of its Odd SQPs, a fact we will make use of when using Partial IDs to build Composite Kernels.

We will say that an ID is critical iff no subset of observables and/or qubits can be deleted such that the remaining set is still an ID.  This definition guarantees that we are working with the minimal elemental ID types, and also guarantees that the states in the joint eigenbasis of the $M$ observables of a critical ID$M^N$ are fully entangled over all $N$ qubits.  There are no critical IDs with $M>N+1$, since at most $N$ commuting observables can have independent eigenvalues in a Hilbert space of $d = 2^N$.

We can also say that a Null ID is critical, iff no subset of observables and/or qubits can be deleted such that the remaining set is still an ID (Null or not).  Critical Null IDs still guarantee the entanglement of the $N$ qubits, and so while we exclude them from listings throughout this text, they may be worth examining elsewhere.

Furthermore, to facilitate cataloging of the different types of IDs that exist within the $N$-qubit Pauli group, we will treat all trivial permutations of an ID as belonging to the same unique ID structure.  So, when we show a unique ID, we actually mean a complete class of IDs that can be obtained from all permutations of the given structure.  Since an ID is a mutually commuting set of observables, the order of the rows is arbitrary, and cannot change anything about the set.  We can also freely permute the order of the qubits, which can result in an ID that is different than the original.  Furthermore, we can freely permute the choice of single-qubit Pauli observables within each SQP to obtain additional variation on the original ID.  This latter point is a bit more subtle, but it works because the set $(Z, X, Y)$ mutually anticommutes.  For SQPs with just 2 different members of the set $(Z, X, Y)$, the 6 permutations are $(ZX, XZ, ZY, YZ, XY, YX)$, as shown using the example SQP in Table \ref{SQP2}.  For SQPs with all 3 members of the set $(Z, X, Y)$, the 6 permutations are $(ZXY, XYZ, YZX, ZYX, YXZ, XZY)$, as shown using the example SQP in Table \ref{SQP3}.  All Odd SQPs are of the latter type, and these permutations may change their sign.  Permutation of Even SQPs cannot change their sign.  It should be clear that these permutations are just the cyclic rotations of the coordinate system for each individual qubit, along with the reflection.
\begin{table}[ht]
\centering
\qquad
\subfloat[][]{
\begin{tabular}{c}
$Z$ \\
$Z$ \\
$X$ \\
$X$ \\
\end{tabular}}
\qquad
\subfloat[][]{
\begin{tabular}{c}
$X$ \\
$X$ \\
$Z$ \\
$Z$ \\
\end{tabular}}
\qquad
\subfloat[][]{
\begin{tabular}{c}
$Z$ \\
$Z$ \\
$Y$ \\
$Y$ \\
\end{tabular}}
\qquad
\subfloat[][]{
\begin{tabular}{c}
$Y$ \\
$Y$ \\
$Z$ \\
$Z$ \\
\end{tabular}}
\qquad
\subfloat[][]{
\begin{tabular}{c}
$X$ \\
$X$ \\
$Y$ \\
$Y$ \\
\end{tabular}}
\qquad
\subfloat[][]{
\begin{tabular}{c}
$Y$ \\
$Y$ \\
$X$ \\
$X$ \\
\end{tabular}}
\qquad
\caption[SQP Permutations (a)]{The 6 permutations of an SQP with only two different members of the set $(Z, X, Y)$.  The pattern of commutation and anticommutation for all 6 permutations is identical, so they can be freely exchanged within a unique ID to form a wide variety of IDs.}\label{SQP2}
\end{table}
\begin{table}[ht]
\centering
\qquad
\subfloat[][]{
\begin{tabular}{c}
$Z$ \\
$X$ \\
$Y$ \\
\end{tabular}}
\qquad
\subfloat[][]{
\begin{tabular}{c}
$X$ \\
$Y$ \\
$Z$ \\
\end{tabular}}
\qquad
\subfloat[][]{
\begin{tabular}{c}
$Y$ \\
$Z$ \\
$X$ \\
\end{tabular}}
\qquad
\subfloat[][]{
\begin{tabular}{c}
$Z$ \\
$Y$ \\
$X$ \\
\end{tabular}}
\qquad
\subfloat[][]{
\begin{tabular}{c}
$Y$ \\
$X$ \\
$Z$ \\
\end{tabular}}
\qquad
\subfloat[][]{
\begin{tabular}{c}
$X$ \\
$Z$ \\
$Y$ \\
\end{tabular}}
\qquad
\caption[SQP Permutations (b)]{The 6 permutations of an SQP with all 3 members of the set $(Z, X, Y)$.  The pattern of commutation and anticommutation for all 6 permutations is identical, so they can be freely exchanged within a unique ID to form a wide variety of IDs.  The first three permutations of these Odd SQPs have product $+i I$, while the last three have product $-i I$.  The same 6 permutations would work for an Even SQP, but would all have the same sign.}\label{SQP3}
\end{table}

It should now be clear that in searching for and listing unique IDs, we need only ever use SQPs with the first permutations in Tables \ref{SQP2} and \ref{SQP3}.  Once we have a unique critical ID, we can obtain any member of the general class of critical IDs it describes by permuting the order of the qubits, and by independently permuting $(Z, X, Y)$ within each SQP.  It should be noted that the latter set of permutations do not always generate $6^N$ distinct IDs, since some choices of these permutations simply put the original observables into a different order, which actually changes nothing.  The same thing can happen when permuting the order of the qubits.

For most, if not all critical IDs, it is possible to permute the SQPs such that every observable includes $Y$ for an even number of qubits.  When this is done, all of the observables in the ID are fully real-valued (in the standard basis used for the Pauli Observables), as are the rays of their eigenbasis.  If Kernels and Observable-based KS proofs are built using only real IDs, then they too will of course be real.  To see a clear contrast on this issue, consider that the Mermin Square we examined in Chapter \ref{sec:MerminSquare} was chosen to be real-valued, though other permutations of the qubits would have resulted in its being complex.  For the 2-qubit Whorl of Chapter \ref{sec:q2Whorl} however, even though it is built from the same real Kernel, there is no choice of permutations that will make this entire Observable-based KS proof real.  Thus sets of IDs can be irreducibly complex.  There may also exist individual IDs like this as well, though all of the cases we have examined in detail can be made fully real by a suitable choice of permutations.

There is only one unique 2-qubit ID, as shown in the upper half of Table \ref{q2kernel}.  There are two unique 3-qubit IDs, shown in Table \ref{q3ID4s}.  The 9 unique 4-qubit IDs are shown in Table \ref{q4IDs}.

As a final note on the usefulness of critical IDs to characterize entanglement, consider that we know with certainty that the eigenstates of a critical ID$M^N$ are fully entangled simply because the ID cannot be factored into the direct product of two IDs.  We do not need to examine the full set of $2^N -1$ mutually commuting observables that share this eigenstate (the stabilizer group) to see this, since a well chosen subset of just $M\leq N+1$ of them can suffice.  A critical ID is then a sort of minimal {\it stamp} of entanglement for an eigenbasis of the $N$-qubit Pauli group.  A given stabilizer group for an entangled $N$-qubit state must contain at least one critical ID$^N$, and may contain IDs of several different unique types.  This issue is explored further in Chapter \ref{IDEntanglement}.

\pagebreak

\subsection{Kernels and Composite Kernel Structures (CKSs)} \label{sec:Kernels}

In this section we will formalize the definitions of Kernels and Composite Kernel Structures (CKS), and what makes each critical.  

A Kernel is any set of one or more IDs that satisfy two properties.  First, the set contains an odd number of Negative IDs.  Second, for each qubit, the observables $(Z, X, Y)$ each appear an even number of times throughout all of the IDs in the set.  These properties guarantee that the Kernel proves the Strong KS theorem, and generates (usually many) Observable-based KS proofs.  Kernels divide into two distinct classes:  A Single-ID Kernel contains just one Negative Whole ID.  A Composite Kernel contains multiple Partial IDs, such that an odd number are Negative.

The Strong KS theorem is an extension of the usual KS theorem, wherein we require noncontextual truth-values $\pm 1$ to be assigned to every single-qubit observable $(Z, X, Y)$ in every ID.  Any Kernel gives a proof that quantum mechanics rules out the existence of these truth-values, which can be seen from the fact that on the one hand, because every single-qubit observable occurs an even number of times, the overall product of the IDs in the Kernel must be +1, while on the other hand, because an odd number of the IDs are negative, the quantum prediction (and experimental result) will always be -1.  Positive Whole (Null) IDs fail to prove the Strong KS theorem, since the quantum prediction will always be +1, in full agreement with the noncontextual truth-values.

Critical Single-ID$M^N$ Kernels can also be used to prove the GHZ theorem using more than two spacelike separated parties.  A source repeatedly produces a particular joint eigenstate of the ID, and sends one particle to each of $N$ parties.  Each party randomly chooses one of the four measurement bases, $Z$, $X$, $Y$, or $I$ ($I$ means they do nothing), and measures their qubit, such that the choice and measurement for all $N$ qubits are spacelike separated.  In any run where the $N$ parties randomly measure one of the $M$ different $N$-qubit observables of the ID, the product of their individual results is correlated, and must equal the eigenvalue of that $N$-qubit observable in the prepared state.  The product of all $M$ such eigenvalues is -1 in any prepared eigenstate, because the ID is negative.  Because the measurements are chosen randomly, a local hidden variables theory requires that a truth-value $\pm 1$ be preassigned to all single-qubit observables that might be measured.  Furthermore, locality also guarantees that no party's measurement result cannot be affected by any other party's choice of measurement basis, meaning the single-qubit observables in the ID must be assigned noncontextual truth-values.  If we define $B$ as the product of the IDs, then because each single-qubit observable appears an even number of times in the ID, the truth-values assigned to each one will be raised to an even value, and so the overall product will always be $B_L = +1$.  But we know already that the experimental value is $B_Q = -1$ for any prepared eigenstate, which rules out local hidden variable theories without using probabilities, and proves the GHZ theorem for $N$ parties.

It is also possible to prove the GHZ theorem for $N$ parties using a noncritical Single-ID Kernel by making use of an appropriate $N$-qubit entangled state, as in the GHZ experiment of Walther et al.\cite{Walther2005} using a Cluster State.  As in that experiment, the stabilizer group of the state contains other IDs that are critical, guaranteeing the needed entanglement.

Kernels can also be used to generate complete Observable-based KS proofs.  The process always involves adding new IDs to the set, and there are typically many ways that this can be done such that the result satisfies the requirements to be an Observable-based KS proof.  Numerous examples have been given throughout this text.

Single-ID Kernels are just Negative Whole IDs, as can be seen in Tables \ref{GHZKernel} and \ref{ID5_q4_1}, and so the complete process for finding them and establishing their criticality is encompassed by the process of finding critical IDs.

The set of Partial IDs in a Composite Kernel must be chosen such that every Odd SQP in one ID is paired with an Odd SQP in another ID, and such that an odd number are Negative IDs.  This then guarantees that the combined set has the properties of a Kernel.  Explicit examples of Composite Kernels can be seen in Tables \ref{q2kernel}, \ref{Kite3Kernel}, and \ref{Wheel3Kernel}.

We will say that a Composite Kernel is critical iff no subset of IDs and/or qubits can be deleted to result in a smaller Kernel.  As we will see, verifying that a Composite Kernel is critical is a subtle business that we will have to consider very carefully.  It can also be checked computationally by a direct brute force approach.

A Composite Kernel is built by assigning an appropriate set of Partial IDs to a Composite Kernel Structure (CKS).  The Composite Kernel Structure only dictates how the Odd SQPs within the set of IDs must fit together, and so the only qubits listed in a CKS are Odd qubits - where any qubit for which some IDs have an Odd SQP is an Odd qubit.  A CKS is shown as a table of `O's and `I's, where each column shows the Odd qubits of a different ID, and each column is a different qubit in the set.  Each column of a CKS must contain an even number of `O's.  We say that a CKS is critical iff no subset of IDs and/or qubits can be deleted such that the remaining set is a smaller CKS.  The complete list of critical CKSs for $N=2,3,4$ is given in Table \ref{CompStructs234}.  We have used a computer to exhaustively generate all unique critical CKSs for up to $N=7$ Odd qubits, where by unique we mean unique up to permutations of the qubits.  For $N=5$ there are 10 critical CKSs, as shown in Table \ref{CompStructs5}.  For $N=6$ there are 109 critical CKSs, and for $N=7$ there are 1,521 critical CKSs.  A complete listing of these can be found on our website \cite{MainWebsite}.

\begin{table}[ht]
\centering
\qquad
\subfloat[][]{
\begin{tabular}{ccccc}
O & O & O & O & I \\
O & O & O & I & O \\
I & I & I & O & O \\
\end{tabular}\label{CKS5_3}}
\qquad
\subfloat[][]{
\begin{tabular}{ccccc}
O & O & O & O & I \\
O & O & O & I & O \\
O & O & I & O & O \\
O & O & I & I & I \\
\end{tabular}\label{CKS5_4_1}}
\qquad
\subfloat[][]{
\begin{tabular}{ccccc}
O & O & O & O & I \\
O & O & O & I & O \\
O & I & I & O & I \\
O & I & I & I & O \\
\end{tabular}\label{CKS5_4_2}}
\qquad
\subfloat[][]{
\begin{tabular}{ccccc}
O & O & O & O & I \\
O & O & I & I & I \\
I & I & O & I & O \\
I & I & I & O & O \\
\end{tabular}\label{CKS5_4_3}}
\qquad
\subfloat[][]{
\begin{tabular}{ccccc}
O & O & O & O & I \\
O & O & O & I & O \\
O & O & I & O & O \\
O & I & O & O & O \\
I & O & O & O & O \\
\end{tabular}\label{CKS5_5_1}}
\qquad
\subfloat[][]{
\begin{tabular}{ccccc}
O & O & O & O & I \\
O & O & O & I & O \\
O & O & I & O & O \\
O & I & O & I & I \\
I & O & O & I & I \\
\end{tabular}\label{CKS5_5_2}}
\qquad
\subfloat[][]{
\begin{tabular}{ccccc}
O & O & O & O & I \\
O & O & O & I & O \\
O & O & I & I & I \\
O & I & I & O & I \\
I & O & I & I & O \\
\end{tabular}\label{CKS5_5_3}}
\qquad
\subfloat[][]{
\begin{tabular}{ccccc}
O & O & O & O & I \\
O & O & I & I & I \\
O & I & O & I & I \\
O & I & I & I & O \\
I & I & I & O & O \\
\end{tabular}\label{CKS5_5_4}}
\qquad
\subfloat[][]{
\begin{tabular}{ccccc}
O & O & O & O & I \\
O & I & I & I & O \\
I & O & I & I & O \\
I & I & O & I & O \\
I & I & I & O & O \\
\end{tabular}\label{CKS5_5_5}}
\qquad
\subfloat[][]{
\begin{tabular}{ccccc}
O & O & I & I & I \\
I & O & O & I & I \\
I & I & O & O & I \\
I & I & I & O & O \\
O & I & I & I & O \\
\end{tabular}\label{CKS5_5_6}}
\qquad
\caption{The 10 Critical Composite Kernel Structures for $N=5$ qubits.}\label{CompStructs5}
\end{table}

Any set of IDs with correct Oddnesses can be assigned to a CKS.  The IDs are permuted so that their Odd qubits are assigned to the `O's, and their Even or Trivial qubits are assigned to the `I's.  The result of this process is always a Composite Kernel, though it will not necessarily be critical.

As the final piece of this section we will examine the issue of critical Composite Kernels, and show some interesting examples of how to build them.  The main issue in guaranteeing that a Composite Kernel is critical has to do with the criticality of the IDs we assign to the CKS.  As we have seen, the IDs assigned to a CKS need not be critical in order for the resulting Composite Kernel to be critical, but they nevertheless link certain qubits together in an important way.

The general procedure for establishing whether a Composite Kernel is critical requires that we construct what we call the {\it Criticality Network} of the proof.  

Before attempting to state rules for constructing the Criticality Network of a proof, we need to take a closer look at the properties of noncritical IDs.  We can always form noncritical IDs by combining critical IDs side by side and/or by adding in Trivial SQPs (direct products), though we will not bother with IDs made up entirely of Trivial SQPs.  Even within such an ID, the SQPs of a critical sub-ID can still be linked together in the sense that they can only be removed as a complete unit such that the remaining set forms a smaller ID.  For IDs composed of a single critical ID combined with some number (possibly zero) of Trivial SQPs, this is always the case.  For noncritical IDs composed of multiple critical IDs side by side, determining if these fixed units exist is much more subtle and is usually best checked by a computer program.  Either way, if two qubits in an ID belong to such a unit, then we say that they are {\it Critically Linked} together (which also incidentally means that those qubits are entangled in the eigenstates of the ID).  So a noncritical ID may contain multiple separate groups of Critically Linked qubits.

The Criticality Network is built by starting at an arbitrary SQP `O' of the Composite Kernel, and connecting other SQPs throughout the Kernel according to a particular set of rules.  There are two rules for connecting one SQP in the Kernel to another:  First, if an SQP is an `O', it can be connected to any one other `O' in the same column.  Second, all Critically Linked SQPs in the same row can be connected.  Either or both of these rules are applied as many times as needed until the complete Criticality Network has been formed.  Finally, if all of the qubits and IDs are connected by the Criticality Network, then the Composite Kernel is critical.

This process may sound enigmatic, but the explicit examples of the Tables \ref{CritStruct} and \ref{NonCritStruct}, which both use the same CKS, should help to clarify it.
\begin{table}[ht]
\centering
\subfloat[][]{
\begin{tabular}{cccc}
O & O & O & O \\
\textbf{O} & O & I & I \\
I & I & O & O \\
\end{tabular}}
\qquad
\subfloat[][]{
\begin{tabular}{cccc}
\textbf{O} & O & O & O \\
\textbf{O} & O & I & I \\
I & I & O & O \\
\end{tabular}}
\qquad \qquad \qquad \qquad \qquad \qquad
\subfloat[][]{
\begin{tabular}{cccc}
\textbf{O} & \textbf{O} & \textbf{O} & \textbf{O}  \\
\textbf{O} & O & I & I \\
I & I & O & O \\
\end{tabular}}
\qquad
\subfloat[][]{
\begin{tabular}{cccc}
\textbf{O} & \textbf{O} & \textbf{O} & \textbf{O}  \\
\textbf{O} & \textbf{O} & I & I \\
I & I & \textbf{O} & \textbf{O} \\
\end{tabular}}
\caption[Example Criticality Network (a)]{This sequence of tables shows the steps of building the Criticality Network (shown in bold font) for a Composite Kernel, with a critical ID$^4_4$ assigned to the first row, and critical ID$^2_2$s with Trivial SQPs assigned to the other two rows.  Because the Criticality Network reaches all IDs and qubits of the Composite Kernel, it is critical.}\label{CritStruct}
\end{table}

\begin{table}[ht]
\centering
\subfloat[][]{
\begin{tabular}{ccccc}
O & O & \textbf{O} & O & E \\
O & O & I & I & I \\
I & I & O & O & I \\
\end{tabular}}
\qquad
\subfloat[][]{
\begin{tabular}{ccccc}
O & O & \textbf{O} & \textbf{O} & \textbf{E} \\
O & O & I & I & I \\
I & I & O & O & I \\
\end{tabular}}
\qquad
\subfloat[][]{
\begin{tabular}{ccccc}
O & O & \textbf{O} & \textbf{O} & \textbf{E} \\
O & O & I & I & I \\
I & I & \textbf{O} & \textbf{O} & I \\
\end{tabular}}
\caption[Example Criticality Network (b)]{This sequence of tables shows the steps of building the Criticality Network (shown in bold font) for a Composite Kernel with the following properties.  The first row is assigned a noncritical ID composed of a critical ID$^2_2$ for the first two qubits and a critical ID$^3_2$ for the last three qubit, such that the first two qubits are Critically Linked together, as are the last three qubits.  The other two rows are assigned critical ID$^2_2$s combined with Trivial SQPs.  Because the Criticality Network fails to reach all IDs and qubits of the Composite Kernel, it is noncritical.}\label{NonCritStruct}
\end{table}

\pagebreak

\subsection{Observable-Based KS Proofs}\label{sec:ObKSproofs}

In this section, we formalize the definition of an Observable-based KS proof, first described in Chapter \ref{sec:MerminSquare}. The symbol $O_x - I_y$ denotes a set of $I$ IDs containing $O$ different observables.  Each observable appears in $x$ of the IDs, and each ID contains $y$ observables.  If a set contains IDs of differing sizes, or Observables that appear in differing numbers of IDs, the symbol on either side of the dash may be repeated.  If all values of $x$ are even, and the number of Negative IDs in the set is odd, then it is also an Observable-based KS proof.

To see that this is true, consider that any noncontextual hidden variables theory (NCHVT) must preassign truth-values $\pm 1$ to the $O$ observables in the set.  If we define $A$ to be the product of all IDs in the proof, then because each observable appears in an even number of IDs, all of these truth-values are raised to some even power, meaning that $A_{NC} = +1$.  However, because an odd number of the IDs are negative, we have $A_Q = -1$, which is consistent with experimental results.  This contradiction rules out NCHVTs, regardless of locality, and proves the Kochen-Specker theorem.

Any $N$-qubit Observable-based KS proof can also be extended into a proof of the GHZ theorem.  To do this, a source prepares $N$ pairs of qubits in the correlated Bell State, and then sends one member of each pair to Alice, and the other to Bob.  Both Alice and Bob choose randomly among the IDs in the Observable-based KS proof, and then perform that measurement on their $N$ qubits, such that the two random choices and the two measurements are spacelike separated.  Because of the correlated Bell states, whenever Alice and Bob both measure the same observable as part of the ID they choose, the results always agree.  In any local hidden variables theory, we must then preassign-truth values to all of the $N$-qubit observables that might be measured.  Because each observable occurs an even number of times throughout the Observable-based KS proof, the product of all of its IDs must be +1.  But an odd number of the IDs are Negative, which means their product must be -1 in order to agree with all of the experimental results.  This contradiction rules out local hidden variable theories without using probabilities, proving the GHZ theorem.  

As a final simplification of this experiment, we note that Alice and Bob need not both choose between all of the IDs in the proof in order for the proof to obtain.  The IDs in the proof can be loosely divided into two groups, such that Alice chooses randomly from one group, and Bob from the other.  Some IDs may need to appear in both groups, such that Alice's IDs contain all of the observables in the proof, as do Bob's.  Provided they can each individually measure every observable, and together measure all of the IDs, locality still requires a noncontextual truth-value assignment to the entire set.

We will further say that an Observable-based KS proof is critical if no subset of IDs and/or qubits can be deleted such that the remaining set is still an Observable-based KS proof.  There is one important subtlety to this statement regarding proofs containing noncritical IDs like the ID$7$s discussed in Chapter \ref{sec:q3IDs}, that can be broken up into the product of an ID$3$ and an ID$4$ in 7 different ways.  In cases like this, we must consider the deletion of any of these sub-IDs, since another sub-ID would still remain.  This complication has been entirely avoided in the cases we have considered by using only IDs that are critical in the specific sense that they cannot be broken into the product of two IDs.  Note that these are different than the noncritical IDs we obtain by placing two critical IDs side by side, or by adding Trivial SQPs.  All of the Observable-based KS proofs we have introduced are critical.

We have developed a general format for diagrams of Observable-based KS proofs, which allow the structure of how the IDs and observables fit together to be seen.  In these diagrams each observable appears inside a small circle, and all of the observables in an ID are connected by a continuous line, arc, or circle.  Thin lines indicate that the ID is Positive, while thick lines indicate that it is negative.  These diagrams transparently show how the set proves the KS theorem, since one can see that each observable appears on an even number of lines, and that an odd number of lines are thick.  Examples of such diagrams can be seen in Figures \ref{q2Whorl}, \ref{MerminStar}, \ref{q3Kite}, \ref{Wheel3}, \ref{Wheel3_Exp}, \ref{q3Square_K2}, \ref{q4Star}, \ref{q4Whorl}, and \ref{q4Windmill}.

As we discussed in Chapter \ref{sec:q2}, the 2-qubit Pauli group contains only two unique Observable-based KS proofs, the Mermin Square and the 2-qubit Whorl.  There are several possible versions of each, obtained by permutations of qubits and/or SQPs, but only these two have unique diagrams - in the specific sense of isomorphism discussed in Chapter \ref{sec:Wheel3}.  For more qubits, the number of Observable-based KS proofs grows beyond being easily counted, and so we are only able to construct and consider particular cases, as we have done throughout this text.

We have conjectured that all Observable-based KS proofs must contain, and in some sense be generated by, a Kernel.  We have never found any counterexample of this in our extensive searches.  It follows that the best way to obtain some idea of the many unique structures of proofs is by considering the Kernels that generate them.  Furthermore, all Kernels can be obtained using IDs in conjunction with CKSs, both of which we have fully enumerated for up to 5 qubits (and partially enumerated for more qubits, as can be seen on our website \cite{MainWebsite}).  This means we have building blocks with which to construct the vast variety of Observable-based KS proofs.  It might even be possible to count them all for 3 qubits, though we have made no concerted effort to do this.

Let us return to and review the notion that Observable-based KS proofs are generated by Kernels.  A Kernel is a set of one of more IDs, with each single-qubit observable appearing an even number of times, and an odd number of IDs negative.  An Observable-based KS proof is also a set of ID$^N$s with an odd number of the IDs negative, but now it is each $N$-qubit observable that must appear an even number of times.  To obtain an Observable-based KS proof from a Kernel, one adds some set of IDs, an even number of which must be negative, in such a way that each $N$-qubit observable that appeared an odd number of times in the Kernel is paired with another ID containing that observable.  In general, there are many ways to obtain an Observable-based KS proof from the same Kernel, as we have seen throughout this text.

Here we will describe a general method (the same as in Chapter \ref{sec:Kite3}) by which any critical Kernel can be used to generate at least one Observable-based KS proof, in such a way that no two IDs in the set will share more than one $N$-qubit observable in common.  For each observable that appears an odd number of times in the Kernel, we will generate a new Whole Positive (Null) ID by supplementing that observable with its own qubit-decomposition.  In many cases, this is the single-qubit-decomposition, so for example, we would multiply a Kernel observable $ZZIX$ by the new observables $ZIII$, $IZII$, and $IIIX$ to form a new ID$4^4$.  If however two of the observables that appeared an odd number of times in the Kernel share more than one element in common, then those common elements need not be decomposed in the two new Positive IDs they generate.  For example, if we have Kernel IDs containing observables $ZZZIX$, and $ZZZIY$, the portion $ZZZII$ is common to both, and so the two new ID$3^5$s will be ($ZZZIX$, $ZZZII$, $IIIIX$) and ($ZZZIY$, $ZZZII$, $IIIIY$).  This generation process always works as desired by exploiting the even number of single-qubit observables in a Kernel to obtain an even number of $N$-qubit observables in an Observable-based KS proof.  The fact that we started with a critical Kernel also guarantees that this Observable-based KS proof will be critical.

The final benefit of this method is that if we start with a Kernel in which no observable appears more than twice, we always obtain an Observable-based KS proof of the form $O_2 - I$ in which no two IDs share more than one common observable.  Proofs of this type generate a particularly simple $R-B$ set that contains exactly $2^O$ critical parity proofs, as we will discuss in Chapter \ref{sec:RBSets}.

Next we give a prescription for using any Observable-based KS proof in an experimental test of quantum contextuality, which is a simple generalization of the one given in \cite{cabello2008experimentally}.  We begin with an Observable-based KS proof $A$ with symbol $O-I$, and define for it the quantity
\begin{equation}
\alpha = \sum_{+I_i \in A} (\prod_{O_j \in +I_i}O_j) - \sum_{-I_i \in A} (\prod_{O_j \in -I_i}O_j),
\end{equation}
where the summations are taken over the Positive and Negative IDs in the set respectively.  Each term in $\alpha$ is a product of mutually commuting observables, and so its expectation value is given by
\begin{equation}
\langle\alpha\rangle = \sum_{+I_i \in A} \langle\prod_{O_j \in +I_i}O_j\rangle - \sum_{-I_i \in A} \langle\prod_{O_j \in -I_i}O_j\rangle.
\end{equation}
Because each ID has a fixed sign, quantum mechanics predicts this expectation in any state will be $\langle\alpha_{QM}\rangle = I$.  However, a simple counting argument shows that if noncontextual truth-values $\pm 1$ are assigned to all $O$ observables in the set, then $\langle\alpha_{NC}\rangle \leq I-2$.  The experimental violation of this inequality is then a demonstration of quantum contextuality.  It is worth noting that that Kite family of proofs discussed in Chapter \ref{sec:Kites} always have $I=6$, regardless of the number of qubits, and thus give the strongest violation of this inequality.
\pagebreak

\subsection{The Rays Generated by Observable-Based KS Proofs}\label{sec:RBSets}

In this section, we will explore the sets of rays and bases that are generated by individual IDs, as well as entire Observable-based KS proofs.  We say that rays in the joint eigenbasis of all of the observables in an ID are the rays that ID generates.  Because a given ID$M^N$ has a fixed sign, only $M-1$ of its eigenvalues are independent, meaning that there are only $2^{M-1}$ rays in the eigenbasis, each of rank $2^{N-M+1}$.  Rays span a dimension equal to their rank, and contain internal degrees of freedom if that rank is greater than one.  In this way, critical ID$M^N$s with $M<N+1$ generate continuous classes of entangled rank-1 projectors which are guaranteed to be entangled $N$-qubit states for any choice of values for these internal parameters.

For the sake of simplicity, we can express each ray simply as a list of the eigenvalues of its $M$ observables.  For example, an eigenstate of a Positive ID$3$ with the commuting observables $\{A,B,C\}$ could be expressed as $|\psi\rangle = \{ A(-1),B(+1),C(-1) \}$.  Two rays are orthogonal iff they share a common observable for which they each have a different eigenvalue.  This makes it very easy to see if two rays expressed in this form are orthogonal, but there is one very important detail that must be addressed.  Any additional observable that can be obtained as a product of the observables in the ID also has a fixed eigenvalue for each eigenstate.  Suppose that $D=AB$ is another observable not in the original ID of our example.  We would then need to expand our expression for the eigenstate to include $D$, whose eigenvalue is the product of the eigenvalues of $A$ and $B$, giving us $|\psi\rangle = \{ A(-1),B(+1),C(-1),D(-1) \}$.  The complete expression for an eigenstate must include all observables that can be obtained as products of the observables in the ID.  For any critical ID$M^N$ with $M=N+1$, the full set of products is always a maximal commuting set of $2^N -1$ observables - the stabilizer group.

This is essential because, without it, we would not see that the rays $|\psi\rangle = \{ A(-1),B(+1),C(-1),D(-1) \}$ and $|\phi\rangle = \{D(+1),E(-1),F(-1) \}$ are orthogonal, which is one of the most important issues as far as the Kochen-Specker theorem is concerned.

Finally, it should be noted that rays with rank greater than one may be `optionally' orthogonal to other rays.  This is because there may be additional observables that commute with those in the ID, but are not obtained as their products.  These observables represent the internal degrees of freedom of the ray, since their eigenvalues are not fixed.  If we chose to fix them, we would reduce the rank of the ray, but the ray also becomes orthogonal to any other ray with different eigenvalue for a new observable.  This is because another ray may not be orthogonal to the entire subspace spanned by a rank-$r$ ray, but it might still be orthogonal to particular rays within that subspace.  We do not make use of any of these extra orthogonalities in the proofs we present, so nothing further need be said on this point.

Now that we have a good way to express the eigenstates of a given ID, we can start to consider the sets of rays and bases that are formed by a set of multiple IDs.  To begin, we consider the case of two IDs that share just one observable in common.  Let the observables of some ID$K^N$ be $\{A, B, \ldots \}$, and the observables of an ID$L^N$ be $\{A,C,\ldots \}$ such that $A$ is the only observable they share (even with products taken into account).  The eigenbasis of the ID$K^N$ is shown in Table \ref{Eig1}, where state shown actually represents a group of $2^{K-3}$ mutually orthogonal eigenstates.  Likewise, the eigenbasis of the ID$L^N$ is shown in Table \ref{Eig2}, where each state shown actually represents a group of $2^{L-3}$ mutually orthogonal eigenstates.
\begin{table}[ht]
\centering
\subfloat[][]{
\begin{tabular}{c}
$|\psi_1\rangle = \{ A(+1),B(+1),\ldots \}$,  \\
$|\psi_2\rangle = \{ A(+1),B(-1),\ldots \}$, \\
$|\psi_3\rangle = \{ A(-1),B(+1),\ldots \}$,  \\
$|\psi_4\rangle = \{ A(-1),B(-1),\ldots \}$, \\
\end{tabular} \label{Eig1}}
\qquad
\subfloat[][]{
\begin{tabular}{c}
$|\phi_1\rangle = \{ A(+1),C(+1),\ldots \}$, \\  
$|\phi_2\rangle = \{ A(+1),C(-1),\ldots \}$, \\
$|\phi_3\rangle = \{ A(-1),C(+1),\ldots \}$,  \\
$|\phi_4\rangle = \{ A(-1),C(-1),\ldots \}$, \\
\end{tabular} \label{Eig2}}
\qquad
\subfloat[][]{
\begin{tabular}{c}
$|\psi_1\rangle = \{ A(+1),B(+1),\ldots \}$,  \\
$|\psi_2\rangle = \{ A(+1),B(-1),\ldots \}$, \\ 
$|\phi_3\rangle = \{ A(-1),C(+1),\ldots \}$,  \\
$|\phi_4\rangle = \{ A(-1),C(-1),\ldots \}$, \\
\end{tabular}\label{Hyb1} }
\qquad
\subfloat[][]{
\begin{tabular}{c}
$|\phi_1\rangle = \{ A(+1),C(+1),\ldots \}$,   \\
$|\phi_2\rangle = \{ A(+1),C(-1),\ldots \}$, \\
$|\psi_3\rangle = \{ A(-1),B(+1),\ldots \}$,  \\
$|\psi_4\rangle = \{ A(-1),B(-1),\ldots \}$, \\
\end{tabular} \label{Hyb2}}
\qquad
\caption[Eigenbases and Hybrid Bases for Two IDs with One Common Observable]{The 4 bases generated by a pair of IDs with exactly one common observable $A$.  The eigenbases are shown in \subref{Eig1} and \subref{Eig2}, and the hybrid bases are shown in \subref{Hyb1} and \subref{Hyb2}.}
\end{table}
It is easy to see that these groups mix to form only the two hybrid bases shown in Tables \ref{Hyb1} and \ref{Hyb2}.  It can also be verified that regardless of the number rays in each group, or of their relative ranks, these hybrid bases always span the space.  To make this clearer, work out the case where both IDs are ID$3^2$s.  It is also worth noting that this set of 4 bases and 8 rays is `saturated,' in the sense that every orthogonal pair of rays appears within one of these bases.  It should be easy to see that this saturation will then extend to the entire set of rays and bases generated by any set of IDs.  This means that the complete table of bases is equivalent to the Kochen-Specker diagram of the rays - a type of diagram in which every ray is a vertex, and lines join pairs of orthogonal rays.

Here we have obtained a very important rule that we have made use of throughout this text: If two IDs share exactly one observable in common, then the rays of their two eigenbases mix to form two complementary hybrid bases - complementary in the sense that the two hybrids have no rays in common, and together include all of the rays of both eigenbases.  This rule is especially important because the generalized method we have introduced in Chapter \ref{sec:ObKSproofs} for generating an Observable-based KS proof from a critical Kernel always generates a set in which no two IDs share more than one observable in common.

A more elaborate general calculation can be performed for the case of two IDs that share $s$ observables in common.  In this case, the rays of their two eigenbases mix to form $h = 2^{2^s} - 2$ hybrid bases.  These still occur in complementary pairs, but the hybrid bases need no longer be equally spanned by the rays of the two eigenbases.  This will be essential in understanding the family of Kites, though most of the other families we consider only involve $s\leq 1$.  The saturation rule is of course still satisfied in these cases.

Hybrid bases formed by rays from the eigenstates of two IDs are not the only types of hybrid bases that can exist.  Of course we can also have hybrid bases composed of rays from more than two IDs, as in the case of Ruuge's Observable-based KS proof \cite{Ruuge, Ruuge_Err}.  None of the cases we are considering contain any hybrid bases of this sort, and so we have not explored the issue of how many are generated by an arbitrary group of IDs that are joined by common observables, nor have we fully explored the issue of saturation for these cases.

The full set of rays and bases generated by a set of IDs will be denoted $R-B$, for a set of $R$ rays which form $B$ bases.  We will also use the expanded symbol $R^r_m - B_n$ for a set of $R$ rays of rank $r$, each of which occurs in $m$ bases, and $B$ bases each containing $n$ rays.  If the set contains rays of differing $r$ or $m$, or bases of differing $n$, the symbol on either side of the dash can be repeated.  These parameters all follow directly from the characteristics of the hybrid bases, which can be determined directly from the diagram (though this becomes more elaborate for sets with $s>1$).  In general, the total number of rays (ignoring rank) is
\begin{equation}
R = \displaystyle\sum\limits_{I_y}^{} 2^{y-1},
\end{equation}
where the sum is taken over all IDs in the $O-I$ set.

We now see that the structure of the $R-B$ set is entirely determined by the structure of the diagram for the Observable-based KS proof that generates it, and this has one very important consequence: If two different Observable-based KS proofs have the same diagram structure, which is to say that they are isomorphic, then they also generate isomorphic $R-B$ sets.  The two sets will have the same structure of orthogonality between rays, since this depends only on how the observables and IDs fit together.  The only possible difference between two such sets is an overall rank factor for all $R$ rays.  As we will see, it is particularly easy to construct Observable-based KS proofs with certain diagrams, and so we will actually find many isomorphic sets of rays.

Finally, if an $R-B$ set is generated by an Observable-based KS proof, then the $R-B$ set also proves the KS theorem, which can be seen most easily by finding parity proofs within the set.  A parity proof is any $R_m - B$ set for which all values of $m$ are even, and $B$ is odd.  This proves the KS theorem, because any NCHVT must assign truth-value 1 to one of the rays in each bases and 0 to all other rays in that basis, but each ray appears in an even number of bases and there are an odd number of bases, so this assignment is impossible.  Every $R-B$ set generated by an Observable-based KS proof from the $N$-qubit Pauli group appears to contain parity proofs of the KS theorem.  We cannot prove this, but we have never found a counterexample.

In particular, we have identified, though not derived, a pattern that seems to generate all $2^O$ parity proofs that are present within the $R-B$ set generated by an $O_2-I$ Observable-based KS proof with $s \leq 1$.   Such sets contain $O$ complementary pairs of hybrid bases along with their $I$ eigenbases, for a total of $B = I + 2O$ bases.  To obtain all of the proofs, we begin by choosing one hybrid basis from each of the $O$ complementary pairs, which can be done in $2^O$ different ways.  For a given choice of $O$ bases, find all of the rays that occur in an odd number of them, and then add the eigenbases that generated those rays to the set.  This process of beginning with $O$ hybrid bases and adding in the missing eigenbases results in all $2^O$ parity proofs that exist within the set.  A clear example of this process is described for the $24-24$ set of Peres in Chapter \ref{sec:MerminSquare}.  It should be noted that smallest parity proof within the set always has $B = O$ or $B=O+1$, whichever makes $B$ odd.  Numerous other examples of sets like this can be found on our website \cite{MainWebsite}.

We do not know conclusively that other parity proofs are impossible within these sets, but exhaustive computer searches of many of them have never revealed any.

Next we give a prescription for using any parity proof in an experimental test of quantum contextuality, which is again a simple generalization of the one given in \cite{cabello2008experimentally}.  We begin with a parity proof $A$ with symbol $R-B$, and define for it the quantity
\begin{equation}
\alpha = \sum_{B_i \in A} (\prod_{R_j \in B_i}\chi_j),
\end{equation}
with expectation value
\begin{equation}
\langle\alpha\rangle = \sum_{B_i \in A} \langle\prod_{R_j \in B_i}\chi_j\rangle,
\end{equation}
where an observable $\chi_j = 2|\psi_j\rangle\langle\psi_j| - 1$ is defined for each of the $R$ rays in $A$.  Quantum mechanics predicts this expectation in any state will be $\langle\alpha_{QM}\rangle = B$.  However, a simple counting argument shows that if noncontextual truth-values $\pm 1$ are assigned to all observables $\chi_j$ in the set, then $\langle\alpha_{NC}\rangle \leq B-2$.  The experimental violation of this inequality is then a demonstration of quantum contextuality.  It is worth noting that that Kite family of proofs discussed in Chapter \ref{sec:Kites} always contain a critical parity proofs with $B=9$, regardless of the number of qubits, and thus give the strongest violation of this inequality as well.

\pagebreak

\subsection{Continuous Classes of Entanglement from Critical IDs} \label{IDEntanglement}

In this section we discuss the continuous classes of entangled states generated by critical ID$M^N$s with $M<N+1$.  These classes only exist for $N\geq4$ qubits, but the number of degrees of freedom grows with $N$ at a stupendous rate.  This section should also help to elucidate the nature of projectors with rank greater than 1 associated with these ID$M^N$s.

First let us consider the simplest case: ID$4^4$s like the ones in Table \ref{q4IDs}.   We label the 4 observables of the ID as $\{A,B,C,D\}$.  Since the ID is critical, it must be true that we can find one more observable that commutes with those in the ID, but is independent of them in the sense that it cannot be obtained as a product of the other observables in the set.  For the 4-qubit case, we will call this independent observable $E$.  Using the compact notation of the previous section, we can then write the eigenstates of the ID$4^4$ as
\begin{equation}
|\psi\rangle = a_{+1}|A(\lambda_A),B(\lambda_B),C(\lambda_C),D(\lambda_D),E(+1)\rangle + a_{-1}|A(\lambda_A),B(\lambda_B),C(\lambda_C),D(\lambda_D),E(-1)\rangle,
\end{equation}
where the product $\lambda_A \lambda_B \lambda_C \lambda_D$ is fixed by the sign of the ID, and with $a_{+1}$ and $a_{-1}$ any pair of complex numbers such that $|a_{+1}|^2 + |a_{-1}|^2 = 1$.

For the general case of an ID$M^N$ with $M<N+1$, let $\{O\}$ be the set of $M$ observables in the ID, and $\{P\}$ be the set of $\eta = N+1-M$ independent observables, such that all of the observables in $\{O\} \cup \{P\}$ mutually commute.  We can then write the eigenstates as
\begin{equation}
|\psi\rangle =  \sum_{\mu_1 = \pm 1} \cdots \sum_{\mu_\eta = \pm 1}   a_{\mu_1 \cdots \mu_\eta}|O_1(\lambda_1),\cdots,O_M(\lambda_{M}),P_1(\mu_1),\cdots,P_\eta(\mu_\eta)\rangle,
\end{equation}
where the product $\lambda_{1} \cdots \lambda_{M}$ is fixed by the sign of the ID, and with $a_{\mu_1 \cdots \mu_\eta}$ any complex coefficients such that
\begin{equation}
\sum_{\mu_1 = \pm 1} \cdots \sum_{\mu_\eta=\pm 1}   |a_{\mu_1 \cdots\mu_\eta}|^2 = 1.
\end{equation}

In all of these cases, it is the criticality of the IDs that certifies $N$-qubit entanglement for the entire continuous class of states.  We can also think of the entire class of states as a single entangled rank-$r$ projector, where $r = 2^{N-M+1}$ is the number of free coefficients $a_{\mu_1 \cdots \mu_\eta}$.

For critical ID$M^N$s, the maximum value of $N$ grows much more rapidly than $M$.  Table \ref{ShortKites} shows examples of a critical ID$5^7$, a critical ID$6^{11}$, and a critical ID$7^{16}$, with the rank of the projectors growing at an accelerating rate.

Clearly the possibility of using critical IDs as a way of characterizing multiparticle entanglement warrants further exploration.

%

%\end{document}

%\documentclass[12pt]{article}
%\usepackage{graphicx}
%\usepackage{epsfig}
%\usepackage{amsfonts}
%\usepackage[lofdepth,lotdepth]{subfig}
%
%\begin{document}
%

\section{Structure Families}\label{sec:Families}

In this Chapter we introduce and examine several remarkable families of structures that give proofs of the KS theorem and/or the GHZ theorem for all numbers of qubits.  To begin we will introduce the related families of Wheel and Whorl proofs for $N$ (odd and even respectively) qubits, and discuss some of the $R-B$ sets they generate.  Next we will move on to discuss the families of Star proofs for odd and even $N$, which prove both the KS and GHZ theorems.  Finally, we will introduce the more general family of Kite proofs and discuss the smallest known proofs of the KS theorem in the Hilbert Space of $N$ qubits.

\subsection{The Wheel and Whorl Structure}\label{sec:Wheel_Whorl}
\begin{figure}[ht]
  \centering
  \includegraphics[width=2.5in]{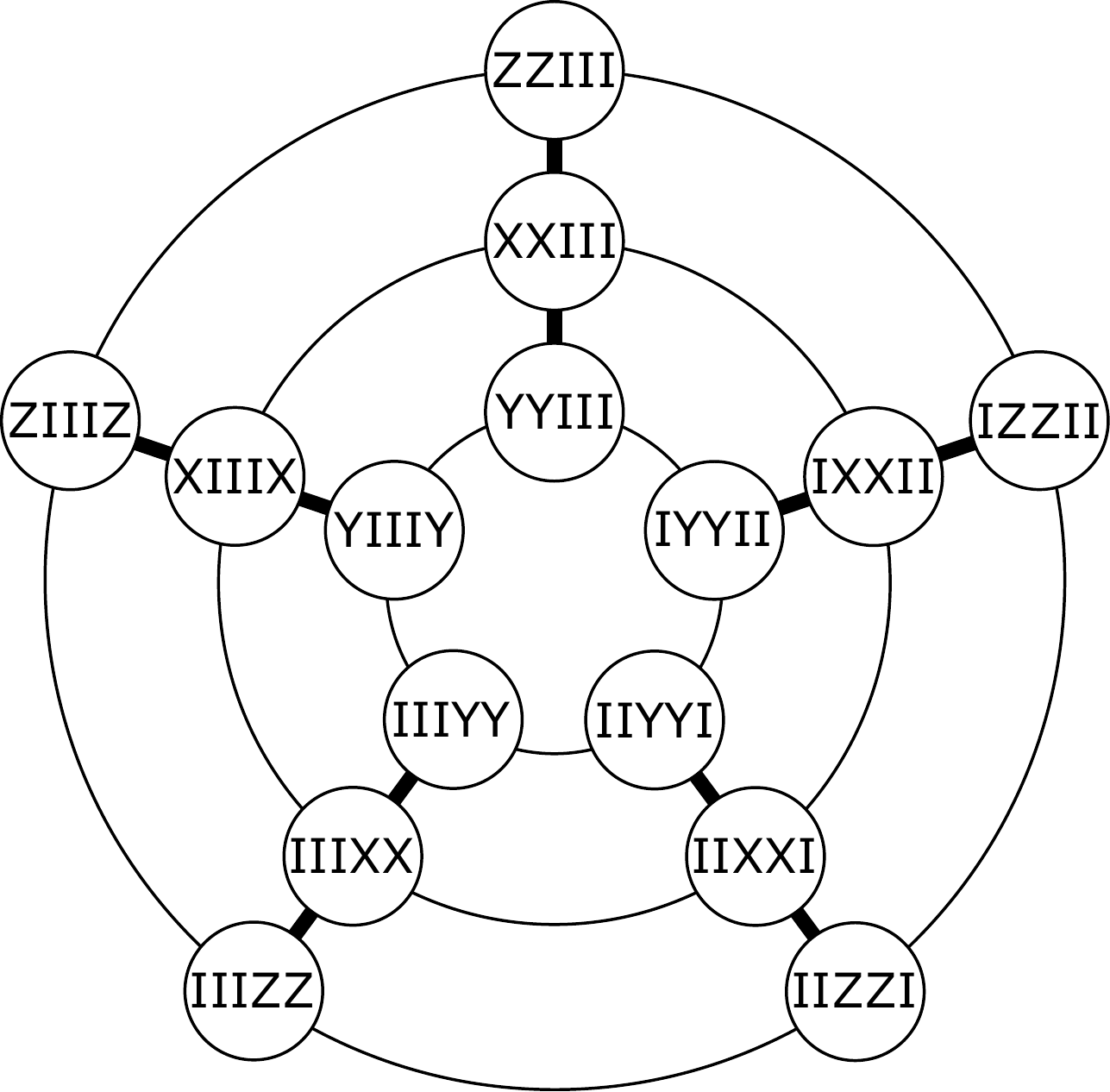}\\
  \caption[5-qubit Wheel]{The 5-qubit Wheel, a $15_2 -3_5 5_3$ Observable-based KS proof for 5 qubits.  Each line or large circle connects the observables of an ID.  The thin lines denote Positive IDs, while the thick lines denote Negative IDs.  The 5 ID$3^5_2$ of the Kernel form the `spokes' of the Wheel.}\label{Wheel5}
\end{figure}
\begin{figure}[ht]
  \centering
  \includegraphics[width=2.5in]{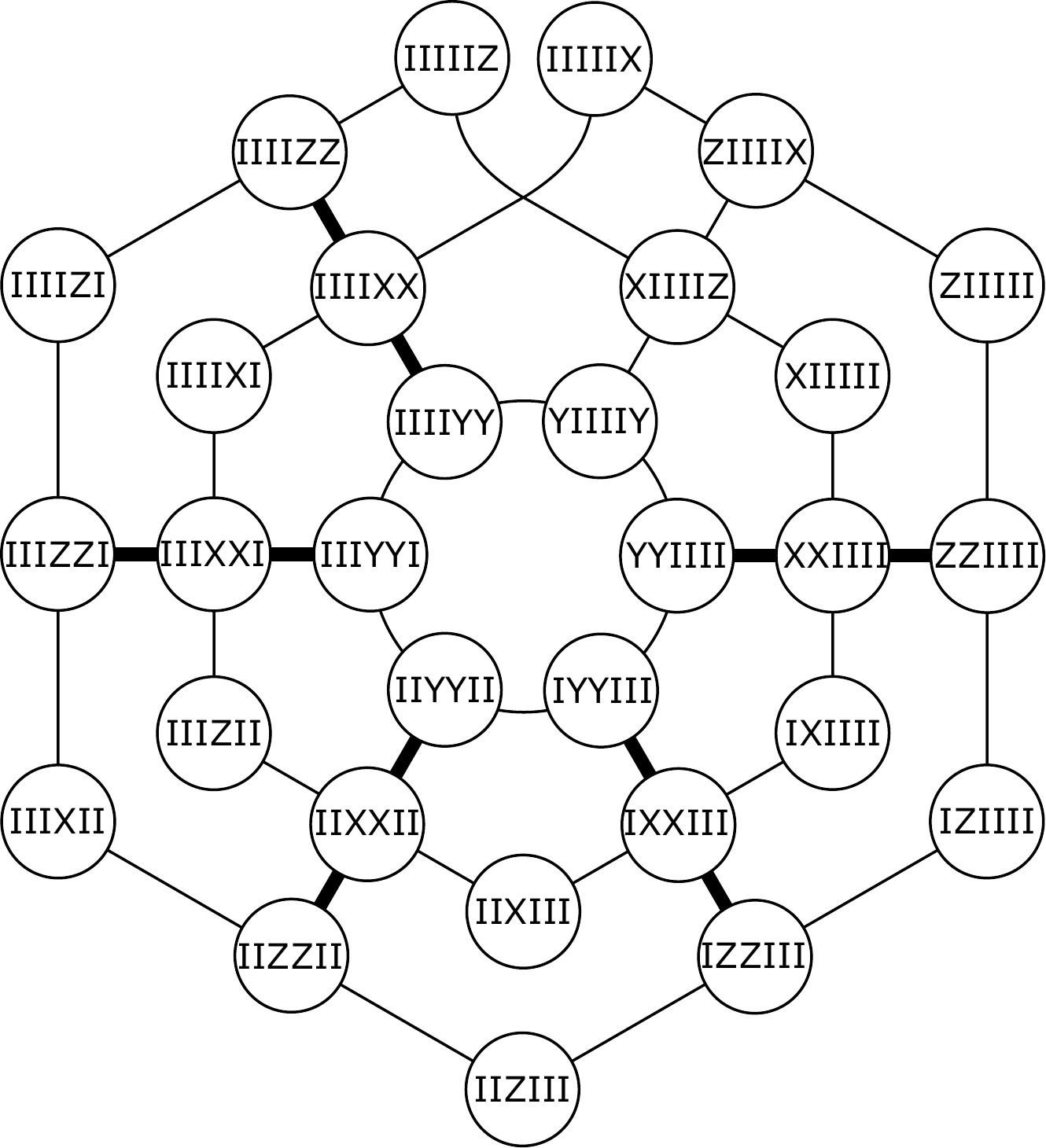}\\
  \caption[6-qubit Whorl]{The 6-qubit Whorl, a $30_2 -1_6 18_3$ Observable-based KS proof for 6 qubits.  Each line or large circle connects the observables of an ID.  The thin lines denote Positive IDs, while the thick lines denote Negative IDs.  The 6 ID$3^6_2$ of the Kernel form the `spokes' of the Whorl.}\label{Whorl6}
\end{figure}

In this section, we will review the properties of a general family of $N$-qubit CKSs for $N\geq 3$, and the simplest Kernels they generate, which are somewhat different for odd and even $N$.  We refer to these as the Wheel (odd $N$) and Whorl (even $N$) because of the structure of the Observable-based KS proof diagrams that show them.  The family of CKSs is shown in Table \ref{CKSWheels}.
\begin{table}[ht]
\centering
\qquad
\subfloat[][]{
\begin{tabular}{ccc}
O & O & I \\
I & O & O \\
O & I & O \\
\end{tabular}\label{CKSWheel3}}
\qquad
\subfloat[][]{
\begin{tabular}{cccc}
O & O & I & I \\
I & O & O & I \\
I & I & O & O \\
O & I & I & O \\
\end{tabular}\label{CKSWheel4}}
\qquad
\subfloat[][]{
\begin{tabular}{ccccc}
O & O & I & I & I \\
I & O & O & I & I \\
I & I & O & O & I \\
I & I & I & O & O \\
O & I & I & I & O \\
\end{tabular}\label{CKSWheel5}}
\qquad
\subfloat[][]{
\begin{tabular}{cccccc}
O & O & I & I & I & I \\
I & O & O & I & I & I \\
I & I & O & O & I & I \\
I & I & I & O & O & I \\
I & I & I & I & O & O \\
O & I & I & I & I & O \\
\end{tabular}\label{CKSWheel6}}
\qquad
\caption[Wheel/Whorl CKS Family]{The first 4 critical CKSs of the Wheel/Whorl family.  The critical Composite Kernels are formed by assigning the critical ID$3^2_2$ combined with trivial ($I$ only) SQPs to each row of the CKS.}\label{CKSWheels}
\end{table}
In order to form the composite Kernels from these CKSs, we assign the simplest noncritical ID$3_2$, which is composed of the critical ID$3^2_2$ with additional trivial SQPs (containing only $I$).  For odd $N$, we can choose our permutations so that all $N$ critical ID$3^2_2$s are identical and Negative.  For even $N$, we must permute some odd number of the IDs to be positive in order to obtain a Kernel, meaning that not all of the IDs can be identical.  This is the crucial difference between the Wheel and Whorl families, which ultimately makes the Whorl family somewhat more constrained, as we will see.

The IDs of the Kernel form the `spokes' of the Wheel or Whorl.  In the Wheel cases, where the IDs are all Negative, we see that we can always form 3 more Positive ID$N^N$s around the spokes to form a complete Wheel diagram.  The 3-qubit Wheel is shown in Figure \ref{Wheel3}, and the 5-qubit Wheel is shown in Figure \ref{Wheel5}.  In the Whorl case, because some of the IDs are of opposite sign, we will only be able to form one positive ID$N^N$s around the spokes.  To complete the diagram, we must instead form $2N$ ID$3^N$s using $2N$ new single-qubit observables.  These ID3s connect the spokes in a sort of m\"{o}bius ring, that circles the spokes twice to return to where it begins.  The 4-qubit Whorl is shown in Figure \ref{q4Whorl}, and the 6-qubit Whorl in Figure \ref{Whorl6}.

It is worth noting that the 2-qubit Whorl, while technically not a member of this family, because it has 3 spokes, still has the same m\"{o}bius pattern as the family of Whorls, which is why we give it this name.  The Mermin Square can also be seen as a sort of 2-qubit Wheel with 3 spokes.

It should now be easy to see how this family generalizes to all $N \geq 3$, and that it is critical in each case.  Furthermore, for each $N$, there are actually several related diagrams that we can build trivially from the same Kernel, where we have focused so far on the most compact member in each case.  The general rule is that for each ID$N^N$ circle in the diagram, we can replace that ID by a new ring of $N$ ID$3^N$s by introducing $N$ new single-qubit observables.  The case where all three circles of the 3-qubit Wheel in Figure \ref{Wheel3} have been replaced in this way is shown in Figure \ref{Wheel3_Exp}.  Because each Wheel contains three ID$N^N$ circles, which can be replaced in any combination, this gives a total of 8 different diagrams from a given Wheel Kernel.  Because each Whorl contains just one ID$N^N$, this gives just 2 different diagrams for a given Whorl Kernel.  There are of course many other diagrams that can be built from these Kernels - these are just a particularly trivial set.

The fully expanded Observable-based KS proofs in this family also have one other nice feature, which is that they only contain 2-qubit measurement bases.  Once all of the circle ID$N^N$s have been replaced, every ID3 in the proof contains $I$ for all but 2 of its qubits, so in any given run of an experiment using these proofs, only 2 qubits would need to be measured.

Because all of the diagrams we are discussing are of the form $O_2-I$, with no two IDs sharing more than one common observable, we can easily construct the complete $R-B$ set for each of these diagrams.  Of particular interest are the fully expanded forms of each Wheel and Whorl, because they are composed entirely of ID3s.  We examine them here in detail.  For $N$ qubits, the fully expanded Wheel or Whorl has $I_3 = 4N$ ID3s, each of which generates 4 rays, giving a total of $R^r = 16N$ rays of rank $r = 2^N-2$.  It also has $O_2 = 6N$ observables, and so there are $B = I + 2O = 16N$ bases in the set, each with 4 elements.  Because each ID in this $(6N)_2-(4N)_3$ set shares a common observable with 3 other IDs, each ray occurs in 3 hybrid bases as well as one eigenbasis, giving us finally an $R-B$ set with expanded symbol $(16N)_4^{2^{N-2}} - (16N)_4$.  Furthermore, these sets all contain $2^O =2^{6N}$ distinct critical parity proofs, with the smallest containing $6N+1$ bases, and the largest containing $10N-1$ bases.  These proofs divide up into complementary pairs in the sense that the two proofs in a pair have no bases in common, and together contain all $16N$ bases.

\pagebreak

\subsection{The Star Structure}\label{sec:StarFam}
\begin{figure}[ht]
  \centering
  \includegraphics[width=2.5in]{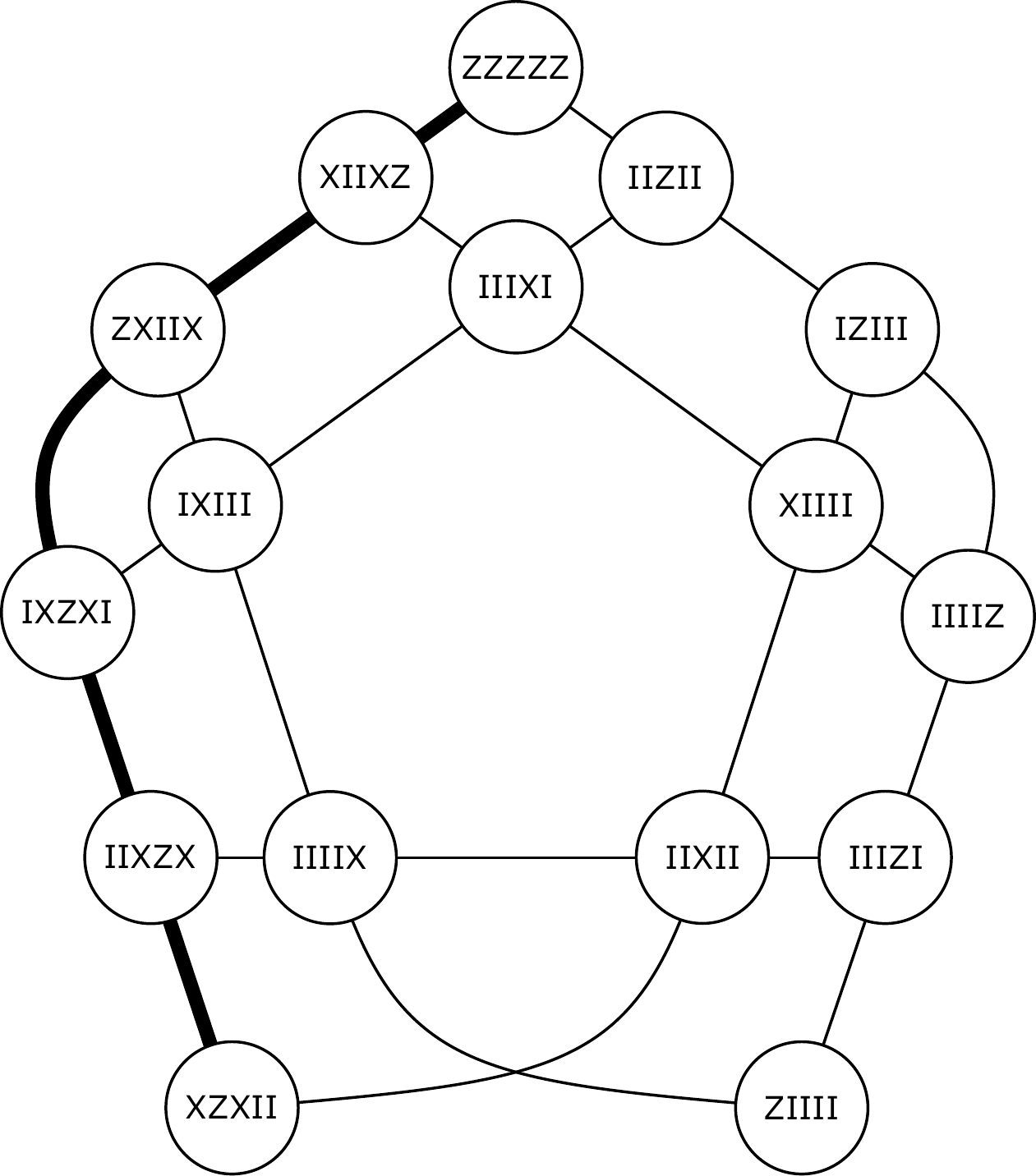}\\
  \caption[5-qubit Star]{The 5-qubit Star, a $11_2 -2_6 5_4$ Observable-based KS proof.  Each line or arc connects the observables of an ID.  The thin lines denote Positive IDs, while the thick line denotes a Negative ID, which in this case is also the Kernel.}\label{Star_DiVinc}
\end{figure}
\begin{figure}[ht]
  \centering
  \includegraphics[width=2.5in]{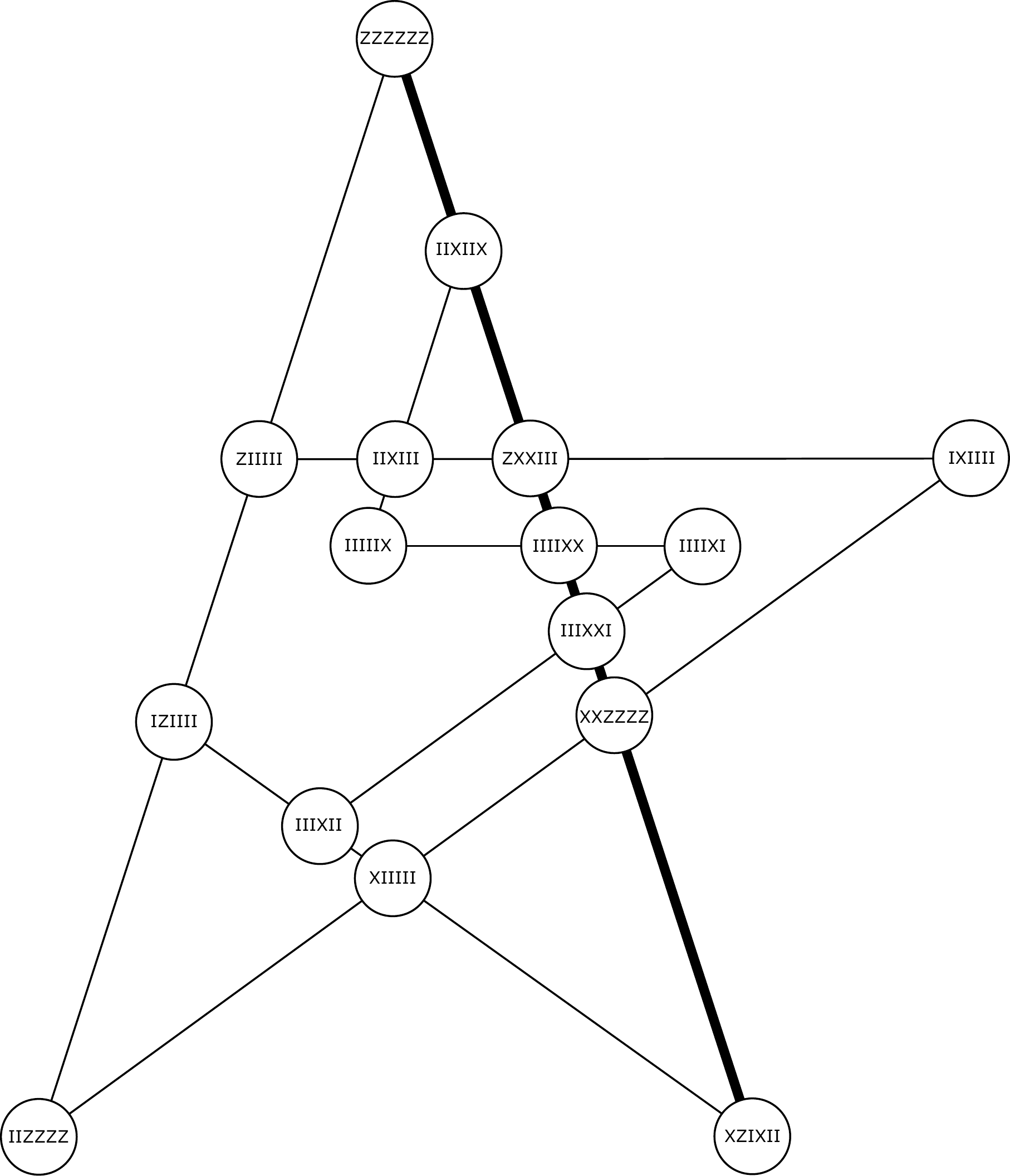}\\
  \caption[6-qubit Star]{The 6-qubit Star, a $16_2 -1_7 4_4 3_3$ Observable-based KS proof.  Each line connects the observables of an ID.  The thin lines denote Positive IDs, while the thick line denotes a Negative ID, which in this case is also the Kernel.}\label{Star6}
\end{figure}
In this section, we discuss two different families of single-ID Kernels, one for $N$ odd, and the other for $N$ even.  We call both of these Star families because of the structure of their simplest diagrams - Figures \ref{MerminStar} and \ref{q4Star}, even though they both grow much more elaborate as $N$ increases, as can be seen in Figures \ref{Star_DiVinc} and \ref{Star6}.  In both cases, the Observable-based KS proof is generated from the Single-ID Kernel using exactly the generalized method described in Chapter \ref{sec:ObKSproofs}, and thus to show these families, we will show how the single ID of the Kernel generalizes to all $N$.  The case for odd $N$, shown in Table \ref{OddStar}, was first given by Aravind, and generalizes a result of DiVincenzo and Peres \cite{Peres_DiVinc} based on 5-qubit error correcting codes (Table \ref{OddStar5}).  The case for even $N$, shown in Table \ref{EvenStar} was introduced recently in our paper \cite{WA_4qubits}.
\begin{table}[ht]
\centering
\qquad
\subfloat[][Mermin Star Kernel]{
\begin{tabular}{ccccccc}
& & $Z$ & $Z$ & $Z$ &  &  \\
& & $X$ & $Z$ & $X$ &  &   \\
& & $X$ & $X$ & $Z$ &  &   \\
& & $Z$ & $X$ & $X$ &  &   \\
\end{tabular}\label{OddStar3}}
\qquad
\subfloat[][DiVincenzo-Peres Kernel]{
\begin{tabular}{ccccccc}
& $Z$ & $Z$ & $Z$ & $Z$ & $Z$ &  \\
& $X$ & $Z$ & $X$ & $I$ & $I$ & \\
& $I$ & $X$ & $Z$ & $X$ & $I$ & \\
& $I$ & $I$ & $X$ & $Z$ & $X$ & \\
& $X$ & $I$ & $I$ & $X$ & $Z$ & \\
& $Z$ & $X$ & $I$ & $I$ & $X$ & \\
\end{tabular}\label{OddStar5}}
\qquad
\subfloat[][Odd $N$ Kernel]{
\begin{tabular}{cccccc}
$Z$ & $Z$ & $Z$ & $Z$ & $\cdots$ & $Z$  \\
$X$ & $Z$ & $X$ & $I$ & $\cdots$ & $I$  \\
$I$ & $X$ & $Z$ & $X$ & $\ddots$ & $\vdots$  \\
$\vdots$ & $\ddots$ & $\ddots$ & $\ddots$ & $\ddots$ & $I$  \\
$I$ & $\cdots$ & $I$ & $X$ & $Z$ & $X$  \\
$I$ & $I$ & $\cdots$ & $I$ & $X$ & $Z$  \\
$Z$ & $X$ & $I$ & $\cdots$  & $I$ & $X$  \\
\end{tabular}\label{OddStarN}}
\qquad
\caption[Star Kernels for Odd $N$]{The pattern of Star Kernels for odd $N$.  The $N=3$ \subref{OddStar3}, $N=5$ \subref{OddStar5} cases are shown explicitly, while \subref{OddStarN} shows how the pattern generalizes to arbitrary odd $N$.  The ellipses in \subref{OddStarN} always indicate repetition of the observable found at either end.}\label{OddStar}
\end{table}
\begin{table}[ht]
\centering
\qquad
\subfloat[][4-qubit Star Kernel]{
\begin{tabular}{cccccc}
& $Z$ & $Z$ & $Z$ & $Z$ &  \\
& $X$ & $X$ & $Z$ & $Z$ & \\
& $Z$ & $X$ & $X$ & $I$ & \\
& $X$ & $Z$ & $I$ & $X$ & \\
& $I$ & $I$ & $X$ & $X$ & \\
\end{tabular}\label{EvenStar4}}
\qquad
\subfloat[][6-qubit Star Kernel]{
\begin{tabular}{cccccccc}
& $Z$ & $Z$ & $Z$ & $Z$ & $Z$ & $Z$ & \\
& $X$ & $X$ & $Z$ & $Z$ & $Z$ & $Z$ & \\
& $Z$ & $X$ & $X$ & $I$ & $I$ & $I$ & \\
& $X$ & $Z$ & $I$ & $X$ & $I$ & $I$ & \\
& $I$ & $I$ & $I$ & $X$ & $X$ & $I$ & \\
& $I$ & $I$ & $I$ & $I$ & $X$ & $X$ & \\
& $I$ & $I$ & $X$ & $I$ & $I$ & $X$ & \\
\end{tabular}\label{EvenStar6}}
\qquad
\subfloat[][Even $N$ Kernel]{
\begin{tabular}{cccccccc}
$Z$ & $Z$ & $Z$ & $Z$ & $Z$ & $Z$ & $\cdots$ & $Z$ \\
$X$ & $X$ & $Z$ & $Z$ & $Z$ & $Z$ & $\cdots$ & $Z$ \\
$Z$ & $X$ & $X$ & $I$ & $I$ & $I$ & $\cdots$ & $I$ \\
$X$ & $Z$ & $I$ & $X$ & $I$ & $I$ & $\cdots$ & $I$ \\
$I$ & $I$ & $I$ & $X$ & $X$ & $I$ & $\cdots$ & $I$ \\
$I$ & $I$ & $I$ & $I$ & $X$ & $X$ & $\ddots$ & $\vdots$ \\
$\vdots$ & $\vdots$ & $\vdots$ & $\vdots$ & $\ddots$ & $\ddots$ & $\ddots$ & $I$ \\
$I$ & $I$ & $I$ & $I$ & $\cdots$ & $I$ & $X$ & $X$ \\
$I$ & $I$ & $X$ & $I$ & $\cdots$ & $I$ & $I$ & $X$ \\
\end{tabular}\label{EvenStarN}}
\qquad
\caption[Star Kernels for Even $N$]{The pattern of Star Kernels for even $N$.  The $N=4$ \subref{EvenStar4}, $N=6$ \subref{EvenStar6} cases are shown explicitly, while \subref{EvenStarN} shows how the pattern generalizes to arbitrary even $N$.  The ellipses in \subref{EvenStarN} always indicate repetition of the observable found at either end.}\label{EvenStar}
\end{table}

Because the ID is so symmetric it is fairly easy to see that each of the IDs shown in Table \ref{OddStar} are critical, in the usual sense that no subset of qubits and/or observables can be deleted such that the remaining set is a smaller ID.  For the IDs of Table \ref{EvenStar}, seeing the criticality is a bit more subtle.  The best way to see it is to first separate the ID of Table \ref{EvenStar4} into a `head' and a `tail,' where the head is simply the first 4 observables, and the tail is the last.  Then one can see that the head structure is internally critical, and needs the tail to be an ID.  If we consider the general ID of Table \ref{EvenStarN}, we can see that the head generalizes in a very simple way, and the tail also generalizes in a simple and symmetric way.  It is then fairly easy to see that both the tail and head are internally critical, and that both are necessary to form the ID.  Unfortunately this demonstration is cumbersome at best, but we have verified it for up to the $N=10$ case using a computer.

Between the two families given here, we can prescribe an experiment using $N$ qubits to prove the GHZ theorem with $N$ spacelike separated parties (see Chapter \ref{sec:Kernels}) for any value of $N$, and without requiring the use of any ancillary qubits.  To our knowledge, the proofs of Table \ref{EvenStar} are the only known family for even $N$, and the 4 party experiment has never been conducted.

As a final note on this set, it is worth mentioning that the eigenstates of the Star IDs belong to the family of fully entangled Web States, which we will discuss in more detail elsewhere.  There are two distinct families of Web States within the $N$-qubit Pauli group, both of which possess a particularly robust type of entanglement, and appear to be local-unitary-equivalent to GHZ states.  If any one qubit is measured in the product basis on an $N$-qubit Web state, the remaining $N$ qubits are always left in a (fully entangled) $(N-1)$-qubit Web state.  This entanglement persists as successive qubits are measured, until the last 2 qubits are left in a Bell state.

\pagebreak

\subsection{The Kite Family}\label{sec:Kites}
\begin{figure}[h!]
  \centering
  \includegraphics[width=3in]{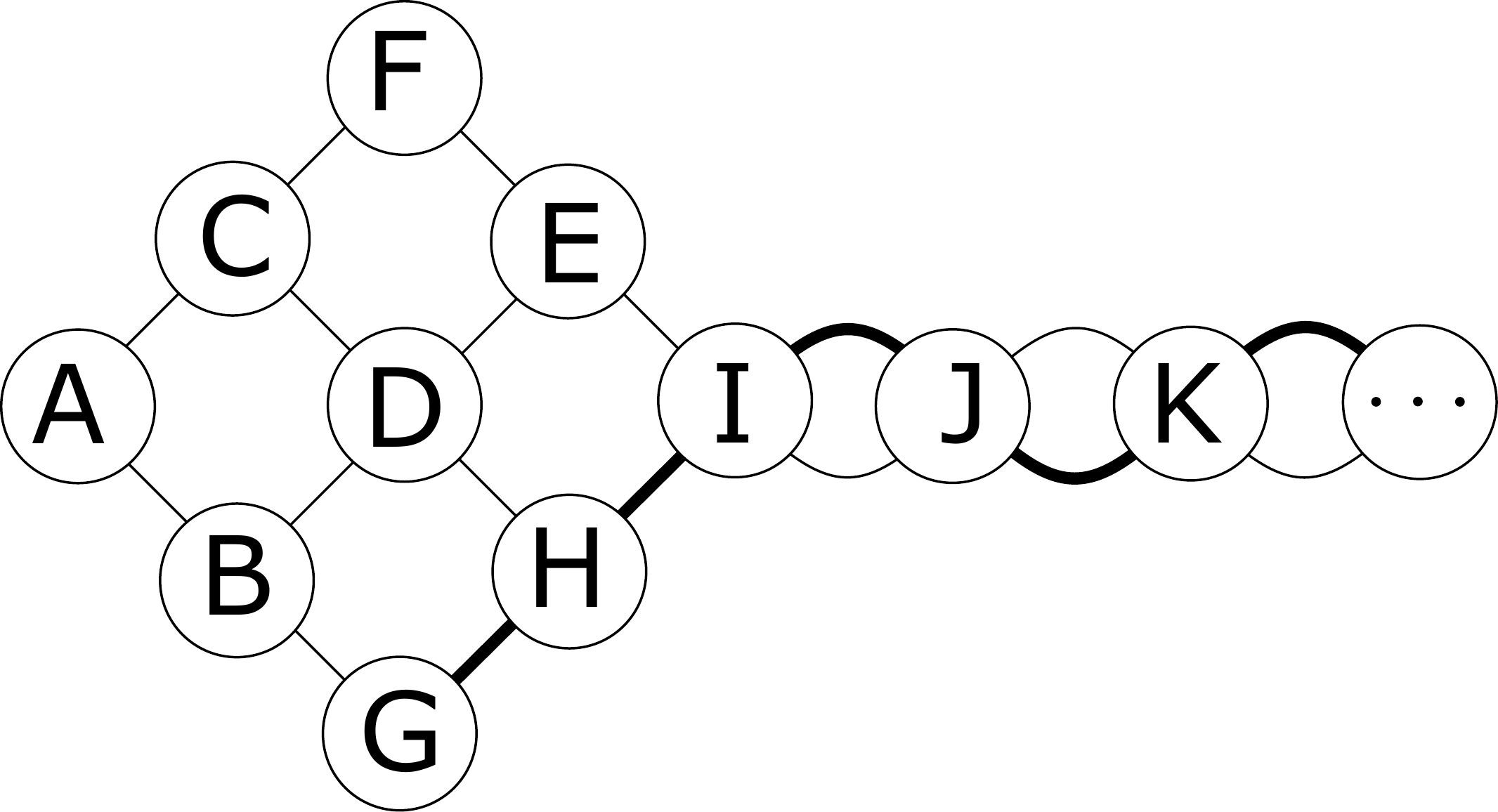}\\
  \caption[General Kite]{This is the general Kite diagram that we can use for Kites with tails of all lengths.  Each letter is an observable, while the lines and arcs represent IDs, with thin lines for Positive IDs and thick lines for Negative IDs.  The two IDs of the tail are the usual Kernel of the Kite.}\label{KiteGen}
\end{figure}
\begin{table}[h!]
\begin{center} {
\begin{tabular}{cc}
$|\psi_1\rangle = \{ A(+1),B(+1),G(+1) \}$  & $|\psi_{10}\rangle = \{ C(+1),D(-1),H(-1) \}$ \\
$|\psi_2\rangle = \{ A(+1),B(-1),G(-1) \}$  & $|\psi_{11}\rangle = \{ C(-1),D(+1),H(-1) \}$ \\
$|\psi_3\rangle = \{ A(-1),B(+1),G(-1) \}$  & $|\psi_{12}\rangle = \{ C(-1),D(-1),H(+1) \}$ \\
$|\psi_4\rangle = \{ A(+1),C(+1),F(+1) \}$  & $|\psi_{13}\rangle = \{ E(+1),F(-1),I(-1) \}$ \\
$|\psi_5\rangle = \{ A(-1),C(+1),F(-1) \}$  & $|\psi_{14}\rangle = \{ E(-1),F(+1),I(-1) \}$ \\
$|\psi_6\rangle = \{ A(-1),C(-1),F(+1) \}$  & $|\psi_{15}\rangle = \{ E(-1),F(-1),I(+1) \}$ \\
$|\psi_7\rangle = \{ B(+1),D(+1),E(+1) \}$  & $|\psi_{16}\rangle = \{ G(+1),H(+1),I(-1) \}$ \\
$|\psi_8\rangle = \{ B(-1),D(+1),E(-1) \}$  & $|\psi_{17}\rangle = \{ G(+1),H(-1),I(+1) \}$ \\
$|\psi_9\rangle = \{ B(-1),D(-1),E(+1) \}$  & $|\psi_{18}\rangle = \{ G(-1),H(+1),I(+1) \}$ \\
\end{tabular} }
\end{center}
\caption[Rays of the Critical Parity Proof generated by the General Kite]{The rays of an $18-9$ critical parity proof associated with the $M=3$ Kite.  Each time $M$ increases by one, rays 1-12 are unchanged, but rays 13-18 split into two orthogonal rays, as shown in Table \ref{Proof9Pattern}.  Each Kite contains 16 replicas of this proof under symmetry, which remains true as $M$ increases.}
\label{Proof18}
\end{table}
 \begin{table}[h!]
\begin{center} {
\begin{tabular}{c}
$\{  |\psi_{1}\rangle,  |\psi_{2}\rangle,  |\psi_{5}\rangle,  |\psi_{6}\rangle  \}$ \\
$\{  |\psi_{1}\rangle,  |\psi_{3}\rangle,  |\psi_{8}\rangle,  |\psi_{9}\rangle  \}$ \\
$\{  |\psi_{4}\rangle,  |\psi_{5}\rangle,  |\psi_{11}\rangle,  |\psi_{12}\rangle  \}$ \\
$\{  |\psi_{7}\rangle,  |\psi_{8}\rangle,  |\psi_{10}\rangle,  |\psi_{12}\rangle  \}$ \\
$\{  |\psi_{2}\rangle,  |\psi_{3}\rangle,  |\psi_{16}\rangle,  |\psi_{17}\rangle  \}$ \\
$\{  |\psi_{4}\rangle,  |\psi_{6}\rangle,  |\psi_{13}\rangle,  |\psi_{15}\rangle  \}$ \\
$\{  |\psi_{7}\rangle,  |\psi_{9}\rangle,  |\psi_{14}\rangle,  |\psi_{15}\rangle  \}$ \\
$\{  |\psi_{10}\rangle,  |\psi_{11}\rangle,  |\psi_{16}\rangle,  |\psi_{18}\rangle  \}$ \\
$\{  |\psi_{13}\rangle,  |\psi_{14}\rangle,  |\psi_{17}\rangle,  |\psi_{18}\rangle  \}$ \\
\end{tabular} }
\end{center}
\caption[The 9 Bases of the Critical Parity Proof generated by the General Kite]{The 9 bases of the $18-9$ critical parity proof.  As $M$ increases, rays 1-12 are unchanged, and rays 13-18 each represent the full group of rays which that ray splits into.  The 9 bases are always the ones shown here for all $M$.}
\label{Bases9}
\end{table}
\begin{table}[h!]
\centering
\subfloat[][]{
\begin{tabular}{c}
$|\psi_{13}\rangle = \{ E(+1),F(-1),I(-1) \}$ \\
$|\psi_{14}\rangle = \{ E(-1),F(+1),I(-1) \}$ \\
$|\psi_{15}\rangle = \{ E(-1),F(-1),I(+1) \}$ \\
$|\psi_{16}\rangle = \{ G(+1),H(+1),I(-1) \}$ \\
$|\psi_{17}\rangle = \{ G(+1),H(-1),I(+1) \}$ \\
$|\psi_{18}\rangle = \{ G(-1),H(+1),I(+1) \}$ \\
\end{tabular} \label{Rays6}}
\qquad
\subfloat[][]{
\begin{tabular}{c}
$|\psi_{13_a}\rangle = \{ E(+1),F(-1),I(+1),J(-1) \}$ \\
$|\psi_{13_b}\rangle = \{ E(+1),F(-1),I(-1),J(+1) \}$ \\
$|\psi_{14_a}\rangle = \{ E(-1),F(+1),I(+1),J(-1) \}$ \\
$|\psi_{14_b}\rangle = \{ E(-1),F(+1),I(-1),J(+1) \}$ \\
$|\psi_{15_a}\rangle = \{ E(-1),F(-1),I(+1),J(+1) \}$ \\
$|\psi_{15_b}\rangle = \{ E(-1),F(-1),I(-1),J(-1) \}$ \\
$|\psi_{16_a}\rangle = \{ G(+1),H(+1),I(+1),J(-1) \}$ \\
$|\psi_{16_b}\rangle = \{ G(+1),H(+1),I(-1),J(+1) \}$ \\
$|\psi_{17_a}\rangle = \{ G(+1),H(-1),I(+1),J(+1) \}$ \\
$|\psi_{17_b}\rangle = \{ G(+1),H(-1),I(-1),J(-1) \}$ \\
$|\psi_{18_a}\rangle = \{ G(-1),H(+1),I(+1),J(+1) \}$ \\
$|\psi_{18_b}\rangle = \{ G(-1),H(+1),I(-1)J(-1) \}$ \\
\end{tabular} \label{Rays12}}
\qquad
\caption[Ray Splitting for the Critical Parity Proofs within the General Kite]{This shows how rays 13-18 each split into two orthogonal rays every time $M$ is increased by one.  The two new rays have opposite eigenvalues for the last observable of the original ray, and the eigenvalue of the new observable is fixed by the sign of the ID (negative for rays 16, 17, and 18).  This splitting process is applied to this entire set of rays each time $M$ increases.  For all values of $M$, rays 1-12 combined with this set of $6 \cdot 2^{M-3}$ rays from the other group always form exactly 9 bases, and a critical parity proof.}\label{Proof9Pattern}
\end{table}
\begin{table}[h!]
\centering
\subfloat[][ID$5^7_2$]{
\begin{tabular}{ccccccc}
$\textbf{Z}$ & $Z$ & $Z$ & $Z$ & $Z$ & $I$ & $I$ \\
$\textbf{X}$ & $X$ & $Z$ & $X$ & $X$ & $Z$ & $Z$ \\
$Y$ & $I$ & $X$ & $Z$ & $Z$ & $X$ & $X$ \\
$I$ & $Y$ & $I$ & $X$ & $I$ & $Z$ & $X $ \\
$I$ & $I$ & $X$ & $I$ & $X$ & $X$ & $Z $ \\
\end{tabular} \label{M5N7}}
\qquad
\subfloat[][ID$6^{11}_2$]{
\begin{tabular}{ccccccccccc}
$ I$ & $I$ & $Z$ & $Z$ & $Z$ & $Z$ & $Z$ & $Z$ & $Z$ & $Z$ & $Z  $ \\
$ \textbf{Z}$ & $I$ & $Z$ & $Z$ & $Z$ & $X$ & $X$ & $I$ & $X$ & $X$ & $I $ \\
$ I$ & $Z$ & $X$ & $X$ & $I$ & $Z$ & $Z$ & $Z$ & $I$ & $I$ & $I $ \\
$ \textbf{X}$ & $I$ & $X$ & $I$ & $X$ & $X$ & $I$ & $X$ & $Z$ & $X$ & $X $ \\
$ I$ & $X$ & $I$ & $X$ & $I$ & $I$ & $X$ & $I$ & $I$ & $Z$ & $Z $ \\
$ Y$ & $Y$ & $I$ & $I$ & $X$ & $I$ & $I$ & $X$ & $X$ & $I$ & $X $ \\
\end{tabular} \label{M6N11}}
\qquad
\subfloat[][ID$7^{16}_{16}$]{
\begin{tabular}{cccccccccccccccc}
$\textbf{Z}$ & $Z$ & $Z$ & $Z$ & $Z$ & $Z$ & $Z$ & $Z$ & $I$ & $I$ & $I$ & $I$ & $I$ & $I$ & $I$ & $I $ \\
$\textbf{X}$ & $X$ & $X$ & $X$ & $I$ & $I$ & $I$ & $I$ & $Z$ & $Z$ & $Z$ & $Z$ & $I$ & $I$ & $I$ & $I $ \\
$Y$ & $I$ & $I$ & $I$ & $X$ & $X$ & $X$ & $I$ & $X$ & $I$ & $I$ & $I$ & $Z$ & $Z$ & $I$ & $I $ \\
$I$ & $Y$ & $I$ & $I$ & $Y$ & $I$ & $I$ & $I$ & $I$ & $X$ & $X$ & $X$ & $X$ & $I$ & $Z$ & $I $ \\
$I$ & $I$ & $I$ & $I$ & $I$ & $Y$ & $I$ & $X$ & $Y$ & $Y$ & $I$ & $I$ & $I$ & $I$ & $X$ & $Z $ \\
$I$ & $I$ & $Y$ & $I$ & $I$ & $I$ & $I$ & $Y$ & $I$ & $I$ & $Y$ & $I$ & $Y$ & $X$ & $I$ & $X $ \\
$I$ & $I$ & $I$ & $Y$ & $I$ & $I$ & $Y$ & $I$ & $I$ & $I$ & $I$ & $Y$ & $I$ & $Y$ & $Y$ & $Y $ \\
\end{tabular}\label{Kite7_16}}
\qquad
\caption[Maximally Compact IDs]{Examples of critical Partial IDs with the largest known $N$ for a given value of $M$.  In each case, a Kite Kernel can be formed by taking the given ID along with a second ID obtained by transposing the two bold observables.}\label{ShortKites}
\end{table}
\begin{table}[h!]
\centering
\subfloat[][ID$4^3_2$]{
\begin{tabular}{ccc}
$\textbf{Z}$ & $I$ & $Z$ \\
$I$ & $Z$ & $Z$ \\
$\textbf{X}$ & $X$ & $X$ \\
$Y$ & $Y$ & $X$ \\
\end{tabular} \label{KiteOdd3}}
\qquad
\subfloat[][ID$5^4_4$]{
\begin{tabular}{ccccc}
$\textbf{Z}$ & $I$ & $I$ & $I$ & $Z$ \\
$I$ & $Z$ & $I$ & $I$ & $Z$ \\
$I$ & $I$ & $Z$ & $I$ & $Z$ \\
$I$ & $I$ & $I$ & $Z$ & $Z$ \\
$\textbf{X}$ & $X$ & $X$ & $X$ & $X$ \\
$Y$ & $Y$ & $Y$ & $Y$ & $X$ \\
\end{tabular} \label{KiteOdd5}}
\qquad
\subfloat[][ID$(N+1)^N_N$]{
\begin{tabular}{ccccc}
$\textbf{Z}$ & $I$ & $\cdots$ & $I$ & $Z$ \\
$I$ & $Z$ & $\ddots$ & $\vdots$ & $\vdots$ \\
$\vdots$ & $\ddots$ & $\ddots$  & $I$ & $Z$ \\
$I$ & $\cdots$  & $I$ & $Z$ & $Z$ \\
$\textbf{X}$ & $\cdots$ & $X$ & $X$ & $X$ \\
$Y$ & $\cdots$ & $Y$ & $Y$ & $X$ \\
\end{tabular} \label{KiteOddN}}
\qquad
\caption[A Family of Critical Partial IDs for Odd $N=M+1$]{A family of critical Partial IDs, with $M=N+1$ for odd $N$.  In each case, a Kite Kernel can be formed by taking the given ID along with a second ID obtained by transposing the two bold observables.  The ellipses in \subref{KiteOddN} indicate repetition of the observable found at either end.}\label{KiteOdd}
\end{table}
\begin{table}[h!]
\centering
\subfloat[][ID$5^4_2$]{
\begin{tabular}{cccc}
$\textbf{Z}$ & $Z$ & $Z$ & $Z$ \\
$Y$ & $Y$ & $Z$ & $Z$ \\
$\textbf{X}$ & $I$ & $X$ & $I$ \\
$I$ & $X$ & $I$ & $X$ \\
$I$ & $I$ & $X$ & $X$ \\
\end{tabular} \label{KiteEven4}}
\qquad
\subfloat[][ID$7^6_2$]{
\begin{tabular}{cccccc}
$\textbf{Z}$ & $Z$ & $Z$ & $Z$ & $Z$ & $Z$ \\
$Y$ & $Y$ & $Z$ & $Z$ & $Z$ & $Z$ \\
$\textbf{X}$ & $I$ & $X$ & $I$ & $I$ & $I$ \\
$I$ & $X$ & $I$ & $X$ & $I$ & $I$ \\
$I$ & $I$ & $X$ & $I$ & $X$ & $I$ \\
$I$ & $I$ & $I$ & $X$ & $I$ & $X$ \\
$I$ & $I$ & $I$ & $I$ & $X$ & $X$ \\
\end{tabular} \label{KiteEven6}}
\qquad
\subfloat[][ID$(N+1)^N_2$]{
\begin{tabular}{cccccc}
$\textbf{Z}$ & $Z$ & $Z$ & $Z$ & $\cdots$ & $Z$ \\
$Y$ & $Y$ & $Z$ & $Z$ & $\cdots$ & $Z$ \\
$\textbf{X}$ & $I$ & $X$ & $I$ & $\cdots$ & $I$ \\
$I$ & $X$ & $I$ & $X$ & $\ddots$ & $\vdots$ \\
$\vdots$ & $\ddots$ & $\ddots$ & $\ddots$ & $\ddots$ & $I$ \\
$I$ & $\cdots$ & $I$ & $X$ & $I$ & $X$ \\
$I$ & $\cdots$ & $I$ & $I$ & $X$ & $X$ \\
\end{tabular} \label{KiteEvenN}}
\qquad
\caption[A Family of Critical Partial IDs for Even $N=M+1$]{A family of critical Partial IDs, with $M=N+1$ for even $N$.  In each case, a Kite Kernel can be formed by taking the given ID along with a second ID obtained by transposing the two bold observables.  The ellipses in \subref{KiteEvenN} indicate repetition of the observable found at either end.}\label{KiteEven}
\end{table}
\begin{table}[h!]
\centering
\subfloat[][ID$4^4_4$]{
\begin{tabular}{cccc}
$\textbf{Z}$ & $Z$ & $Z$ & $I$ \\
$Y$ & $Y$ & $I$ & $Z$ \\
$\textbf{X}$ & $I$ & $Y$ & $Y$ \\
$I$ & $X$ & $X$ & $X$ \\
\end{tabular} \label{KiteEven24}}
\qquad
\subfloat[][ID$5^6_6$]{
\begin{tabular}{cccccc}
$\textbf{Z}$ & $Z$ & $Z$ & $I$ & $I$ & $I$ \\
$Y$ & $Y$ & $I$ & $Z$ & $Z$ & $I$ \\
$\textbf{X}$ & $I$ & $Y$ & $Y$ & $I$ & $Z$ \\
$I$ & $X$ & $X$ & $I$ & $Y$ & $Y$ \\
$I$ & $I$ & $I$ & $X$ & $X$ & $X$ \\
\end{tabular} \label{KiteEven26}}
\qquad
\subfloat[][ID$(\frac{N}{2}+2)^N_N$]{
\begin{tabular}{cccccc}
$\textbf{Z}$ & $Z$ & $Z$ & $I$ & \underline{$I$} & \underline{$I$} \\
$Y$ & $Y$ & $I$ & \underline{$Z$} & \underline{$Z$} & $I$ \\
$\textbf{X}$ & $I$ & \underline{$Y$} & \underline{$Y$} & $I$ & $Z$ \\
$I$ & \underline{$X$} & \underline{$X$} & $I$ & $Y$ & $Y$ \\
\underline{$I$} & \underline{$I$} & $I$ & $X$ & $X$ & $X$ \\
\end{tabular} \label{KiteEven2N}}
\qquad
\caption[A Family of Critical Partial IDs for Even $N = 2M-4$]{A family of critical Partial IDs, with $M=N/2+2$ for even $N$.  In each case, a Kite Kernel can be formed by taking the given ID along with a second ID obtained by transposing the two bold observables.  The underlines in \subref{KiteEven2N} indicate repetition of those pairs of observables along the slant diagonal, such that $N$ increases by 2 each time $M$ increases by one.}\label{KiteEven2}
\end{table}

In this section we discuss Kites (for an alternate discussion, see our paper \cite{WA_Nqubits}), the most general family of Observable-based KS proofs based on the simplest Composite Kernel structure.  As we will see, the Kites give rise to what are most likely the smallest $R-B$ parity proofs of the KS theorem that are possible in the Hilbert space of $N$ qubits.

A Kite is a Composite Kernel, based on a very simple class of CKSs - those with just two IDs, and all `O' entries.  A Kite can be formed using any Partial ID$M^N$ that contains at least one of a certain type of Odd SQP.  The SQP must contain two of the Pauli observables only once, and the third Pauli observable can appear any number of times.  Then this ID$M^N$ can be permuted to have a different sign, simply by transposing those two observables within that SQP.  Once this has been done, the two Partial ID$M^N$s - the original and permuted version - together form a Kernel.  Following the general method discussed in Chapter \ref{sec:ObKSproofs}, we can extend this Kernel into an Observable-based KS proof by adding 4 ID$3^N$s.  This proof will always be a Kite, with the length of the tail determined by $M$.  The 2 unique observables in each ID always combine with 4 new observables to form the head of the Kite.  For example, in Tables \ref{ShortKites}, \ref{KiteOdd}, \ref{KiteEven}, and \ref{KiteEven2}, the unique observables are the two with the bold elements, along with the two others obtained by swapping the bold elements.

There are a huge variety of unique Partial ID$M^N$s that can be used to form Kites in this way, and since any two Kites of the same tail-length are isomorphic, so too are the $R-B$ sets they generate.  As we will see, all of the relevant properties of the structures found within a Kite depend only on $M$.  Only the rank of the rays is determined by $N$, and the Observable-based KS proof has symbol $(6+M)_2 - 2_M 4_3$

To give some idea of the variety of unique critical Partial ID$M^N$s, we will list the ranges of $N$ for the first few values of $M$.  For $M=3$, we get $N=2,3$, which are actually the Mermin Square and 3-qubit Wheel.  For $M=4$, we get $N=3,4$.  For $M=5$, $N=4,5,6,7$, for $M=6$, $N=5,6,\ldots,11$, and for $M=7$, $N=6,7,\ldots,16$.  An example of each of the largest cases is shown in Table \ref{ShortKites}.

We also specify a few families of critical Partial IDs for all values of $N$, which do not have minimal $M$ for the given $N$, but still show that the Kite family exists throughout the $N$-qubit Pauli Group.  These are shown in Tables \ref{KiteOdd}, \ref{KiteEven}, and \ref{KiteEven2}, with the SQPs permuted such that each set is fully real.

Each Kite generates $16$ rays of rank $r = 2^{N-2}$, and $2^M$ rays of rank $r=2^{N-M+1}$, for a total of $R=16 + 2^M$ rays.  The two ID$M^N$s that form the tail share $M-2$ observables in common, and so they generate $2^{2^{M-2}} - 2$ hybrid bases between them.  The other 8 observables are each common to just 2 observables, and so give us another 16 hybrid bases.  These together with the 6 eigenbases give us a total count of $B = 20+2^{2^{M-2}}$ bases in the set.

Because for $M>3$ this set contains a pair of ID$M^N$s that share more than one common observable, the total number of parity proofs within the set is difficult to determine.  We have found all such proofs for $M=4$, as discussed in Chapter \ref{sec:Kite3} \cite{WA_3qubits}.  For $M=5$, an exhaustive computational search of the $48-276$ is a bit outside our reach.

Nevertheless, we have discovered a pattern by which we can generate a group of 16 critical parity proofs from a Kite of any $M$, such that the proof always uses just 9 bases.  First we will introduce the general Kite in Figure \ref{KiteGen}, so that we can follow the labeling of observables given there.

 We begin explicitly with an $18-9$ parity proof from within the $M=3$ Kite (The Mermin Square and 3-qubit Wheel are of this type).  The 18 rays are shown in Table \ref{Proof18}, and the 9 bases in Table \ref{Bases9}.  As the tail grows longer, the first 12 rays are unchanged.  However, for every observable that is added to the tail, the last 6 rays each split into 2 orthogonal rays, as shown in Table \ref{Proof9Pattern}.  The bases remain unchanged as $M$ increases, containing all members of each split group according to the given index.  This process always results in $R = 12 + 6\cdot 2^{M-3}$ rays, which form an $R_2 - 9$ critical parity proof.  There are 4 bases with 4 rays, 4 bases with $2+2\cdot2^{M-3}$ rays, and 1 basis with $4\cdot2^{M-3}$ rays.   We expect these to be the most compact proofs of the KS theorem within the $N$-qubit Pauli group, though we cannot prove this conclusively.  To be explicit, the ID$7^{16}$ shown in Table \ref{Kite7_16} can be used to form a Kite which contains 16 $108_2 - 9$ critical parity proof, with expanded symbol $12_2^{16384} 96_2^{1024} - 1_{64} 4_{34} 4_4$, which we expect is the most compact critical parity proof in a Hilbert space of $d=2^{16}=65,536$.

\subsection{Miscellaneous Structures}

In this section we will review a few other selected cases that we find exceptional.  To begin, we will introduce a few more Kernels and Observable-based KS proofs for more than 4 qubits.  Following this, we will discuss some of ways that CKSs can be combined with IDs to create critical Composite Kernels for large numbers of qubits.

\subsubsection{An Alternate 5-qubit Star}
\begin{figure}[ht]
  \centering
  \includegraphics[width=2.5in]{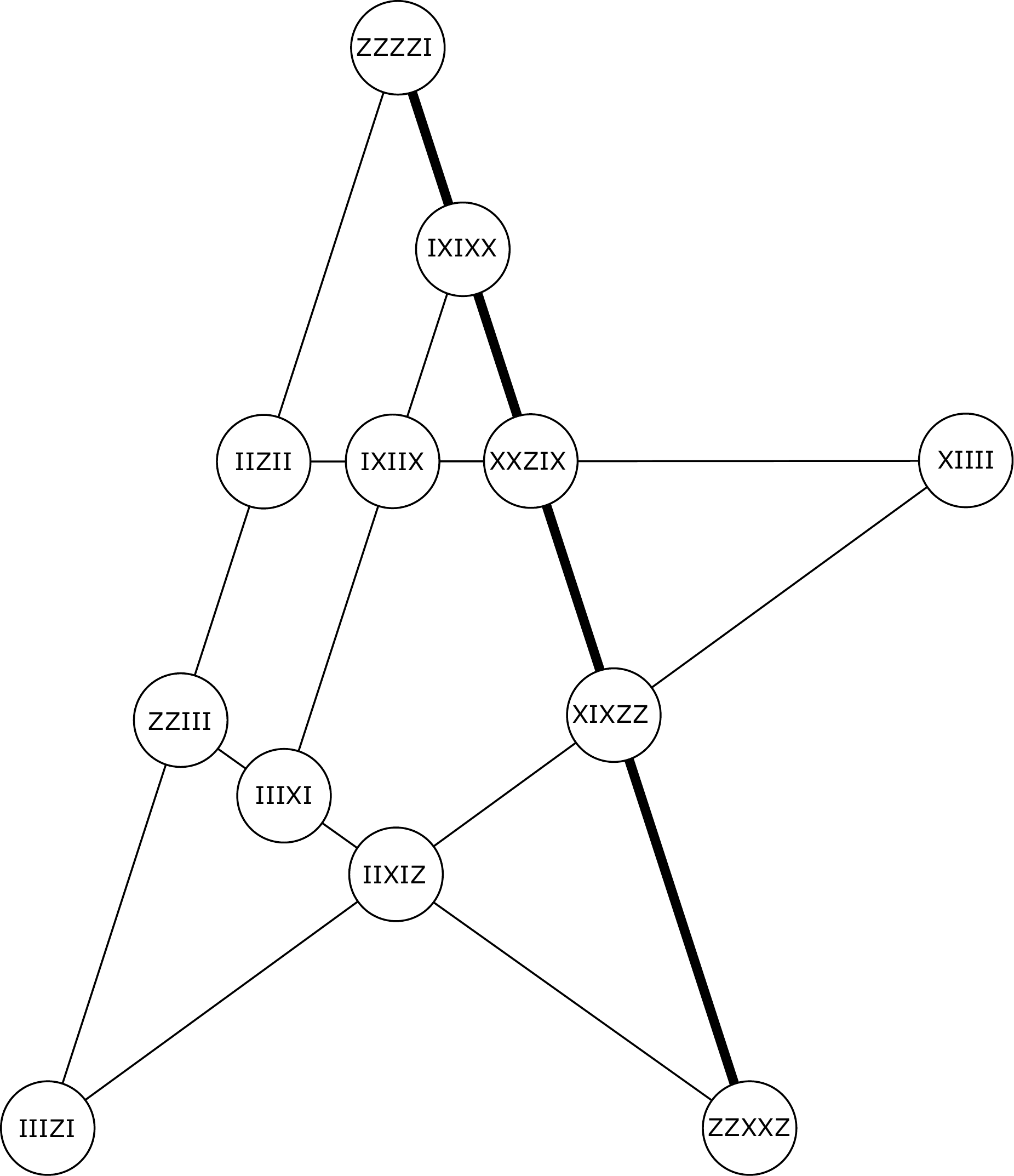}\\
  \caption[Alternate 5-qubit Star]{The alternate 5-qubit Star, an Observable-based KS proof with symbol $12_2 - 1_5 4_4 1_3$.  The negative ID is the Kernel.  This diagram is isomorphic to the 4-qubit Star of Figure \ref{q4Star}.}\label{q5Star_Alt}
\end{figure}

Here we introduce the only unique ID$5^5_0$ that exists in the 5-qubit Pauli group.  This is the smallest number of qubits for which there exists a single-ID Kernel with $M<N+1$, and we expect it to give the most compact proof of the GHZ theorem possible for 5 qubits.  The Observable-based KS proof this Kernel generates is isomorphic to the 4-qubit Star of Chapter \ref{sec:q4Star}, as shown in Figure \ref{q5Star_Alt}.  It also contains an isomorphic $52-30$ set with expanded symbol $16^2_6 32^4_5 4^4_8 - 1_{16} 8_{12} 2_{10} 14_8 4_6 1_4$ (All rays are double the rank of the 4-qubit case), which contains $2^{12}$ critical parity proofs.

\subsubsection{The 6-qubit Arch and Arrow}
\begin{figure}[ht]
  \centering
  \includegraphics[width=2.5in]{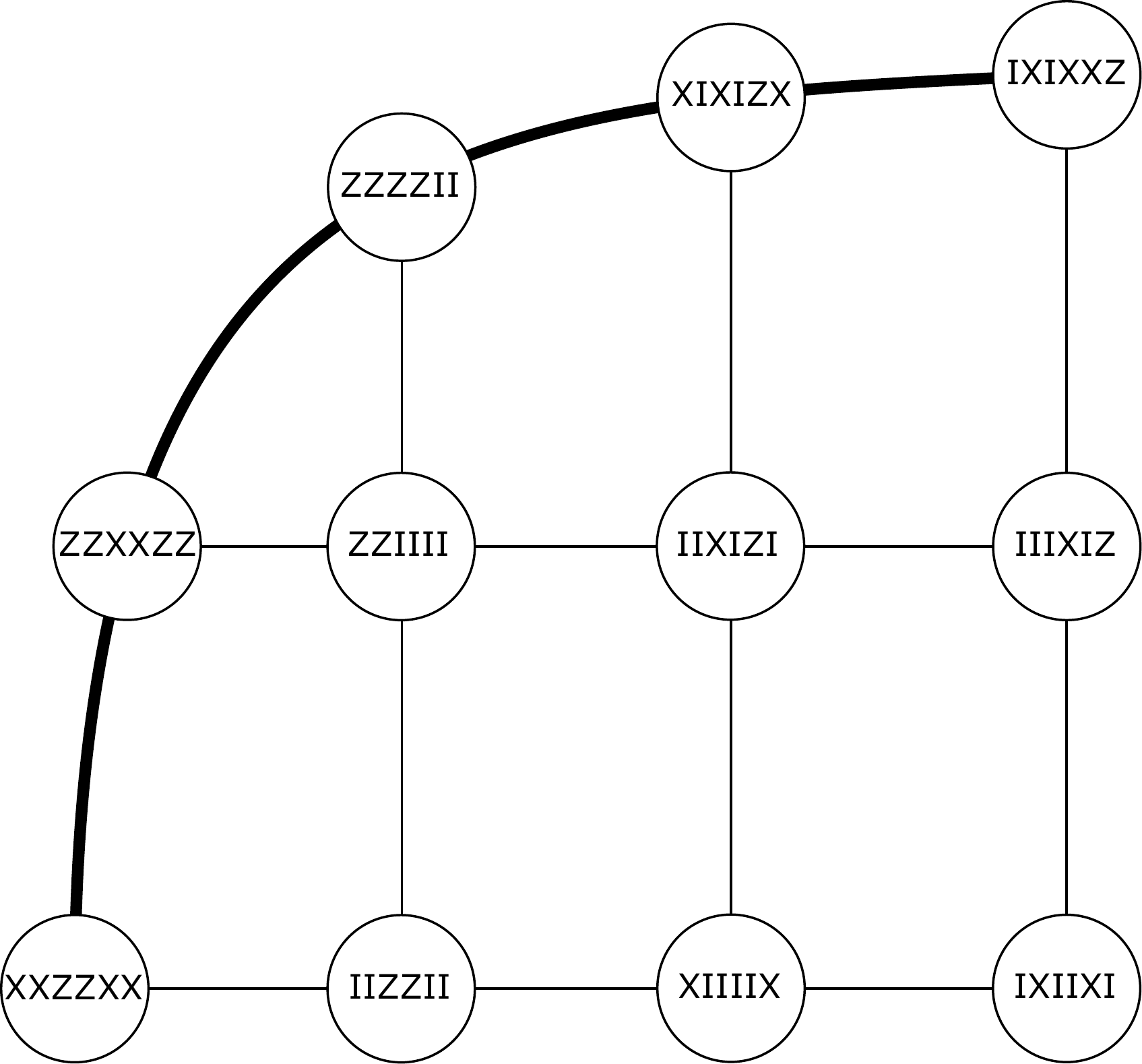}\\
  \caption[6-qubit Arch]{The 6-qubit Arch, a $11_2 - 1_5 2_4 3_3$ Observable-based KS proof.  The negative ID is the Kernel.}\label{q6Arch}
\end{figure}
\begin{figure}[ht]
  \centering
  \includegraphics[width=2.5in]{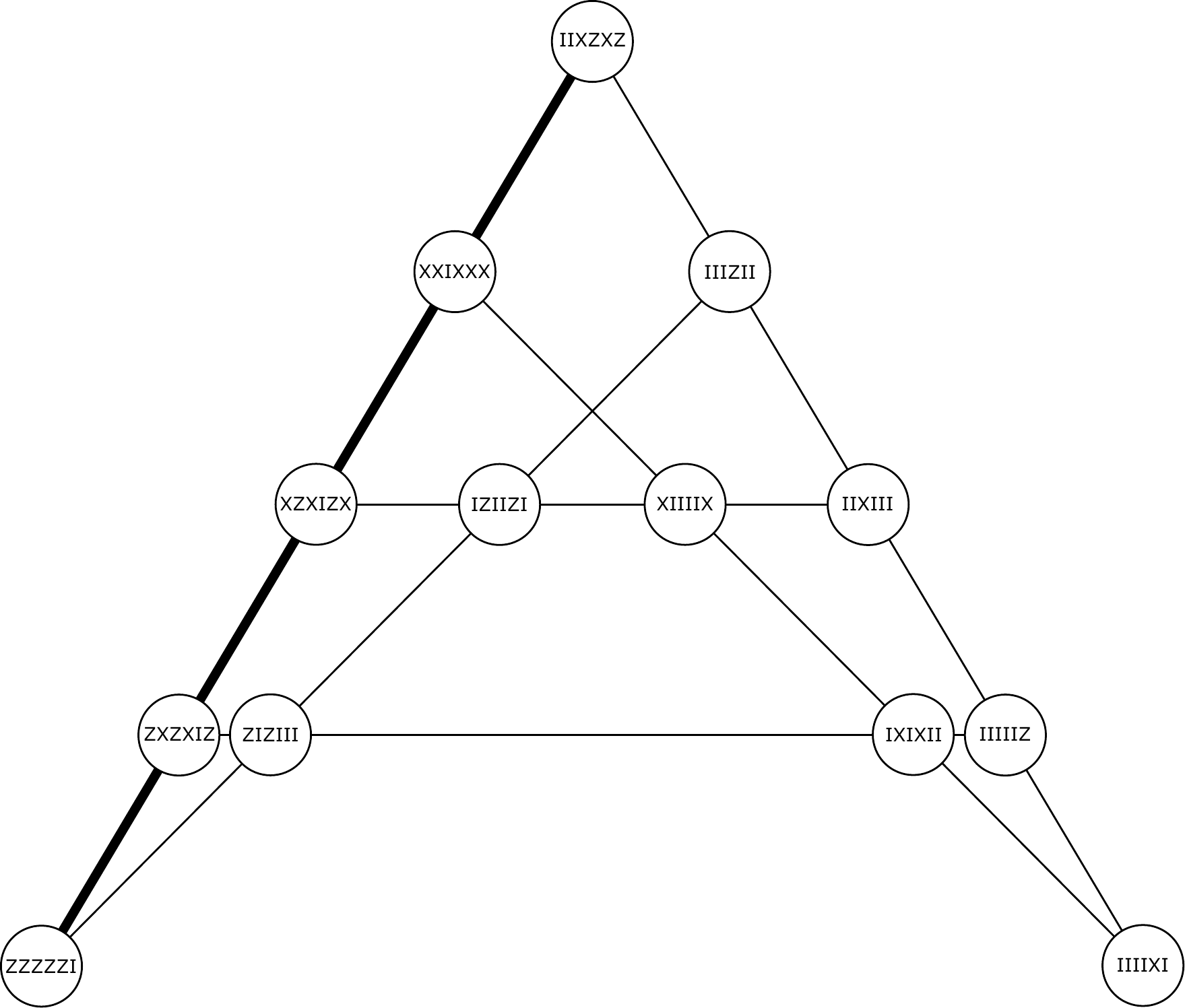}\\
  \caption[6-qubit Arrow]{The 6-qubit Arrow, a $13_2 - 2_5 4_4$ Observable-based KS proof.  The negative ID is the Kernel.}\label{q6Arrow}
\end{figure}

Here we introduce the two unique ID$5^6_0$s that exist in the 6-qubit Pauli group.  Both of these single-ID Kernels give rise to remarkably compact proofs of the GHZ theorem, given that they only generate rays of rank 4 and greater.  In both cases, an Observable-based KS proof can be generated that belongs to the special simple class discussed in Chapter \ref{sec:RBSets}, as shown in Figures \ref{q6Arch}, \ref{q6Arrow}.

The Arch generates a $44-28$ set with expanded symbol $16^4_6 16^8_5 12^{16}_4 - 1_{16} 4_{12} 6_{10} 2_8 12_6 3_4$, and contains $2^{11}$ critical parity proofs.

The Arrow generates a $64-32$ set with expanded symbol $32^4_6 32^8_5 - 4_{16} 16_{12} 12_8$, and contains $2^{13}$ critical parity proofs.

\subsubsection{Exploiting CKSs to Prove the KS Theorem for many qubits}
\begin{table}[ht]
\centering
\qquad
\subfloat[][Saw Kernel]{
\begin{tabular}{cccc}
O & O & E & I \\
O & O & I & E \\
\end{tabular}\label{SawKernel}}
\qquad
\subfloat[][Pinwheel Kernel]{
\begin{tabular}{cccccc}
O & O & I & E & I & I \\
I & O & O & I & E & I \\
O & I & O & I & I & E \\
\end{tabular}\label{PinwheelKernel}}
\qquad
\caption[Maximal Composite Kernels]{The critical Composite Kernels formed using permutations of the critical ID$4^3_2$ of Table \ref{q3PartialID} combined with trivial SQPs.}\label{Saw_Pin}
\end{table}
Now we move on to discussion of ways that we can combine critical IDs and critical CKSs for relatively small numbers of qubits to form critical Composite Kernels for remarkably large numbers of qubits.

The method we will employ was actually shown in a very rudimentary way in Table \ref{CKS2_3}.  The key is that a CKS only tells us how we assign the Odd SQPs in each Partial ID to the qubits.  The Even SQPs in a Partial ID can actually be assigned in whatever way we like - in fact we can assign every Even SQP in every Partial ID in the set its own qubit.  This maximizes the number of qubits we can include using the given CKS and Partial IDs.  To all other qubits in these IDs we simply assign `I'; and then the combined facts that the CKS is critical, and each Partial ID is Critically Linked over all of its Odd and Even SQPs, guarantees that the Composite Kernel is critical as well.

To show a clear example of this process, consider the critical Partial ID$4^3_2$ of Table \ref{q3PartialID} and the critical CKS of Table \ref{CKS2}.  We can combine these two structures to form a critical 4-qubit Composite Kernel, as shown in Table \ref{SawKernel}.  This Kernel generates a $17_2 - 4_4 6_3$ Observable-based KS proof that we call the 4-qubit Saw.  This proof generates a $32^2_5 24^4_4 - 10_8 20_6 14_4$ set, which contains $2^{17}$ critical parity proofs.

As another example, consider the same ID with the critical CKS of Table \ref{CKS3}.  We can combine both of these 3-qubit Structures to form a critical 6-qubit Composite Kernel, as shown in Table \ref{PinwheelKernel}.  This Kernel generates a $16_2-5_4 4_3$ Observable-based KS proof that we call the 6-qubit Pinwheel (clearly we are running out of names).  This proof generates a $40^8_5 16^{16}_4 - 19_8 12_6 10_4$ set, which contains $2^{16}$ critical parity proofs.

 As a more extreme example, take the ID$6^{11}_2$ of Table \ref{M6N11}.  If we assign $N\geq2$ permutations of this ID to the $N$-qubit Wheel CKS (see Table \ref{CKSWheels}), and then assign the 9 Even SQPs from each of the $N$ IDs to its own qubit, we obtain a critical Composite Kernel for a total of $10N$ qubits.  It should be easy to see that given $10N$ qubits, the Observables-based KS proof and $R-B$ set this Kernel generates are extremely compact.  This example in particular demonstrates the power of this method for generating simple structures for even relatively large numbers of qubits, using much more elementary cases as a resource.

\subsection{Graph States}\label{sec:GraphStates}

Now we will take a moment to discuss the class of entangled quantum states known as Graph States, which have generated interest as a promising resource for quantum information processing \cite{BriegelGraphStates}, \cite{YuOhGraphStates}, and how they relate to the general eigenstates of critical IDs.  Simply put, the eigenstates of most, if not all, IDs are Graph States (or are local-unitary-equivalent to Graph States).

To put this into familiar notation, we will recall how a connected $N$-vertex graph generates an $N$-qubit Graph State.  The Graph State is the simultaneous eigenstate of a set of $N$ observables from the $N$-qubit Pauli group with all positive eigenvalues.  Each observable is defined by a vertex $i = 1,\ldots,N$ of the graph as
\begin{equation}
A_i = X_i \prod_{j \in N(i)} Z_j,
\end{equation}
where $N(i)$ is the neighborhood of vertex $i$, and the subscripts on $Z$ and $X$ denote which qubit they belong to (each observable has $I$ for all other qubits).

If we generate one more observable by multiplying all $N$ of these together, we obtain a complete ID$M^N$ with $M=N+1$.  For the graphs for $N=2,3$ qubits, these IDs are particular permutations of the critical ID$3^2_2$, ID$4^3_2$, and ID$4_0^3$ shown in Tables \ref{q2kernel} and \ref{q3ID4s}.

\begin{figure}[ht]
\centering
\subfloat[][]{
\includegraphics[width=0.5in]{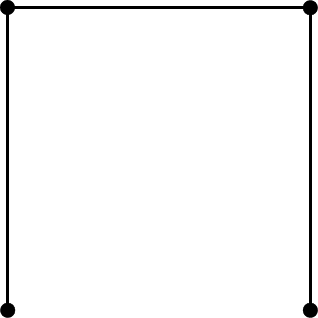}\label{G4_1}}
\qquad
\subfloat[][]{
\includegraphics[width=0.5in]{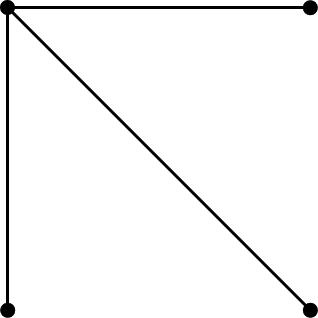}\label{G4_2}}
\qquad
\subfloat[][]{
\includegraphics[width=0.5in]{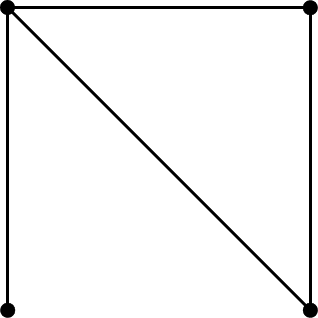}\label{G4_3}}
\qquad
\subfloat[][]{
\includegraphics[width=0.5in]{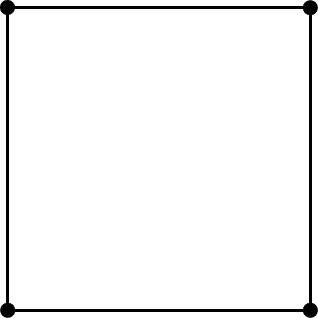}\label{G4_4}}
\qquad
\subfloat[][]{
\includegraphics[width=0.5in]{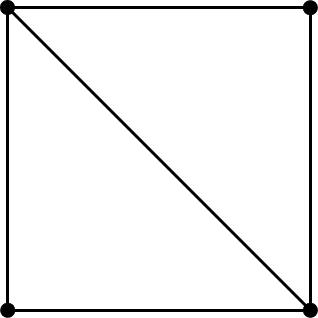}\label{G4_5}}
\qquad
\subfloat[][]{
\includegraphics[width=0.5in]{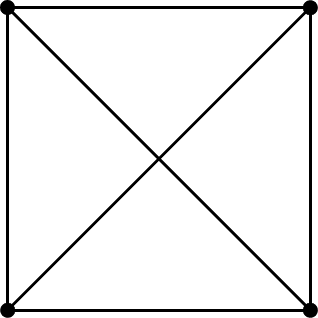}\label{G4_6}}
\qquad
\caption[The 6 Nonisomorphic Connected 4-Vertex Graphs]{}\label{Graphs4}
\end{figure}

For the 4-vertex graphs, there are 6 nonisomorphic graphs, as shown in Figure \ref{Graphs4}, and the situation is more subtle.  The graph of Figure \ref{G4_1} gives a permutation of the critical ID of Table \ref{ID5_q4_3}.  If we generate the complete stabilizer group of $2^N-1 = 15$ mutually commuting observables by taking the products of the ones in this ID, we see that it also contains permutations of every critical ID of Table \ref{q4IDs} except \ref{ID5_q4_2}.  The ID$5^4_2$ generated by the graph is shown in Table \ref{GraphID}, and a single-ID$5^4_0$ Kernel from the stabilizer group is shown in Table \ref{GraphKernel}.
\begin{table}[ht]
\centering
\qquad
\subfloat[][]{
\begin{tabular}{cccc}
$X$ & $Z$ & $I$ & $I$ \\
$Z$ & $X$ & $Z$ & $I$ \\
$I$ & $Z$ & $X$ & $Z$ \\
$I$ & $I$ & $Z$ & $X$ \\
$Y$ & $X$ & $X$ & $Y$ \\
\end{tabular}\label{GraphID}}
\qquad
\subfloat[][]{
\begin{tabular}{cccc}
$Y$ & $X$ & $X$ & $Y$ \\
$Y$ & $X$ & $Y$ & $Z$ \\
$X$ & $Z$ & $I$ & $I$ \\
$X$ & $I$ & $X$ & $Z$ \\
$I$ & $Z$ & $Y$ & $Y$ \\
\end{tabular}\label{GraphKernel}}
\qquad
\caption[Example of an ID generating a Graph State]{The ID generated from the graph of Figure \ref{G4_1} is shown in \subref{GraphID}.  \subref{GraphKernel} shows a single-ID Kernel obtained by taking products of the observables in \subref{GraphID}, and which thus belongs to the stabilizer group of the Graph State.}
\end{table}

The graphs of Figures \ref{G4_2} and \ref{G4_6} give rise to noncritical IDs, but their complete stabilizer groups contain permutations of the critical Partial ID of Table \ref{ID5_q4_2}.

The graphs of Figures \ref{G4_3} and \ref{G4_5} give rise to noncritical IDs, but their complete stabilizer groups contain permutations of every critical ID of Table \ref{q4IDs} except \ref{ID5_q4_2}.

The graph of Figure \ref{G4_4} gives rise to a Null ID, but again the complete stabilizer group contains permutations of every critical ID of Table \ref{q4IDs} except \ref{ID5_q4_2}.

The fact that we can obtain critical IDs for each case verifies that the connected $N$-vertex Graph states are fully entangled $N$-qubit states, and clearly shows two distinct classes of entanglement, involving two mutually exclusive sets of IDs.

 Thus, for every connected graph on $N\leq4$ vertices, we find critical IDs within the stabilizer group of the Graph State.  Furthermore, every unique type of ID is present within the stabilizer for one or more graphs, and it seems very likely that this property will hold for larger $N$.  It should be possible to gain a significant speedup for our existing ID search algorithm by using graph stabilizer groups as a constraint on the search space, but this will have to be done at a later time.

On a related note, the observables of a graph stabilizer group are always real valued, as are the IDs they contain.  Therefore if all unique IDs can be generated by graphs, then irreducibly complex unique IDs would not exist (IDs for which no possible permutation is real valued).

The close connection between IDs and the study of Graph States should make the implementation of the nonclassical structures presented throughout this text much easier, since much of the groundwork has already been laid for the use of Graph States.  In particular, we see that the Graph State generated by Figure \ref{G4_1} can be used to conduct the new 4-party GHZ experiment discussed in Chapter \ref{sec:StarFam}.  This experiment is very similar to the one conducted by Walther et al. in 2005 \cite{Walther2005}, in which a noncritical Single-ID Kernel was measured using an eigenstate whose stabilizer group also contains critical Single-ID Kernels. Preparations for Graph States should also be easy to adapt to virtually all of the KS experiments suggested by our Observable-based KS proofs.  Obviously we hope to see many of these experiments performed in the near future.

%
%\end{document}

%\documentclass[12pt]{article}
%\usepackage{graphicx}
%\usepackage{epsfig}
%\usepackage{amsfonts}
%\usepackage[lofdepth,lotdepth]{subfig}
%
%\begin{document}
%

\section{Computer Algorithms} \label{sec:Algorithms}

In this chapter, we will review the algorithms we used to obtain the various results of this text and the accompanying website.  We have actually constructed a fairly wide library of algorithms that we can use for various quantum information processing applications, but here we will focus on a few particularly relevant ones.  First we will introduce the general tree-search algorithm that we developed, and then we will discuss how it was applied to build programs to check a set for possible truth-value assignments, and to perform exhaustive searches for parity proofs, IDs, and CKSs.  After this, we will discuss the algorithms that are used to establish the criticality of these structures.  We conclude with a discussion of how sets of rays and bases have been generated and examined.

\subsection{The Tree-Search Algorithm}

In this section we will describe the details of how our general tree-search algorithm works, and then we will describe some of the problems we have applied it to.

The tree-search algorithm performs a complete and exhaustive search of the phase space of some particular problem for solution sets that satisfy a particular set of criteria.  For certain problems, this algorithm can be set up to complete in a remarkably small number of iterations, while for other problems it offers no advantage over more direct search algorithms.  The general procedure is to build the desired result from elementary pieces one at a time.  At each step, the set is evaluated, and a set of legitimate choices are used to branch the search tree.  This search algorithm is exhaustive because it explores every possible branch of this tree.

The general algorithm can be broken into two distinct modular functions.  One of these manages the search tree and keeps track of all of the branchings.  The other evaluates a given branch, and returns the list of subsequent legal branchings, based on the logic of the problem at hand.  If there is no practical reason to keep the number of branches at each iteration somewhat small, then the number of computations quickly becomes untenable, so this is the criteria by which we can judge this algorithm's usefulness as applied to a particular problem.

\subsubsection{Kochen-Specker Colorability Search}

The first application we will discuss is the algorithm for determining if a given set of rays and bases can be assigned noncontextual simultaneous truth-values (i.e. colored, in the KS sense).  In general, this problem is always solved most easily using the tree-search algorithm.

For a given set of rays and bases, the question at hand is can the entire set be simultaneously colored, such that each ray is assigned a noncontextual truth-value 0 or 1, and within each basis, one ray is assigned a 1, and all other rays a 0.  Because the entire set must be colored, we can begin our tree on any basis in the set, and so we just take the first one in the list.  For each ray in the chosen basis that could be assigned the value 1, we then have a branch.  If we move down the tree along one of these branches, we see that choosing the given ray to be assigned a 1, means that all of the other rays in bases containing that ray are assigned 0s.  The algorithm then chooses the basis with the most rays already assigned 0 (but none yet assigned 1), and the new branches at this iteration are the remaining unvalued rays that could still be assigned the value 1.  If the algorithm finds that all of the rays in any basis have been assigned 0, then that particular branch represents an illegal assignment to the set.  The branching process continues until every branch is terminated, either by illegality, or by finishing a complete assignment of 0s and 1s to every ray in the set.  Once the full tree has been explored, the program returns it as a table (each row ends on a different branch), along with an accompanying table that shows which branches resulted in illegal assignments.  If every branch shows an illegal assignment, then we have exhaustively shown that the complete set cannot admit a noncontextual truth-value assignment.  On the other hand, the branches that are not illegal are then a complete list of the allowed truth-value assignments to the set.

It should be noted that this algorithm treats two rays as orthogonal only if they both appear in one of the bases in the set.  This is because the bases represent the only measurement contexts we are considering, and we would need additional contexts in order to require additional orthogonality.  We call sets that cannot be assigned truth-values according to these rules basis-noncolorable KS sets.

We have also developed an alternate algorithm that checks the ray-colorability.  This algorithm works like the one described above, except that when a given ray is assigned the value 1, all orthogonal rays are assigned 0, instead of only those rays that share a basis with that ray.  It should be clear that all sets that are basis-noncolorable are also automatically ray-noncolorable.

\subsubsection{Parity Proof Search}

As we discussed in Chapter \ref{sec:MerminSquare}, there exist particular sets of rays and bases for which it is trivial to see that no truth-value assignment can exist, without needing anything so involved as a tree.  These sets contain an odd number of bases, but with each ray appearing in an even number of those bases.  We can see that this set cannot be assigned noncontextual truth-values because if it did, there would be an even number of bases with rays assigned the value 1, but an odd number of bases require this assignment.  This simple odd-even parity not only makes the proof of the Kochen-Specker theorem transparent, but it also makes these proofs particularly easy to find within a larger set using the tree-search algorithm.

In this case, we will start with some set of rays and bases, and search for subsets that give parity proofs of the Kochen-Specker theorem.  The program starts with one of these bases, and builds the tree by adding one basis at a time, generating new branches at each iteration.  The first step includes a branch for each basis in the set.  In a parity proof, every ray must appear in an even number of bases, and we exploit this fact in finding the allowed branches.  The program takes the list of bases for a given branch, and determines which rays appear in the largest odd number of those bases.  It then takes the first of these rays, and creates a new branch for each of the remaining (unchosen) bases in the set that contains that ray.  In this case, some additional logic is used to ensure that identical sets of bases are not built in multiple branches, which speeds up the search process considerably.  If at any stage in this process, there are an odd number of bases on a given branch, and every ray appears in an even number of them, then the set is a parity proof, and it is saved.  Once the full tree has been explored, a complete list of all parity proofs contained within the set is returned.

Even with the speedup of searching only for parity proofs, there are still many sets for which running this algorithm becomes untenable, but we can still use it to obtain some particular proofs, if not an exhaustive listing.  The program can be set to stop after finding some specified number of parity proofs within the set or after running for some specified duration, so that it does not run indefinitely.   We can also speed up the way the program {\it homes in} on a given proof by modifying the branch-selection algorithm, such that we choose in advance the number of rays in the desired proof, and how many bases each will appear in.

Using this method, we can {\it scan} a given set for parity proofs of varying sizes, which is how we found the many example proofs within the $60-75$ set of Chapter \ref{sec:600cell}, and the $60-105$ set of Chapter \ref{sec:q2Pauli}, as seen on the websites \cite{600cellWebsite},\cite{MainWebsite}.  These are by no means complete searches, but they at least give a sense of the variety that exists within the set.

\subsubsection{ID Search}

An ID, as we have discussed in detail in Chapter \ref{sec:IDs}, is a set of $M$ mutually commuting observables from the $N$-qubit Pauli group, whose overall product is $\pm I$ in the space of all $N$ qubits.  All possible unique IDs can be built using the tree-search algorithm, though only the smaller cases are computationally tenable.

In this case, we build the IDs one at a time from SQPs, adding additional branches at each iteration.  The branch-selection process is quite similar to the one used for the Parity Proof Search, but here it is anticommutations that we pair off, rather than rays.  Just as before, a complete list of IDs will be returned by the search.

The SQPs we use are ordered sets of single-qubit observables, whose order matters only with respect to the other SQPs in the set.  Once a complete ID is formed, the set mutually commutes, and so the order of the observables is irrelevant.  The key to building the ID is the mutual commutation, and since each nontrivial SQP contains pairs of single-qubit Pauli observables that anticommute, we need to choose our SQPs so that every anticommuting pair in one SQP is matched with an anticommuting pair in another SQP.  If all anticommutations are paired in this way, then the entire set mutually commutes, and we have an ID.

To search for ID$M^N$s, we begin by generating all of the possible nontrivial SQPs using $M$ elements.  The nontrivial SQPs for $M=3,4$ are shown in Table \ref{SQPs34}.  Next we construct a table which contains a listing of which pairs of elements in each SQP anticommute, which we use to determine the legal branchings at each iteration of the algorithm.  
\begin{table}[ht]
\centering
\subfloat[][]{
\begin{tabular}{c}
$Z$ \\
$X$ \\
$Y$ \\
\end{tabular}}
\qquad
\subfloat[][]{
\begin{tabular}{c}
$Z$ \\
$X$ \\
$Y$ \\
$I$ \\
\end{tabular}}
\qquad
\subfloat[][]{
\begin{tabular}{c}
$Z$ \\
$X$ \\
$I$ \\
$Y$ \\
\end{tabular}}
\qquad
\subfloat[][]{
\begin{tabular}{c}
$Z$ \\
$I$ \\
$X$ \\
$Y$ \\
\end{tabular}}
\qquad
\subfloat[][]{
\begin{tabular}{c}
$I$ \\
$Z$ \\
$X$ \\
$Y$ \\
\end{tabular}}
\qquad
\subfloat[][]{
\begin{tabular}{c}
$Z$ \\
$Z$ \\
$X$ \\
$X$ \\
\end{tabular}}
\qquad
\subfloat[][]{
\begin{tabular}{c}
$Z$ \\
$X$ \\
$Z$ \\
$X$ \\
\end{tabular}}
\qquad
\subfloat[][]{
\begin{tabular}{c}
$Z$ \\
$X$ \\
$X$ \\
$Z$ \\
\end{tabular}}
\caption[Unique SQPs]{The 1 unique nontrivial SQP for $M=3$, and 7 unique nontrivial SQPs for $M=4$.  All IDs for given $M$ must be built from the corresponding set of unique SQPs.  For $M=5$ there are 35 unique SQPs, for $M=6$ there are 155, for $M=7$ there are 721, and for $M=8$ there are 3,227.  It should now be clear why the exhaustive search becomes impossible as the number of unique SQPs grows larger.}\label{SQPs34}
\end{table}

We only need to use one permutation of each SQP for the search algorithm, since all permutations will contain identical pairs of anticommuting elements.  We also need not bother with trivial SQPs, since these never belong to critical IDs.  Finally, as we will show, no critical ID except for the ID$3^2_2$ contains more than one copy of any one SQP.  The reason for this is different for Odd SQPs than it is for Even ones.  For Even SQPs, two copies would share all of the same anticommuting pairs, and have product $+I$ in the space of those two qubits.  This means that both SQPs can always be removed, and the remaining set will still be an ID with the same sign, ergo the original ID was not critical.  For Odd SQPs, two copies again share the same anticommuting pairs.  In this case, all other SQPs can always be discarded along with some of the observables to reduce the set down to the ID$3^2_2$, ergo the original ID was not critical.

With these simplifications, we have performed exhaustive searches for all values of $N$ up to $M=5$, as well as the $N=5,6,7$ cases for $M=6$.  We have also used the scanning approach to find many IDs of $M\leq 8$ and $N\leq 16$, which can be reviewed on our website \cite{MainWebsite}.  With future access to a large cluster, we will be able to obtain results like these for some larger cases.

\subsubsection{CKS Search}

A Composite Kernel Structure (CKS), as we have discussed in detail in Chapter \ref{sec:Kernels}, is an assignment of the Odd SQPs within some set of Partial IDs to $N$ qubits, such that each Odd SQP in any ID is paired with an Odd SQP in another ID in the set.  We say the structure is critical if no subset of qubits and/or ID assignments can be deleted such that the remaining set is still a CKS.  As we have discussed, the Composite Kernel formed by assigning IDs to a critical CKS may not itself be critical, but that is a separate issue.

In this case, we build the CKS one Partial ID at a time, now choosing the branches by pairing off Odd SQPs, rather than pairing off rays or anticommutations as we have done above.  To search for an $N$-qubit CKS, we first build the generalized Partial IDs containing all permutations of the $N$ elements `O' and `I', with all even numbers of `O's $\leq N$.  The complete listing of generalized Partial IDs for $N=4$ is shown in Table \ref{genIDs4} for clarity, and the complete listing of critical CKSs for up to $N = 5$ is given in Tables \ref{CompStructs234} and \ref{CompStructs5}.
\begin{table}[ht]
\centering
\begin{tabular}{cccc}
O & O & O & O \\
O & O & I & I \\
O & I & O & I \\
O & I & I & O \\
I & O & O & I \\
I & O & I & O \\
I & I & O & O \\
\end{tabular}\caption[The 7 Generalized IDs for $N=4$ qubits]{The 7 generalized IDs for $N=4$ qubits.  These can be used to form the unique critical CKSs for $N=4$, as shown in Table \ref{CompStructs234}.}\label{genIDs4}
\end{table}

The complete search results for up to $N=7$ can be found on the website.

\subsection{Criticality Algorithms}

Here we will describe the algorithms we use to verify the criticality of the various structures we have discussed throughout this text.  By critical, we mean that the structure is minimal, in the sense that none of its constituent elements can be deleted to obtain a smaller structure of that class.  In each case, the constituent elements are slightly different, as are the conditions that define the structure.

\subsubsection{Basis-Criticality and Ray-Criticality}

Where we have discussed critical sets of rays and bases throughout this text, we have implicitly meant basis-criticality.  We have algorithms for both cases, and so we briefly define them here for clarity.  When we consider an $R-B$ KS set, we show that the set of rays and bases cannot be assigned noncontextual truth values, such that each basis contains exactly one ray assigned value 1, and all other rays assigned value 0.  If two rays are orthogonal, but the set contains no basis in which they both appear, then their orthogonality is effectively ignored because it is not a necessary part of the proof.  By considering only rays within complete bases to be effectively orthogonal, we are truly considering the set of complete measurement contexts necessary for the proof.  We then define a KS set to be basis-critical if none of the complete bases can be deleted such that the remaining set is still a KS set.  A basis-critical KS set is then a minimal set of measurement contexts that prove the KS theorem.

On the other hand, a ray-noncolorable set is said to be ray-critical if no rays can be deleted such that the remaining set of rays, with all orthogonalities taken into account, is ray-noncolorable.

In both cases, the algorithm is quite simple.  For a set of $B$ bases, the $B$ different subsets obtained by deleting one basis from the set are passed to the Basis-Colorability program, and if all $B$ cases are colorable, then the set is basis-critical.  For a set of $R$ rays, the $R$ different subsets obtained by deleting one ray from the set are passed to the Ray-Colorability program, and if all $R$ cases are colorable, then the set is ray-critical.

\subsubsection{ID Criticality}

An ID is critical if no subset of qubits and/or observables can be deleted such that the remaining set is a smaller ID.  The algorithm to check this deletes all possible combinations of the qubits and/or observables in the ID, and checks the set that remains in each case.  If none of these reduced sets is a smaller ID, then the original ID is critical.

\subsubsection{Kernel Criticality}

A single-ID Kernel is critical if the ID is critical, and so we use the previous algorithm to check these cases.

A Composite Kernel is critical if no combination of IDs and/or qubits can be deleted such that the remaining set is a smaller Kernel.  Technically, it is also possible to delete subsets of observables in order to obtain a smaller Kernel, but we avoid this complication by never building Composite Kernels with noncritical IDs that can be divided into the product (not the direct product) of smaller IDs.  On the other hand, we often build critical Composite Kernels from noncritical IDs that are the direct product of smaller IDs and/or trivial SQPs.

The algorithm to check for criticality deletes all possible combinations of qubits and IDs from the Composite Kernel and checks the set that remains in each case.  If none of these reduced sets is a smaller Kernel, then the original Composite Kernel is critical.

It should be noted that criticality can also be established through the use of Criticality Networks discussed in Chapter \ref{sec:Kernels}, but the above algorithm supersedes this method using brute force.

\subsubsection{Composite Kernel Structure Criticality}

The Composite Kernel obtained by assigning a given set of IDs to a CKS may not be critical, and has to be checked by the previous algorithm.  The CKS only determines how the Odd SQPs will be assigned within each ID such that the set will be a Composite Kernel.

We say that a CKS is critical if no combination of qubits and/or generalized IDs can be deleted such that the remaining set is a smaller CKS.  The algorithm to check this deletes every possible combination of generalized IDs and/or qubits from the CKS and checks the remaining set.  If none of these reduced sets is a CKS, then the original CKS is critical.

\subsection{Generation of Observable-based KS proofs, Rays, and Bases}

We have automated the general method described in Chapter \ref{sec:ObKSproofs} for generating an Observable-based KS proof from a Kernel, though this can typically be done quite easily by hand.

We have also automated the process of generating a set of rays from a set of IDs.  The rays are generated using the notation of Chapter \ref{sec:RBSets}, and all additional observables obtained by multiplying those in the ID are generated and assigned an eigenvalue.  It is then easy to determine which pairs of rays are orthogonal, and from there the complete set of bases can be built using another simple program.  It should be noted that for sets where two or more IDs share more than a few observables in common, direct enumeration of the bases becomes computationally untenable, but otherwise it is quite straightforward.

These sets can then be searched for parity proofs using the algorithm described above.  In the special case where no observable appears in more than two IDs, and no two IDs share more than one common observable, the complete set of bases and parity proofs can be easily determined using the method described in Chapter \ref{sec:RBSets}, which we have also automated.

We also have a program that calculates the rank-1 projectors for ID$M^N$ with $M=N+1$ in their explicit product basis form, by simultaneously diagonalizing the set of observables.

We have many other tools that we have developed for a variety of purposes, but we have contented ourselves by describing only the most important ones here.

%
%
%\end{document}

\section{Conclusions} \label{sec:Conclusions}

Throughout this text, we have introduced many varieties of geometric structures in the Hilbert space of $N$ qubits that prove the BKS and GHZ theorems.  We have also introduced the simpler elements with which this zoo of nonclassical structures can be constructed, and developed a method to identify all of these that exist within the $N$-qubit Pauli group.

These results flesh out a family of nonclassical structures of which only the simplest members were previously known.  Each is a unique nonclassical structure that requires only simple operations on $N$ qubits, and which has many obvious and potential applications in quantum information processing \cite{CabelloKSDim, ekert1991quantum, shor2000simple, bennett1992practical, bennett1996mixed, calderbank1997quantum, steane1996multiple, raussendorf2001one, cleve1998quantum, ekert1998quantum}.  Because the variety of these structures is so immense, we have presented the results by first presenting the elementary structures of central importance (IDs), and then showing how these can be used to build an enormous variety of different nonclassical structures (Observable-based KS proofs).  We hope that this text will be a useful resource from which an interested reader can learn to find and use these elements to construct any particular nonclassical structures they may need.

The structure of IDs is also intimately connected with the structure of entanglement, which we hope to explore in more detail in our future work.  Of particular interest are the eigenstates of critical ID$M^N$s for which $M<N+1$, since these rays contain internal degrees of freedom, meaning that they truly describe continuous families of rank-1 projectors which seem to share some type of common entanglement.  We ignored Null IDs throughout this work, but it seems likely we will need to include them for a full analysis of entanglement, since their eigenstates can be entangled while still admitting classical correlations.  Our hope is to identify some useful measures of entanglement for the pure states of the $N$-qubit Pauli group, but more work will be needed to determine if this goal can be accomplished.

We would also like to examine how this class of nonclassical structures and entangled states might be used in the design of new teleportation protocols and quantum computation algorithms.

Finally, we hope that the scope of what we have presented here will be a useful contribution towards a greater and more general understanding of nonclassical phenomena in quantum physics.
\newline

%\textbf{Acknowledgements:}  I thank first and foremost my adviser and friend P.K. Aravind, who first sparked my interest in the work of Bell and others on nonclassical physics, and who has provided me with insightful and supportive guidance throughout my tenure as his pupil.  Since my youth I have been inspired and fascinated by the illusions, patterns, and impossible figures conceived by M.C. Escher, so how fitting to find myself learning the physics of the impossible from a man with Escher prints and geometric baubles adorning his office.
%
%I also thank the other readers of my thesis , Walter Lawrence and David Cyganski,, with whom I have had interesting and enlightening conversations about the mysteries of quantum mechanics, and L. Ramdas Ram-Mohan who has graciously agreed to be part of my defense committee in lieu of Dr. Lawrence.
%
%Next I thank Mladen Pavi\v{c}i\'{c} and Norman Megill for their insights and collaboration at various stages of this research.
%
%Though we have never communicated directly, I would also like to acknowledge Asher Peres, David Mermin, Ad\'{a}n Cabello, and many others, whose work laid the foundations for what we have presented here.
%
%Finally I would like to thank the Physics Department of Worcester Polytechnic Institute for supporting this work.

\bibliographystyle{ieeetr}
%\bibliography{citation_database}
\bibliography{thesis.bbl}

\end{document}